\documentclass[%
  aip,
  11pt,
  reprint, 
  twocolumn,
  showpacs,
  preprintnumbers,
  amsmath,amsfonts,amssymb,
  jcp,
  floatfix]{revtex4-2}
\usepackage{graphicx}
\usepackage{dcolumn} % Align table columns on decimal point
\usepackage{bm} % bold math
\usepackage{bbold}
\usepackage{hhline}
\usepackage{ragged2e}
\usepackage{caption}
\usepackage{subcaption}
\usepackage{color}
\usepackage{natbib}
\usepackage{float}
\usepackage[normalem]{ulem}
\usepackage{dcolumn}
\usepackage{setspace}
\usepackage[utf8]{inputenc}
\usepackage{url}
\usepackage{xr-hyper}
\usepackage{hyperref}
\usepackage{multirow}
\usepackage{pgffor}
\hypersetup{
    colorlinks,
    linkcolor={blue},
    citecolor={blue},
    urlcolor={blue}
}
\usepackage[nolist]{acronym}

\begin{document}
%-------------------------------------------------------------------------------
\title{Variational augmentation of Gaussian continuum basis sets for calculating atomic higher harmonic generation spectra}
\author{Sai Vijay Bhaskar Mocherla}
\affiliation{Tata Institute of Fundamental Research Hyderabad, Hyderabad 500046, India}

\author{Raghunathan Ramakrishnan}
\email{ramakrishnan@tifrh.res.in}
\affiliation{Tata Institute of Fundamental Research Hyderabad, Hyderabad 500046, India}
\begin{abstract}
    We present a variational augmentation procedure to optimize the exponents of Gaussian continuum basis sets for simulating strong-field laser ionization phenomena such as higher harmonic generation (HHG) in atoms and ions using the time-dependent configuration interaction (TDCI) method. We report the distribution of the optimized exponents and discuss how efficiently the resulting basis functions span the variational space to describe the near-continuum states involved in HHG. Further, we calculated the higher harmonic spectra of three two-electron systems---H$^{-}$, He and Li$^{+}$---generated by 800nm driving laser-pulses with pulse-width of 54fs and peak intensities in the tunnel ionization regime of each system. We analyze the performance of these basis sets with an increasing number of higher angular momentum functions and show that up to $g$-type functions are required to obtain qualitatively accurate harmonic spectra. Additionally, we also comment on the impact of electron correlation on the HHG spectra. Finally, we show that by systematically augmenting additional shells we model the strong-field dynamics at higher laser peak intensities.
\end{abstract}
  
\maketitle
  
\section{Introduction} \label{sec:intro}
    The rapid technological advancements in laser physics over the last two decades have catalyzed progress in attosecond science, and have paved the way for ultrafast spectroscopies with an unprecedented time-resolution\cite{nisoli2017attosecond,calegari2016advances}. A crucial aspect in the development of this new frontier of ultrafast science\cite{schultz2013attosecond,maiuri2019ultrafast} has been the generation of table-top XUV and soft X-ray sources using higher harmonic generation (HHG)\cite{brabec2000intense,krausz2009attosecond}, a highly non-linear optical phenomenon in which coherent higher-order harmonics of the driving laser frequency are emitted. HHG has been observed in a variety of targets: gases\cite{wahlstrom1993high}, plasmas\cite{ganeev2007high}, liquids\cite{luu2018extreme}, and solids\cite{ghimire2019high}. In the case of atomic gases, HHG can be explained in terms of a semi-classical model with a three-step mechanism\cite{reed1992loss,corkum1993plasma}. Popularly known as the three-step model (3SM), in which the electron wave packet is postulated to (i) tunnel ionize in the presence of an intense laser field, (ii) accelerated in the continuum, and
    (iii) followed by recombination with its parent ion, finally, resulting in the emission of higher-order harmonics.

    In the case of atomic gases, only odd harmonics are observed due to the inversion symmetry of the target.  Further, the HHG spectrum is characterized by a rapid decline in the intensity of the first few harmonics, followed by a long plateau region that abruptly ends at a certain energy cutoff $E_\text{cutoff}=I_{p} + 3.17U_{p}$. This cutoff is known to be related to the ionization potential $I_{p}$ of the target and the ponderomotive energy $U_{p}$ or the maximum energy picked by the electron during its excursion in the continuum. The simple picture of gas-phase HHG given by 3SM aided in the development of a variety of innovative experiments such as measurement and control of attosecond electron dynamics\cite{hentschel2001attosecond,baltuvska2003attosecond,kraus2015measurement}, molecular orbital tomography\cite{itatani2004tomographic,mcfarland2008high} and higher-harmonic spectroscopy\cite{worner2010following,worner2011conical,krause2015angle}. Yet, over the years as the interest of the community has been shifting towards systems with increasing complexity\cite{calegari2014ultrafast,marangos2016development}, there is a growing need for theoretical methods that have better computational scaling with the increasing number of electrons. Additionally, the 3SM involves many approximations and a variety of numerical and grid based-methods have been developed\cite{milosevic2002numerical,schafer2009numerical,scrinzi2014time} to overcome the shortcomings of this model. But as these methods quickly become computationally unfeasible for larger multi-electron systems, many hybrid basis representations have been developed for strong-field calculations: numerical grids with Gaussian type orbital (GTO) functions\cite{hochstuhl2012time}, discrete-variable representation (DVR) with GTOs\cite{rescigno2005hybrid,yip2008hybrid,jones2016efficient} and B-spline functions mixed with GTOs\cite{marante2014hybrid,gonzalez2015describing}. Notably, there has been a growing interest to adapt time-dependent ab initio methods from quantum chemistry for simulating HHG\cite{coccia2021time}, owing to their ease of handling multi-center multi-electron integrals in complex systems.
    \begin{figure*}
    \centering
    \begin{minipage}[t]{.49\linewidth}
        {\includegraphics[width=.70\linewidth]{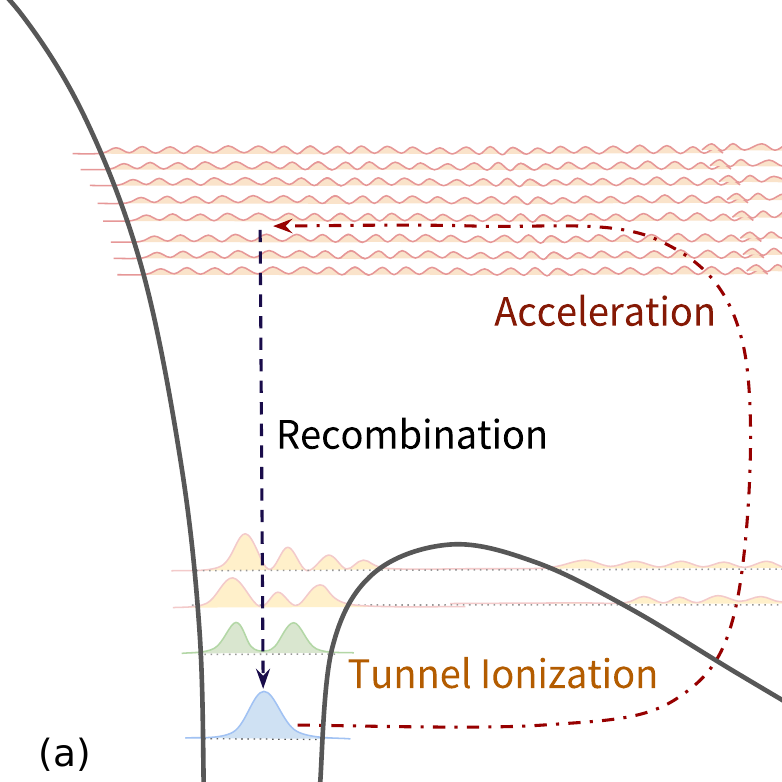}}%
        \refstepcounter{subfigure}
    \end{minipage}
    \hfill
    \begin{minipage}[b]{.49\linewidth}
        {\includegraphics[width=\linewidth]{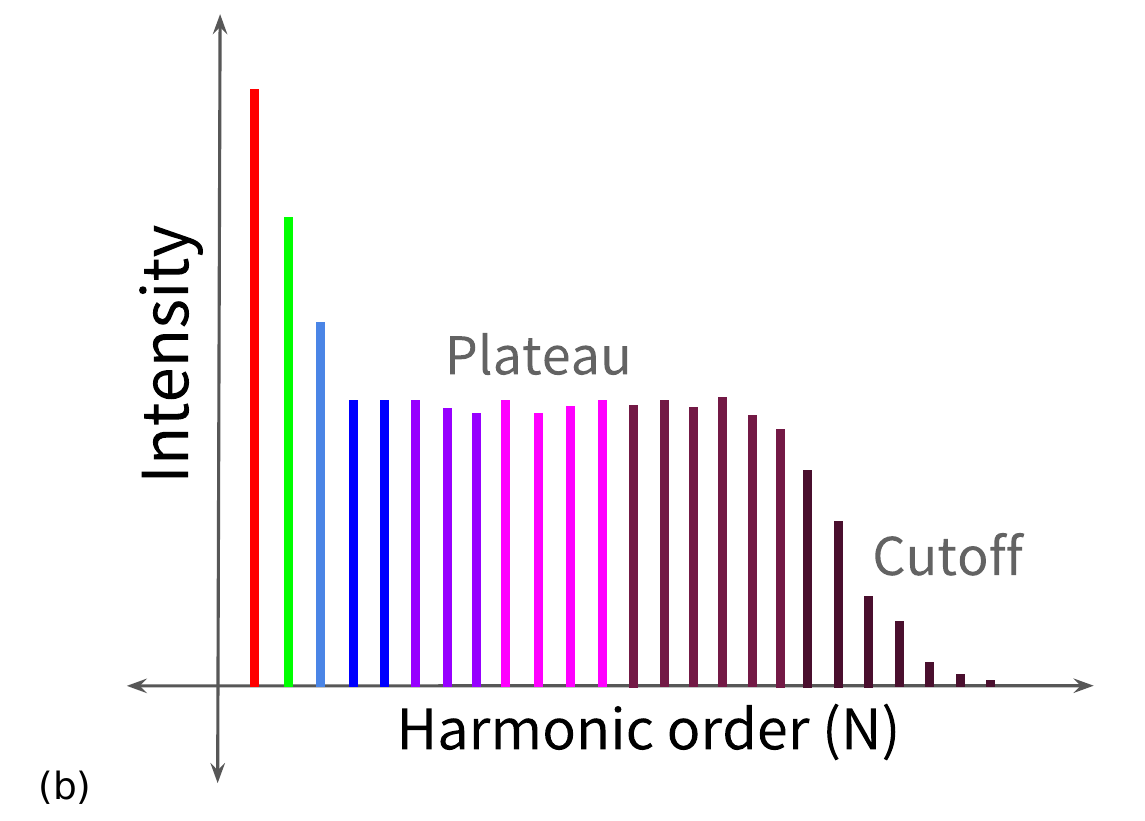}}\\
        \label{fig:spectra_illustion}
      \refstepcounter{subfigure}
        % {\includegraphics[width=.55\linewidth]{images/pulse_illustration.png}}
        % \refstepcounter{subfigure}
    \end{minipage}
    \caption{Schematic illustrations of (a) the three-step model (3SM) and, (b) the spectrum produced by HHG in atomic gases.}% and (c) the time-domain response of HHG, showing the bursts of attosecond pulse trains.}
      \label{fig:hhg_scheme}
   \end{figure*}
  
    Time-domain quantum chemical methods can be broadly classified into two categories\cite{li2020real}: wave function-based and orbital-based methods. Wave function-based methods include time-dependent configuration interaction (TDCI)\cite{klamroth2003laser,krause2007molecular,krause2005time}, time-dependent coupled cluster (TD-CC)\cite{huber2011explicitly,kvaal2012ab,sato2018communication}, time-dependent algebraic diagrammatic construction (TD-ADC)\cite{ruberti2014b,averbukh2018first}, and other multideterminant-based correlation methods. Orbital-based methods, on the other hand, mainly comprise time-dependent Hartree Fock (TD-HF)\cite{zanghellini2003mctdhf, remacle2007laser,nest2007time,haxton2011multiconfiguration,ulusoy2011correlated}, time-dependent density functional theory (TD-DFT)\cite{marques2004time,lopata2011modeling,tussupbayev2015comparison}, and their various adaptations. In this context, one of us has employed the TDCI method with atom-centered Gaussian basis sets to study a variety of charge transfer processes occurring in molecular junctions\cite{ramakrishnan2013electron,ramakrishnan2015charge,ramakrishnan2020charge}. Due to the simplicity of this formalism, there has been a growing interest in the theoretical attosecond science community to use it for simulating laser-driven electron dynamics\cite{saalfrank2020molecular,bedurke2021many}. While being computationally efficient, this approach suffers from a severe shortcoming in its inability to describe the motion of electrons far away from the nuclei in presence of strong laser fields. In spite of some of these limitations, Gaussian basis sets have been shown to be a promising alternative to grid-based methods for calculating HHG spectra\cite{luppi2012computation,luppi2013role,white2016computation}.

    In quantum chemical calculations, along with the preference of method, the choice of basis set plays a crucial role in determining the accuracy of results and the associated computational cost\cite{jensen2013atomic,hill2013gaussian}. Accordingly, many different families of basis sets have been developed with specific objectives in mind: Pople-style $k$-$lmn$G \cite{krishnan1980self,francl1982self,frisch1984self} basis sets, Dunning \textit{et al.}'s split-valence correlation-consistent cc-pV$X$Z basis sets \cite{dunning1989gaussian,woon1993gaussian,peterson2002accurate},  Alrichs \textit{et al.}'s Def2 basis sets \cite{schafer1992fully}, Koga \textit{et al.}'s segmented Sapporo basis sets, Roos \textit{et al.}'s ANO basis sets, and Jensen \textit{et al.}'s pc-$n$ basis sets. Most of these basis sets have been optimized for ground state molecular properties, and due to their inherent local nature, it becomes difficult to accurately describe the excited state properties of atoms and molecules. 
  
    However, a few decades ago Kaufmann \textit{et al.} presented a method for calculating the Rydberg and continuum states with pure L$^{2}$ methods\cite{kaufmann1989universal}. They showed that it was possible to generate an optimized sequence of Gaussian exponents whose linear combinations could ideally imitate Laguerre-Slater functions. More recently, Luppi and co-workers\cite{coccia2016gaussian,coccia2016optimal, labeye2018optimal} re-introduced this idea by combining K-functions with augmented Dunning basis sets\cite{peterson2002accurate} (abbreviated hereafter as aXZ) to get a balanced description of excited states for HHG calculations. They call them $n$-aug-cc-pV$X$Z+$N$K, where $n$ is the number of shells of diffuse functions, $X$ is the cardinal number of the basis set and $N$ is the number of K-functions added for each angular momentum up to X. Following their work, we propose a variational augmentation procedure to prepare hybrid basis sets by combining aXZ basis sets with continuum K-functions. 
  
  \section{Theoretical Methods} \label{sec:theory}
  \subsection{Ab initio Model}
  To simulate the response of atoms and molecules to intense optical laser fields, we begin by considering a Hamiltonian in the semi-classical dipole approximation\cite{craig1998molecular},
  \begin{equation}
    \hat{H}(\{\mathbf{r}, \mathbf{R}\}, t)  = \hat{H}_{0}(\mathbf{r}, \mathbf{R}) - \mathbf{E}(\mathbf{r}, t) \cdot \hat{\mu},
  \end{equation}
  where $\mathbf{E}(\mathbf{r}, t)$ is the electric field vector, $\hat{\mu}$ is the dipole operator and, 
  $\hat{H}_{0}(\mathbf{r}, \mathbf{R})$ is the total electronic Hamiltonian within the Born-Oppenheimer approximation, describing 
  the motion of $N$ electrons in the field of $M$ nuclei treated as point charges\cite{szabo1996modern}.  
  
  \begin{align}
    \hat{H}_{0} = &  \ \hat{T}_{e}(\mathbf{r}) + \hat{U}_{e,n}(\mathbf{r}, \mathbf{R}) + \hat{U}_{e,e}(\mathbf{r}) + \hat{U}_{n,n}(\mathbf{R}) \nonumber \\
    = &  -\sum_{i}^{N} \frac{1}{2} \vec{\nabla}_{i}^{2} - \sum_{i}^{N} \sum_{A}^{M} \frac{Z_{A}}{|r_{i}- R_{A}|} + 
    \sum_{i>j}^{N} \frac{1}{r_{i,j}} \nonumber \\
    & + \sum_{A>B}^{M} \frac{Z_{A}Z_{B}}{R_{A,B}}
  \end{align}
  
  Further, to track the laser-driven real-time electron dynamics of the system, we solve the time-dependent Schrödinger equation (TDSE)
  \begin{equation}
    i\frac{d}{dt}|\Psi(\mathbf{r},t)\rangle  = \hat{H}(\{\mathbf{r}, \mathbf{R}\}, t)  |\Psi(\mathbf{r},t)\rangle 
  \end{equation}
  to obtain the explicitly time-dependent electronic wave function $|\Psi(\mathbf{r},t)\rangle$.
  %, which in general, can be expanded in any given many-body basis as $|\Psi(\mathbf{r},t)\rangle = \sum_{k} c_{k}(t) |\varphi_{k}(\mathbf{r})\rangle$. 
  We then calculate the higher harmonic spectra $S(\omega)$ as the norm-squared Fourier transform of the dipole velocity
  \cite{baggesen2011dipole} expectation value,
  \begin{equation}
  S(\omega) = \left| \frac{1}{t_{f}- t_{i}} \int_{t_{i}}^{t_{f}}\hspace{-0.75em}dt \ \Big\langle \Psi(\mathbf{r},t) 
  \left| \dot{\mu}(t) \right| \Psi(\mathbf{r},t) \Big\rangle e^{i\omega t} \right|^{2},
  \end{equation}
  where $\langle \dot{\mu}(t)\rangle$ is approximated as a derivative of the dipole expectation value $d\langle\mu(t)\rangle / dt$ using the Ehrenfest theorem. It should be noted that this same approximation can not be used in the case of semi-classical models using the strong field approximation (SFA), as it could lead to qualitatively incorrect results\cite{granados2012invalidity}. 
  
  \subsection{Configuration interaction}\label{sec:configint}
  The variational strategy to construct a many-body ansatz $|\Psi\rangle$ is to start with an approximate wave function 
  $|\Phi_{0}\rangle$ and find a set of states $\{|\Phi_{k}\rangle\}$ linearly independent to $|\Phi_{0}\rangle$ such that 
  $|\Psi\rangle = c_{0}|\Phi_{0}\rangle + \sum_{} c_{k}|\Phi_{k}\rangle$,where $c_{k}$'s are parameters that are to be 
  determined along with the states $\{|\Phi_{k}\rangle\}$ as a non-linear variational problem\cite{henderson2019geminal}. 
  Here, we take the configuration interaction (CI) approach\cite{shavitt1977method,
  sherrill1999configuration}, where the wave function is constructed using the linear variational principle from a set of known
  $N$-electron states that are generated from a reference configuration. The CI wave function in terms of excitations from 
  a Hartree-Fock reference (a Slater determinant of spin orbitals $\{\chi_{i}: i=1,2,\cdots N\}$) can be written as
  \begin{equation}
    |\Psi_{\text{\tiny CI}}\rangle = c_{0} |\Phi_{0}^{}\rangle 
    + \underset{a}{\sum^{\text{occ}}}\underset{r}{\sum^{\text{vir}}} c_{a}^{r} |\Phi_{a}^{r}\rangle 
    + \underset{a \leq b}{\sum^{\text{occ}}}\hspace{.25em}\underset{r \leq s}{\sum^{\text{vir}}}
      c_{ab}^{rs}|\Phi_{ab}^{rs}\rangle + \cdots  
  \end{equation}
  where $|\Phi_{a}^{r}\rangle, |\Phi_{ab}^{rs}\rangle, \cdots $ represent the singly, doubly excited configurations and 
  so on, and $c_{a}^{r}, c_{ab}^{rs}, \cdots$ their corresponding amplitudes. Here, $a,b,\cdots$ and $r,s,\cdots$ denote 
  the indices going over occupied and virtual orbitals respectively. This approach tends to be formally exact or otherwise called full CI (FCI), as long as the expansion includes all possible N-electron states. In practice, the FCI approach can hardly be applied for $N>2$ electron systems for time-dependent strong field photoionization calculations, for two reasons: (1) the atomic basis sets required for an appropriate description of the continuum states tend to be  rather large compared to those used for ground-state electronic structure calculations, and (2) the length of the FCI expansion scales exponentially in terms of the number of occupied ($N_{o}$) and virtual ($N_{v}$) orbitals. 
  
  Here, we choose to expand the CI wave function in terms of spin-adapted configuration state functions (CSFs), which are linear combinations of Slater determinants (the bars over orbital indices indicate whether an $\alpha$ or $\beta$ type spatial orbital is involved in the excitation)\cite{szabo1996modern},
  \begin{eqnarray}
    |^{1}\Phi_{a}^{r} \rangle & =& \frac{1}{\sqrt{2}} \left( |\Phi_{a}^{r} \rangle 
                                  + |\Phi_{\bar{a}}^{\bar{r}}\rangle \right) \\ 
    |^{1}\Phi_{aa}^{rr} \rangle &=& |\Phi_{a\bar{a}}^{r\bar{r}} \rangle \\
    |^{1}\Phi_{aa}^{rs} \rangle &=& \frac{1}{\sqrt{2}} \left( |\Phi_{a\bar{a}}^{r\bar{s}} \rangle 
                                  + |\Phi_{a\bar{a}}^{s\bar{r}} \rangle \right) \\  
    |^{1}\Phi_{ab}^{rr} \rangle &=& \frac{1}{\sqrt{2}} \left( |\Phi_{a\bar{b}}^{r\bar{r}}\rangle 
                                  + |\Phi_{b\bar{a}}^{r\bar{r}}\rangle \right)  \\
    |^\text{A}\Phi_{ab}^{rs} \rangle &=& \frac{1}{\sqrt{12}} \left( 2|\Phi_{ab}^{rs}\rangle 
                                      + 2|\Phi_{\bar{a}\bar{b}}^{\bar{r}\bar{s}}\rangle 
                                      - |\Phi_{\bar{a}b}^{\bar{s}r}\rangle \right.\nonumber\\
                                    & & \hspace{2em}\left. - |\Phi_{a\bar{b}}^{s\bar{r}}\rangle 
                                    + |\Phi_{\bar{a}b}^{\bar{r}s}\rangle 
                                    + |\Phi_{a\bar{b}}^{r\bar{s}}\rangle \right) \\
    |^\text{B}\Phi_{ab}^{rs} \rangle &=& \frac{1}{2} \left(|\Phi_{\bar{a}b}^{\bar{s}r}\rangle 
                                      + |\Phi_{a\bar{b}}^{s\bar{r}}\rangle 
                                      + |\Phi_{\bar{a}b}^{\bar{r}s}\rangle + |\Phi_{a\bar{b}}^{r\bar{s}}\rangle \right). 
  \end{eqnarray}
  
  In the case of larger systems, to keep the problem computationally tractable, we truncate the expansion to include all 
  single and only a few selected active space double excitations, similar to the restricted-active-space CI (RASCI) 
  approach taken in quantum chemistry\cite{olsen1988determinant,hochstuhl2012time}.
  
  \subsection{Gaussian continuum basis sets}\label{sec:gbs}
  The one-electron wave functions ({\it i.e.,} molecular orbitals, MOs) in the Slater determinants (or CSFs) are expanded as linear combinations of atom-centered Gaussian basis functions, $\psi_{i}(\mathbf{r}) = \sum_{} d_{i,\mu} \chi_{\mu}$. Here, $\chi_{\mu}$ is a 
  Gaussian-type atomic-orbital (GTO) centered on an atom at ${\bf R}=(X,Y,Z)$, that can be represented in cartesian coordinates 
  as\cite{jensen2017introduction,helgaker2013molecular}
  \begin{equation}
    \chi_{\mu,l}(\mathbf{r}; {\bf R}) = N_{\alpha, l} (x-X)^{l_{x}} (y-Y)^{l_{y}} (z-Z)^{l_{z}}  e^{-\alpha_{\mu} |{\bf r}-{\bf R}|^{2}},
  \end{equation}
  where $N_{\alpha,l}$ is the normalization constant, $\alpha$ is the exponent that provides the radial extent of the 
  function, and $l_{x}, l_{y}, l_{z}$ are non-negative integers whose sum determines the type of the atomic orbital 
  (\textit{i.e.} azimuthal quantum number $l=l_{x}+l_{y}+l_{z}$). 
  
  Traditional Gaussian basis sets used in quantum chemistry cannot accurately describe the bound and continuum-excited states that are required to account for the complex electron dynamics involved in HHG. However, a long time ago Kaufmann \textit{et al.} proposed that a sequence of Gaussian functions obtained by maximizing their overlap with  Slater-type functions characterized by a constant exponent and a variable principal quantum number could describe Rydberg states and a discretized limit of the continuum ({\it i.e.,} the quasi-continuum) states\cite{kaufmann1989universal}. They 
  showed that a sequence of such Gaussian exponents $\{\alpha_{n,l}\}$ would have a general form,  
  \begin{equation}
    \alpha_{n,l} = \frac{\zeta^{2}}{4(a_{l}n + b_{l})^{2}}, \text{ where } n=1,2,3,\cdots 
    \label{eq:exp_sequence}
  \end{equation}
  where $a_{l}$ and $b_{l}$ are free parameters of the sequence. Luppi and co-workers\cite{coccia2016gaussian,coccia2016optimal,coccia2017ab, labeye2018optimal, coccia2019detecting}introduced the idea of combining augmented Dunning basis sets\cite{dunning1989gaussian,peterson2002accurate} containing very diffuse functions with Kaufmann functions (abbreviated as K-functions) to get a balanced description of all the states relevant for the HHG process. They denote these basis sets $n$-aug-cc-pV$X$Z+$N$K, where $n$ is the number of shells of diffuse functions, $X$ is the cardinal number of the selected Dunning basis set and $N$ is the number of K-functions added for each angular momentum up to X. Following their work, we take a slightly different approach. We only consider the singly augmented aug-cc-pV$X$Z (abbreviated hereafter as aXZ) basis sets and add $N-l*c$ continuum functions of each angular momentum $l$ upto $l_{max}$, where $c = \{0,1\}$ is just another parameter. Hereafter we refer to such a hybrid basis set as aXZ+($N$,$l_{max}$,$c$), and in Sec.[\ref{sec:results}] we discuss the criteria involved in optimizing their parameters to obtain a balanced performance in HHG simulations.  
  \subsection{Real-time laser-driven electron dynamics}
  To numerically solve the TDSE, we assume the time evolution to be discrete and ignore the time-dependency of $\hat{H}(t)$ 
  during an infinitesimal time-step $\delta t$ \textit{i.e.}, 
  \begin{equation}
  |\Psi(\mathbf{r},t+\delta t)\rangle = \hat{U}(t+\delta t, t)|\Psi(\mathbf{r}, t)\rangle
  \end{equation}
  where $\hat{U}(t+ \delta t, t) = \exp{-i \hat{H}(t+\delta t) \delta t}$ is the unitary time-evolution operator. Many integration schemes such as Crank-Nicholson\cite{wozniak2021systematic,wozniak2022effects}, Split-Operator\cite{klinkusch2009laser,luppi2012computation,luppi2013role} and the Runge-Kutta\cite{ramakrishnan2013electron,ramakrishnan2015charge,ramakrishnan2020charge} have been used for real-time propagation in TDCI simulations.
  % The Crank-Nicholson method considers the exponential in the propagator as the lowest-order Pade approximation 
  % \begin{equation}
  %   \hat{U}(t+ \delta t, t) = {\left( \mathbb{1} + i\frac{\delta t}{2}\hat{H}(t) \right)}^{-1} 
  %                             {\left( \mathbb{1} - i\frac{\delta t}{2}\hat{H}(t) \right)},
  % \end{equation}
  % where $\mathbb{1}$ is the identity matrix. The Split-Operator technique approximates it as
  % \begin{equation}
  %   {\Phi}(t+{\Delta} t)=\left[ \mathbb{W}^{\dagger} \cdot e^{i\hat{\mathbf{r}}\cdot\vec{\mathbf{E}}(t){\Delta}t} 
  %                               \cdot \mathbb{W} \right] e^{-i {\hat{\mathbf{H}}_{0} {\Delta}t}} \ {\Phi}(t),
  % \end{equation}
  % where $\mathbb{W}$ is a unitary transformation describing the change from dipole to CI eigenbasis. 
  In our calculations, we choose to use the explicit fourth-order Runge-Kutta (RK4) scheme\cite{hairer1993solving,press2007numerical} due to its good accuracy when using small stepsizes ($dt \approx 10^{-4} fs$) at a lower computational cost. For propagating $|\Psi(t)\rangle$ forward in time by $\delta t$, RK4 method involves the following
  intermediate steps:
  \begin{eqnarray}
    |y_{1}\rangle &=& -i \hat{H}(t+\delta t)|\Psi(t)\rangle \nonumber \\
    |y_{2}\rangle &=& -i \hat{H}(t+\delta t)\left[|\Psi(t)\rangle + \frac{1}{2}\delta t|y_{1}\rangle \right] \nonumber \\
    |y_{3}\rangle &=& -i \hat{H}(t+\delta t)\left[|\Psi(t)\rangle + \frac{1}{2}\delta t|y_{2}\rangle \right] \nonumber \\
    |y_{4}\rangle &=& -i \hat{H}(t+\delta t)\Big[|\Psi(t)\rangle + \delta t|y_{3}\rangle \Big] \nonumber 
  \end{eqnarray}
  and the final wave function is calculated as
  \begin{equation}
  |\Psi(t+\delta t)\rangle = |\Psi(t)\rangle + \frac{\delta t}{6} \Big[ |y_{1}\rangle + 2|y_{2}\rangle 
                          + 2|y_{3}\rangle + |y_{4}\rangle \Big] + \mathcal{O}({\delta t}^{5}).
  \end{equation}
  % \footnote{The RK4 method is known to have issues, where the norm is not preserved well for some systems and required much smaller timesteps. In those cases using the Crank-Nicolson method tends to be computationally cheaper and more effective when compared to the Split-Operator method.}
  
  In this work, we examined the HHG spectra produced by linearly polarized laser pulses with an electric field that oscillates as, 
  \begin{equation}
    \mathbf{E}(t) = \mathbf{E}_{0} f(t) \ e^{i\omega_{0} t + \phi}
  \end{equation} 
  where $\omega_{0}$, $\phi$, $\sigma$, and $f(t)$ are the carrier frequency, phase, full-width at half maximum (FWHM) and envelop function of the laser pulse. Here, the laser amplitude, $\mathbf{E}(t)$, reaches a maximum value of $\mathbf{E}_{0}$  at $t_{p}$. All the higher harmonic spectra presented here were computed for a cosine-squared ($\cos^{2}$) envelope defined as 
  \begin{equation}
    f(t) = \begin{cases}
      \cos^{2}(\frac{\pi}{2\sigma} (t - t_{p})) & |t- t_{p}| \leq \sigma,\\
       0 & \text{otherwise}
    \end{cases}.
  \end{equation}
  In our study, we have calculated the higher harmonic spectra by varying the peak laser intensity $\text{I}_{0}=\epsilon_0 c E_0^2/2$ as a control parameter to test the performance of our basis sets.

  \subsection{Finite lifetime models}\label{sec:lifetime_model}
  
  Despite adding optimized continuum basis functions, the incompleteness of the space spanned by the finite basis sets 
  presents many problems, as the CI states above the ionization threshold are simply discrete representations of the 
  continuum, \textit{i.e.} they act as pseudo-continuum states. Therefore, to avoid unphysical reflections of the 
  electronic wavepacket and to be able to treat ionization within the TDCI scheme, we apply the heuristic lifetime model 
  proposed by Klinkusch \textit{et al.}\cite{klinkusch2009laser}. Within this approach, all the CI eigenstates above the 
  ionization potential ($I_{p}$) are treated as non-stationary states, and their energies are adjusted as, 
  \begin{equation}
    E_{k}^{\text{CI}} \rightarrow E_{k}^{\text{CI}} - \frac{i}{2} \Gamma_{k}^{\text{CI}} \hspace{2em}  
    \forall \hspace{2em} E_{k}^{\text{CI}} \geq E_{0} + I_{p}.
  \end{equation}   
  In the imaginary term, $\Gamma_{k}^{\text{CI}}$ can be interpreted as the ionization rate for a state $k$, where the state is considered to be irreversibly depopulated by the laser field with a lifetime $\tau_{k}^{\text{CI}} = 1 / \Gamma_{k}^{\text{CI}}$. In the original heuristic model, the values of $\Gamma_{k}^{\text{CI}}$'s are calculated as weighted sums of CSF amplitudes with one-electron ionization rates ($\gamma_{r}$) of their corresponding virtual orbital. For example, the ionization rates for CIS and CISD eigenstates are given by  
  \begin{align}
  \Gamma_k^{\text{CIS}} = & \sum_{a}^{} \sum_{r}^{}\left|c_{a,k}^{r}\right|^2 \gamma_{r}^{k} \\
  \Gamma_k^{\text{CISD}} = & \sum_{a}^{} \sum_{r}^{} \left(\left|c_{a,k}^{r}\right|^2 \gamma_{r}^{k} + \left|c_{aa, k}^{rr}\right|^2 2 \gamma_{r}^{k}\right)  \nonumber \\
  &+\sum_{a}\sum_{r\leq s}\left|c_{aa, k}^{rs}\right|^2\left(\gamma_{r}^{k}+\gamma_{s}^{k}\right)
  +\sum_{a\leq b}\sum_{r}\left|c_{ab, k}^{rr}\right|^2 2 \gamma_{r}^{k}  \nonumber \\
  &+\sum_{a\leq b}\hspace{.15em}\sum_{r\leq s}\left(\left|^{\text{A}} c_{ab, k}^{rs}\right|^2 
  +\left|^{\text{B}} c_{ab, k}^{rs}\right|^2\right)\left(\gamma_{r}^{k}+\gamma_{s}^{k}\right).
  \end{align}
  
  Here, $\gamma_{r}^{k}$ is calculated by assuming semi-classical interpretation, where the electron in the virtual orbital $r$ is considered to have an escape velocity $v$ and therefore a kinetic energy $\varepsilon_{r} = \frac{1}{2} v_{r}^{2}$ such that the inverse lifetime is given by 
  \begin{equation}
    \gamma_{r} = \frac{1}{\mathcal{\tau}_{r}} = \Theta(\varepsilon_{r})\frac{\sqrt{2\varepsilon_{r}}}{d}
  \end{equation}
  where $\Theta(...)$ is the Heaviside function, and $d$ is a free parameter of the model that can be interpreted as the escape length traveled by the electron in time $\tau_{r}$.
  
  \subsection{Grid-based numerical calculations}
  As a numerical reference for our TDCI-based calculations, we performed grid-based calculations within the single-active-electron (SAE) approximation. The SAE Hamiltonian in the velocity is given by
  \begin{equation}
  \hat{H}_{\tiny\text{SAE}}(\mathbf{r},t) = -\frac{1}{2} \vec{\nabla}^{2} + V_{\tiny\text{SAE}}(\mathbf{r}) - i\vec{A}(t)\frac{\partial}{\partial z} + V_{\tiny\text{CAP}},
  \end{equation}
  where the time-dependence arises from the magnetic vector potential $\vec{A}(t)$ of the electric-field. Here we consider $V_{\tiny\text{SAE}}(\mathbf{r}) = V_\text{long} + V_\text{short}$ \cite{reiff2020single}, where $V_\text{long}$ is the long-range Coulomb term:
  \begin{equation}
      V_\text{long}(r) = - \frac{C_{0}}{r},
  \end{equation}
  and $V_\text{short}$ is a screened (Yukawa) short-range Coulomb potential 
  \begin{equation}
      V_\text{short}(r) = - \frac{Z_{c}e^{-cr}}{r}.
  \end{equation}
  Here, $C_{0}=Z-(N-1)$, where Z is the charge of the nucleus of an $N$-electron atom or ion, $Z_{c}=Z-C_{0}$ for the remaining charge and, c is a parameter used to fit the potential to approximately reproduces the ground state and first few excited states of the system. Additionally, to prevent any nonphysical reflections of the electron wave packet from the boundary regions of the grid we use a complex absorbing potential (CAP) $V_{\tiny\text{CAP}}$. We have used an absorbing potential derived by Manolopoulos\cite{manolopoulos2002derivation}, that has a physical parameter $k_\text{min}$ which corresponds the minimum energy $E_\text{min}=k_{\text{min}}^{2}/2$ at which absorption is needed, and an accuracy parameter $\delta$.

  \section{Computational details}\label{sec:comp}
  
  All the electronic structure calculations were performed using Psi4 (version 1.6)\cite{smith2020psi4} software package. We optimized the exponents of K-functions in aQZ+($N$,$l_{max}$,c) basis sets, by minimizing the ground-state energy as a function of their parameters $\{a_{l}, b_{l}\}$ as given in Eq.[\ref{eq:exp_sequence}]. Psi4's SCF program was used for all ground-state energy minimization calculations. The basis set parameters were optimized using the gradient-based, Broyden-Feltcher-Goldfarb-Shano (BFGS) algorithm\cite{nocedal1999numerical} (as implemented in Python's SciPy package\cite{2020SciPy-NMeth}) with a convergence criterion for gradient tolerance of 10$^{-6}$. While we obtained similar results with the Nelder--Mead algorithm\cite{nelder1965simplex}, for larger basis sets we observed a slower convergence. 
  
  Strong-field electron dynamics simulations were performed using the TDCI approach with in-house codes. First, the matrix elements of the many-body operators were evaluated on a CSF basis constructed using the one-electron wave functions obtained from an initial Hartree--Fock calculation. Then, the CI eigenenergies $\{E_{k}\}$ and their corresponding eigenvectors in the CSF basis were determined by diagonalizing the field-free many-body Hamiltonian matrix ${\bf H}_0$. For calculating the heuristic lifetimes for many-body eigenstates above the ionization threshold, we used the lifetime model described in the section [\ref{sec:lifetime_model}] with a free parameter scheme,   
    \begin{equation}
      d = \begin{cases}
          \text{E}_{0}/\omega_{0}^{2} & E_{0}+I_{p} \leq E_{k} \leq E_\text{cut} \\
          0.1 & E_{k} > E_\text{cut}
      \end{cases}
  \end{equation}
  where we choose the escape length to be equal to the semi-classical quiver amplitude of the electromagnetic field $\text{E}_{0}/\omega_{0}^{2}$, for all the states below $E_\text{cut}$ and, for the rest of the states above $E_\text{cut}$, we set it to a very small value. This improvisation over the original lifetime model allows better retention of contributions originating from the low-lying continuum states and limits those coming from the high-lying states\cite{coccia2016optimal}. All the HHG spectra calculated using the TDCI approach presented in this work were done incorporating the heuristic lifetime model (unless stated otherwise).
  
  For all the HHG simulations, a carrier frequency of $\omega_{0} = 1.550$eV ($\lambda_{0}=800$ nm) was chosen to correspond to a near-IR driving laser. The total pulse duration ($2\sigma$) was set to 53.4 fs (10 optical cycles) and the CI wave function was propagated for a total time of $T = 80 fs$ using the RK4 method with a finite time-step $\delta t = 10^{-4} fs$ (0.004134 a.u.) such that the maximum amplitude of the electric field $\text{E}_{0}$ occurs at $t_{p} = T/2$. As tunnel ionization (TI) is an essential prerequisite for generating higher-order harmonics, we have used different peak intensities $\text{I}_{0}$ that correspond to this regime of ionization. This was mainly determined using the Keldysh parameter,
  \begin{equation}
    \gamma = \sqrt{\frac{I_{p}}{2U_{p}}}
    \label{eq:keldysh_param}
  \end{equation}  
  where $I_{p}$ is the ionization energy of the system and $U_{p}= \text{E}_{0}^{2}/\omega_{0}^{2}$ is the ponderomotive energy picked up by the electron during an excursion away from the target\cite{keldysh1965ionization,popruzhenko2014keldysh}. TI is generally considered to be the dominant mechanism for ionization when $\gamma \lesssim 1$\cite{amini2019symphony}. All these physical parameters that are relevant to the HHG simulations have been reported in Table \ref{table:laser_params}. The HHG calculations within the SFA limit were done in Mathematica\cite{mathematica} using the RBSFA package\cite{pisanty2016electron}\cite{pisanty2020rbsfa} written in Wolfram language. 

  \renewcommand{\arraystretch}{1.2}
  \begin{table}[!ht]
    \centering
    \caption{The physical parameters used in TDCI simulations of
    HHG in He atom, Li$^{+}$ cation and H$^{-}$ anion:
  I$_{0}$ is the peak intensity (in $10^{14}$ W/cm$^{2}$), 
  $\gamma$ is the Keldysh parameter (Eq.\ref{eq:keldysh_param}), 
  $U_{p}$ is the ponderomotive energy (in E$_{h}$), 
  $E_\text{cut}=I_{p}+ 3.17U_{p}$ is the energy cutoff (in E$_{h}$), 
  and $R_\text{max}$ is the semi-classical estimate 
  for the maximum quiver amplitude (in bohr) of the electron.}
    \label{table:laser_params}
    \begin{tabular}{ccccccc}
    \hline
    \hline
      System & I$_{0}$  & $\gamma$ & $U_{p}$ & $E_{\text{cut}}$ & R$_{\text{max}}$ \\
      \hline 
  H$^{-}$  & 4.5$\times10^{12}$ & 1.18 & 0.0099 & 0.0590 & 6.98 \\ 
           & 9.5$\times10^{12}$ & 0.82 & 0.0209 & 0.0939 & 10.14 \\ 
           & 1.5$\times10^{13}$ & 0.65 & 0.0329 & 0.1321 & 12.75 \\ \hline
  He & 2.0$\times10^{14}$ & 1.01 & 0.4392 & 2.2959 & 46.55 \\ 
           & 3.0$\times10^{14}$ & 0.83 & 0.6588 & 2.9921 & 57.01 \\ 
           & 5.0$\times10^{14}$ & 0.64 & 1.0980 & 9.3933 & 73.59 \\ \hline
  Li$^{+}$ & 5.0$\times10^{14}$ & 1.13 & 1.0980 & 6.2605 & 73.59 \\ 
           & 7.0$\times10^{14}$ & 0.95 & 1.5373 & 7.6529 & 87.08 \\ 
           & 9.5$\times10^{14}$ & 0.82 & 2.0863 & 9.3933 & 101.44 \\ \hline
  
        \hline
    \end{tabular}
  \end{table}
  
  \section{Results and Discussions} \label{sec:results}  
  
  \subsection{Composite/Hybrid basis sets}
\begin{figure*}[ht!]
      \centering
      \begin{subfigure}[b]{0.48\textwidth}
          \centering
          \includegraphics[width=\textwidth]{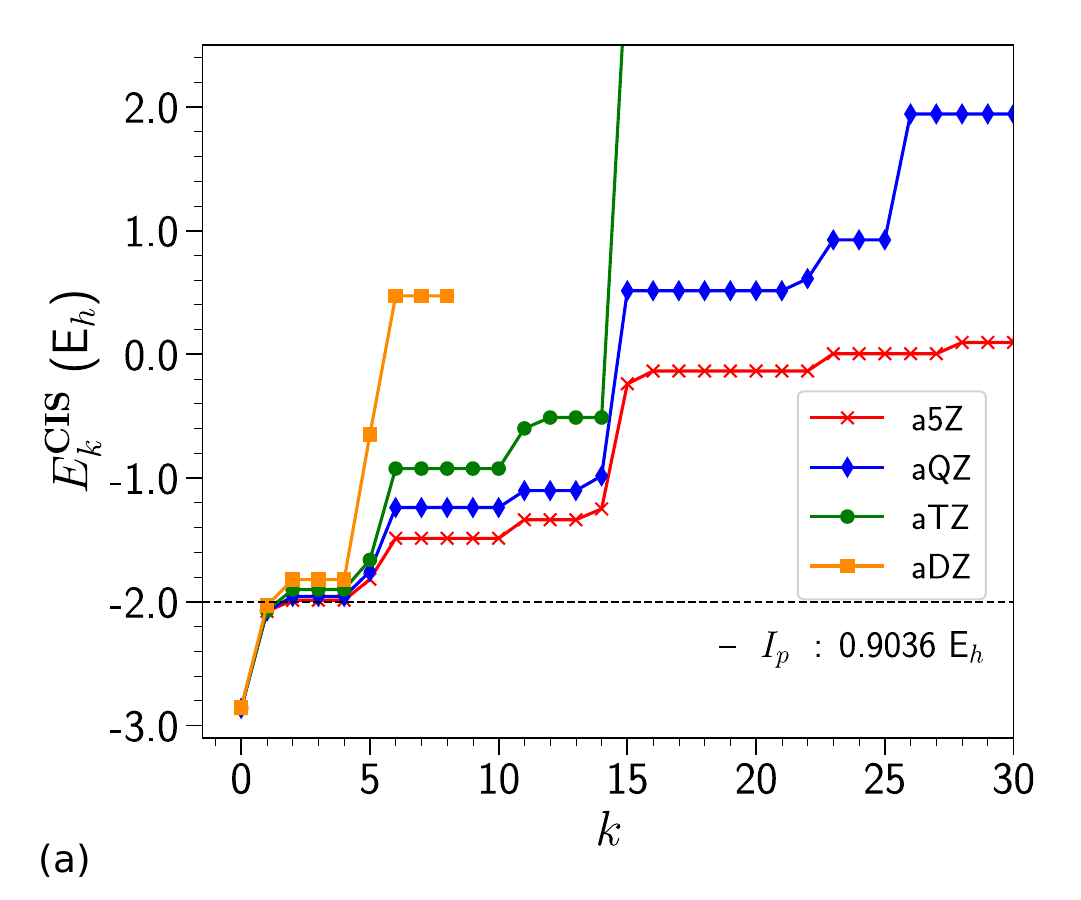}
          \refstepcounter{subfigure}
          \label{fig:he_axz_ecis_comp}
      \end{subfigure}
      \hfill
      \begin{subfigure}[b]{0.49\textwidth}
          \centering
          \includegraphics[width=\textwidth]{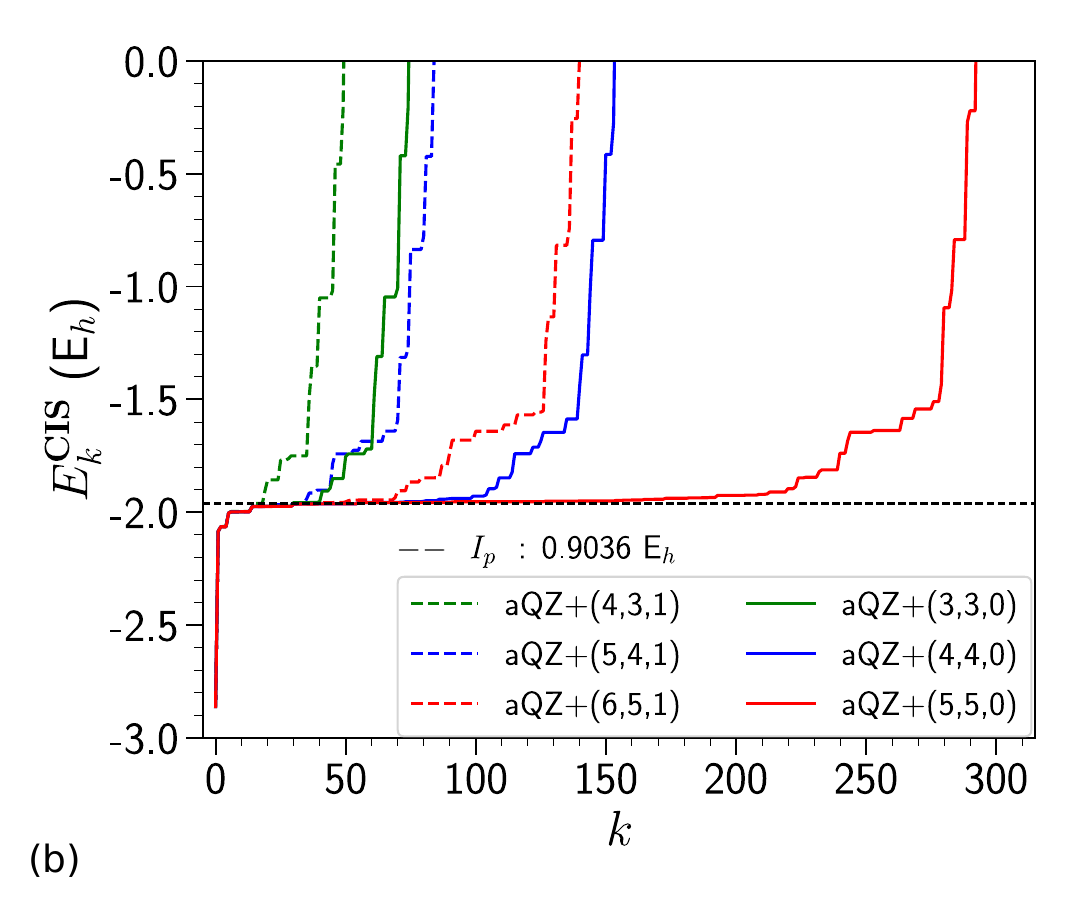}
          \refstepcounter{subfigure}
          \label{fig:he_aqz_plus_ecis_comp}
      \end{subfigure}
      \vspace{-2em}
      \caption{Distribution of CIS energies for Helium atom for (a) augmented Dunning (aXZ) and (b) aQZ+($N$,$l$,$c$) basis sets.}
      \label{fig:he_ecis_comp}
    \end{figure*}

  We begin by discussing the properties of the variationally augmented Gaussian continuum basis sets. In Fig. [\ref{fig:he_axz_ecis_comp}], we compare the energy distribution of the first few CIS states calculated with different aXZ basis sets. While all of them have the same number of states near the ionization threshold, $X=$ Q and 5 also include pseudo continuum states relevant to HHG. This can also be seen in the plots of HHG spectra compared in Fig. [\ref{fig:he_axz_3E14_hhg_comp}], where the spectra generated with $X=$ Q and 5 show features of several higher harmonics above 10 that are absent in the case of $X=$ D or T. 

  \renewcommand{\arraystretch}{1.2}
  \begin{table}[ht!]
    \centering
    \caption{The number of cartesian ($N_\text{crt}$) Gaussian functions, and the highest angular momentum quantum number
    ($l_\text{max}$) of the basis function present in different basis sets for Helium atom. The corresponding number of bound and continuum CIS states, along with the energy maximum $\varepsilon_{max}$ (in $\text{E}_{h}$ Hartree) of the one-electron state are presented.}
    \label{table:cis_spectra}
    \begin{tabular}{llccccc}
    \hline\hline
      System & Basis Sets & $l_{\rm max}$ & N$_{\rm cart}$ & N$_\text{bound}$ & N$_\text{cont.}$ & $\varepsilon_{\rm max}$ \\ \hline
      H$^{-}$ & aTZ           & 2 &   25  &    6  &   19 & 5.84  \\
              & aQZ           & 3 &   55  &    6  &   49 & 28.60 \\ 
              & a5Z           & 4 &  105  &    6  &   99 & 62.54 \\ \cline{2-7}  
             % & aQZ+(3,2,1)  & 3 &  70 &  17 & 53 & 30.15 \\
             % & aQZ+(4,3,1)  & 3 &  89 &  36 & 53 & 31.87 \\
             % & aQZ+(5,4,1)  & 4 & 124 &  65 & 59 & 36.15 \\
             % & aQZ+(3,4,0)  & 4 & 159 &  91 & 68 & 36.75 \\
             % & aQZ+(2,2,0)  & 3 &  74 &  21 & 53 & 31.86 \\
              & aQZ+(3,3,0)  & 3 & 114 &  61 & 53 & 31.96 \\
              & aQZ+(4,4,0)  & 4 & 194 & 126 & 68 & 37.02 \\ \hline 
      He  % & aDZ          & 1 &   9  &   6  &  3  &   3.02 \\ 
          & aTZ          & 2 &  25  &  15  & 10  &  10.72 \\
          & aQZ          & 3 &  55  &  15  & 40  &  41.57 \\
          & a5Z          & 4 & 105  &  23  & 82  & 145.22 \\ \cline {2-7}
          & aQZ+(4,3,1)  & 3 &  90  &  50  & 40  & 45.56 \\
          & aQZ+(5,4,1)  & 4 & 125  &  84  & 41  & 45.86 \\
          & aQZ+(6,5,1)  & 5 & 181  & 140  & 41  & 45.88 \\
          & aQZ+(3,3,0)  & 3 & 115  &  75  & 40  & 45.59 \\
          % & aQZ+(1,4,0)  & 4 &  90  &  50  & 40  & 45.73 \\
          % & aQZ+(2,4,0)  & 4 & 125  &  84  & 41  & 45.86 \\
          % & aQZ+(3,4,0)  & 4 & 160  & 119  & 41  & 45.88 \\
          & aQZ+(4,4,0)  & 4 & 194  & 154  & 40  & 45.21 \\
          % & aQZ+(5,4,0)  & 4 & 228  & 188  & 40  & 45.23 \\
          & aQZ+(5,5,0)  & 5 & 335  & 293  & 40  & 45.23 \\ \hline 
      Li$^{+}$ & aTZ          & 2 & 55  &   54 & 1 & 6.26 \\     
               & aQZ          & 3 & 104  &  103 & 1 & 12.41 \\     
               & a5Z          & 4 & 181  &  176 & 5 & 23.41 \\ \cline{2-7}     
               % & aQZ+(3,3,0)  & 3 &  159  &  157 & 2 & 16.90 \\     
               & aQZ+(4,4,0)  & 4 &  234  &  232 & 2 & 17.02 \\     
               & aQZ+(5,4,0)  & 4 &  268  &  266 & 2 & 17.02 \\     
               & aQZ+(6,4,0)  & 4 &  302  &  300 & 2 & 17.02 \\ \hline     
               % & aQZ+(7,4,0)  & 4 &  336  &  334 & 2 & 17.03 \\ 
               % & aQZ+(5,5,0)  & 5 &  367  &  365 & 2 & 17.02 \\ \hline          
        %% commented out %% 
        % Helium basis set data
        %  DZ &   5 & 1 &  5 (100\%) &  0(0\%)   &   2.52 \\ 
        %  TZ &  15 & 2 &  5  (33\%) & 10(67\%)  &   8.34 \\ 
        %  QZ &  35 & 3 &  6  (17\%) & 29(83\%)  &  34.35 \\ 
        %  5Z &  70 & 4 &  6  (9\%) & 64(91\%)   & 141.78 \\ \hline
        % aDZ &   9 & 1 &  6  (67\%) &  3 (33\%) &   3.02 \\ 
        % aTZ &  25 & 2 & 15  (60\%) & 10 (40\%) &  10.72 \\ 
        % aQZ &  55 & 3 & 15  (27\%) & 40 (73\%) &  41.57 \\ 
        % a5Z & 105 & 4 & 23  (22\%) & 82 (78\%) & 145.22 \\ \hline 
        % d-aDZ & 13 & 1 & 10 (77\%) & 3(23\%)   &   3.09 \\ 
        % d-aTZ & 35 & 2 & 25 (71\%) & 10(29\%)  &  11.33 \\ 
        % d-aQZ & 75 & 3 & 35 (47\%) & 40(53\%)  &  43.11 \\ 
        % d-a5Z & 140 & 4 & 50 (36\%) & 90(64\%) & 145.93 \\ \hline 
     \end{tabular}
  \end{table}
  
\begin{figure}
    \includegraphics[width=.46\textwidth]{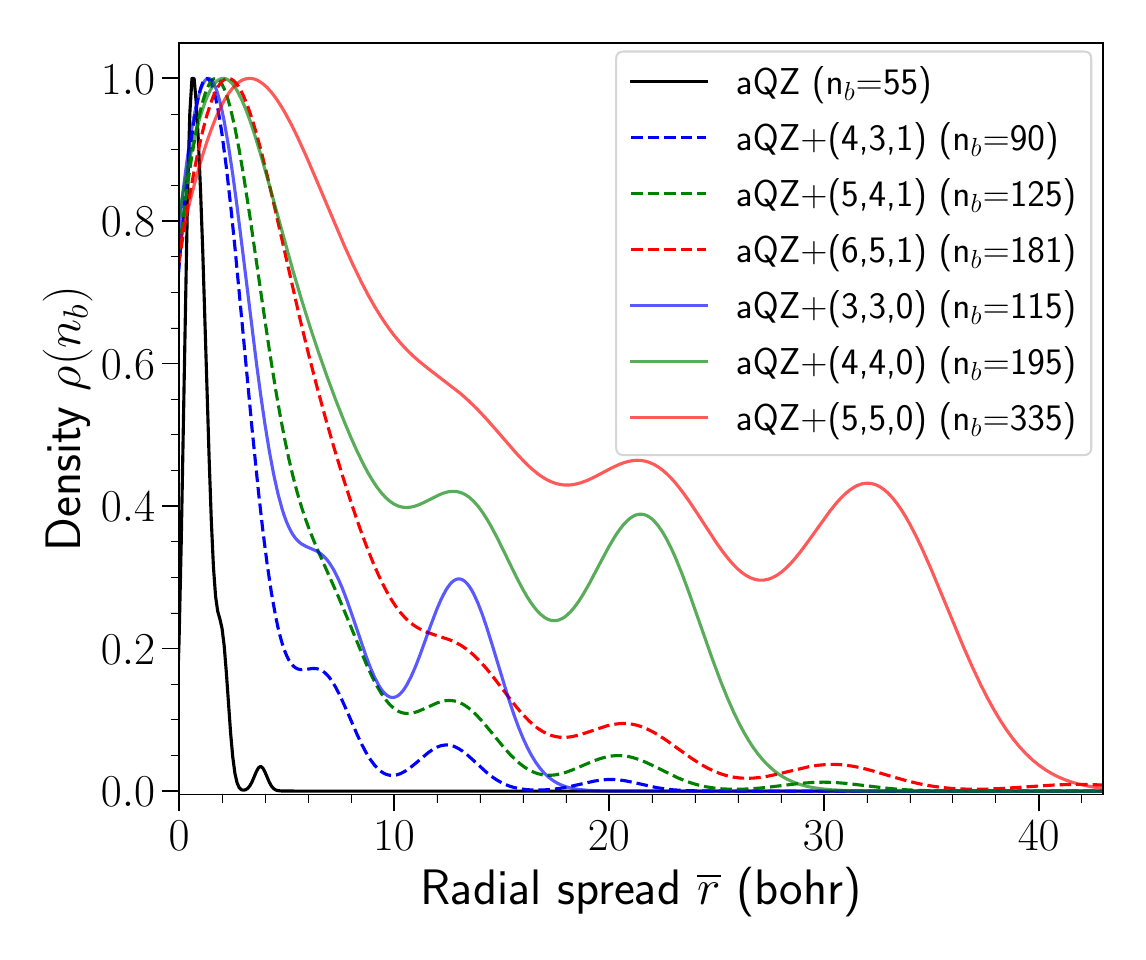}
    \caption{Density of basis functions $\rho(n_{bf})$ as function of their average radial spread $\overline{r}=\sqrt{\ln{2}}\alpha^{-1/2}$.}
    \label{fig:he_aqz_nlc_dist}
  \end{figure}

  However, even the largest basis set fails to capture the characteristic features of an HHG spectrum: a distinct plateau region, followed by a sharp cutoff. To accurately model these qualitative features the description of electronic states around the ionization threshold has to be improved. This can be done by systematically incorporating uncontracted K-functions to a principal basis set such as aXZ\cite{coccia2016gaussian}. The scheme used in this work for preparing such hybrid basis sets is described in Sec.[\ref{sec:gbs}]. In essence, the fitness of a basis set to simulate the strong-field dynamics up to a certain I$_{0}$ could be explained in terms of its ability to spatially span the region up to R$_\text{max}$. To examine the spatial composition of various basis sets, we calculated the density of basis functions $\rho(n_{bf})$ as a function of their average radial spread ($r_\text{avg} = \sqrt{\ln{2}}\alpha^{-1/2}$). Fig. [\ref{fig:he_aqz_nlc_dist}] shows that $c=0$ basis sets afford a higher proportion of diffuse functions, and with increasing $N$ the region covered by the basis set also expands. Therefore, to optimize a basis set for a range of laser intensities it is enough to find a composition of the aQZ+(N,l,0) basis set that spans the extended region in space that is required for strong-field electron dynamics simulations. 
  % \textcolor{blue}{TO DO: add some discussion the main problem with L2-based methods for strong-field calculations, and discuss the energy spectrum in this context. Further, provide more details from MOTECC book chapter, and discuss the strategies to improving the standard Gaussian basis sets with Kaufmann functions, in this context, discuss the table[2] and density plots. Finally, end with the recommendation that aQZ+(N,l,0) may be better suited, and give reasons.}
  
  \subsection{HHG spectra with aQZ+(N,l,0)}
  \begin{figure*}
      \centering
      \begin{subfigure}[b]{.49\textwidth}
          \centering
          \includegraphics[width=\textwidth]{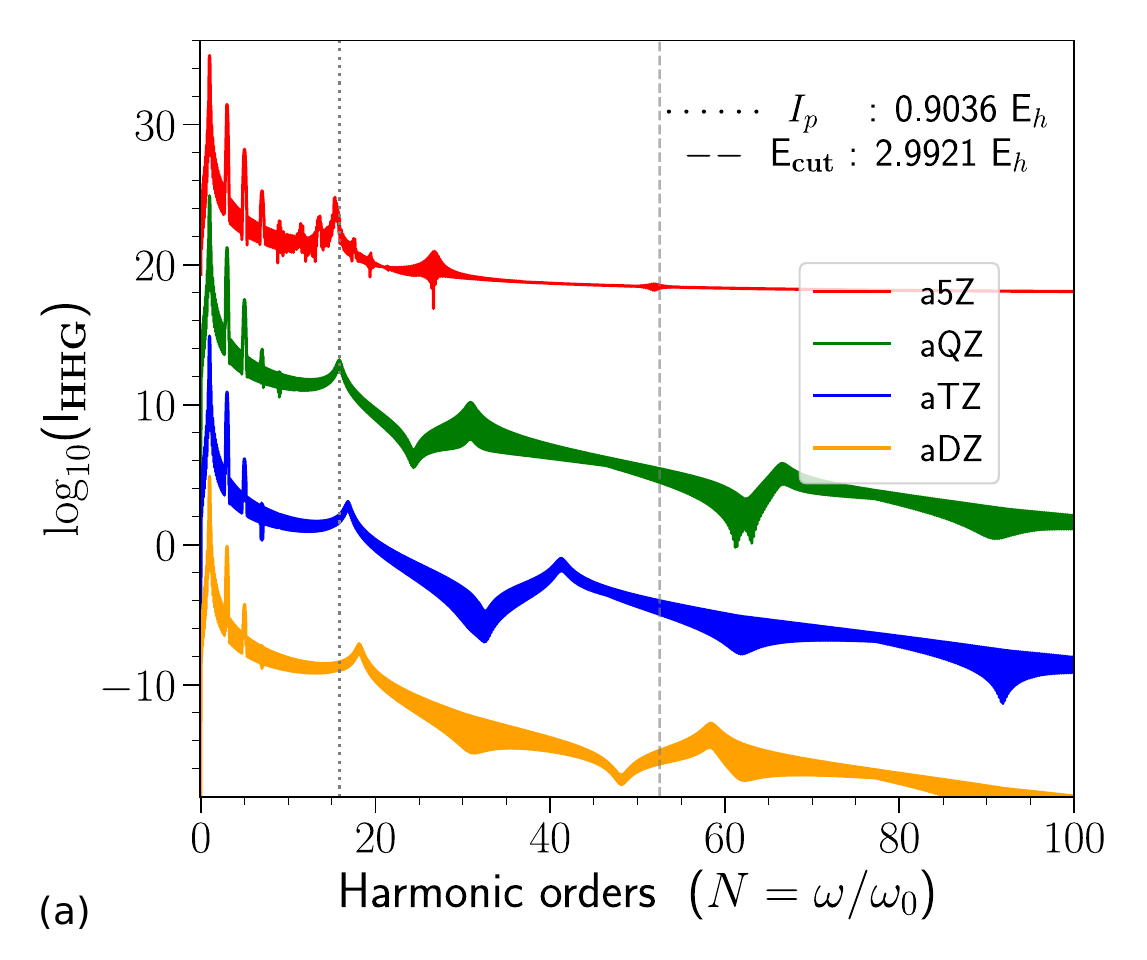}
          \refstepcounter{subfigure}
          \label{fig:he_axz_3E14_hhg_comp}
      \end{subfigure}
      \hfill
      \begin{subfigure}[b]{.49\textwidth}
          \includegraphics[width=\textwidth]{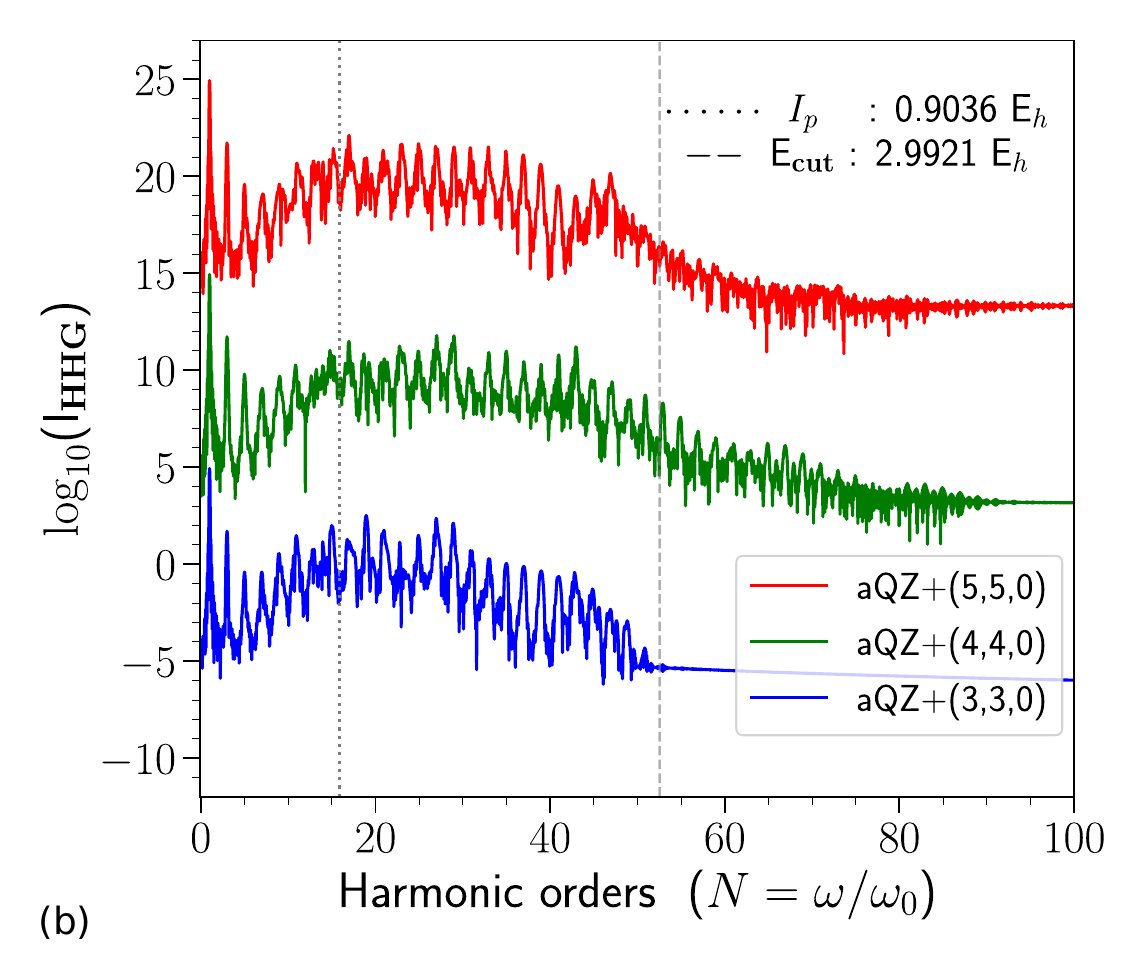}
          \refstepcounter{subfigure}
          \label{fig:he_aqz_nn0_3E14_hhg_comp}
      \end{subfigure}
      \vspace{-2em}
      \caption{Comparison of HHG spectra of Helium atom generated by pulse with $\text{I}_{0} = 3\times10^{14}$W/cm$^{2}$, calculated with TD-CIS with (a) aXZ basis sets and (b)  aQZ+($N$,$l_{max}$,0) basis sets where $l_{max} = N$. The spectra obtained with different basis sets have been systematically upshifted by +10, for the sake of clarity in comparison.}
      \label{fig:he_basisset_hhg_comp}
  \end{figure*}
  Now, we consider the HHG spectra of Helium atom obtained for three different peak laser intensities, $\text{I}_{0} = 2,3,5 \times10^{14} \text{ W/cm}^{2}$ whose parameters have been presented in Table[\ref{table:laser_params}]. The spectra calculated using TD-CIS with aQZ+($N$,$l_{max}$,0) basis sets for different $l_{max}$ values where $N=l_{max}$ are presented in Fig. [\ref{fig:he_aqz_nn0_3E14_hhg_comp}]. For the lowest intensity ($2\times10^{14} \text{ W/cm}^{2}$), all three basis sets, capture the cutoff behavior at the 41st harmonic and the intensity dip in the plateau region around the 37th harmonic. But for higher intensities ($\{2, 3\}\times10^{14} \text{ W/cm}^{2}$), the smaller aQZ+(3,3,0) basis set fails to capture the extended plateau region and predicts an incorrect harmonic cutoff. We also note that only minor differences are observed between the aQZ+(4,4,0) and aQZ+(5,5,0). This suggests that adding diffuse functions beyond $l=4$ does not significantly improve the HHG spectra. Further, to understand how the basis set composition varies with $N$, we calculated the HHG spectra of aQZ+($N$,4,0) basis sets for $N=\{1,2,3,4,5\}$ and the results are presented in Fig. []. We found that increasing the number of shells added per each angular momentum improves the number of harmonic orders that are recovered.

\begin{figure}
    \centering
    \includegraphics[width=.49\textwidth]{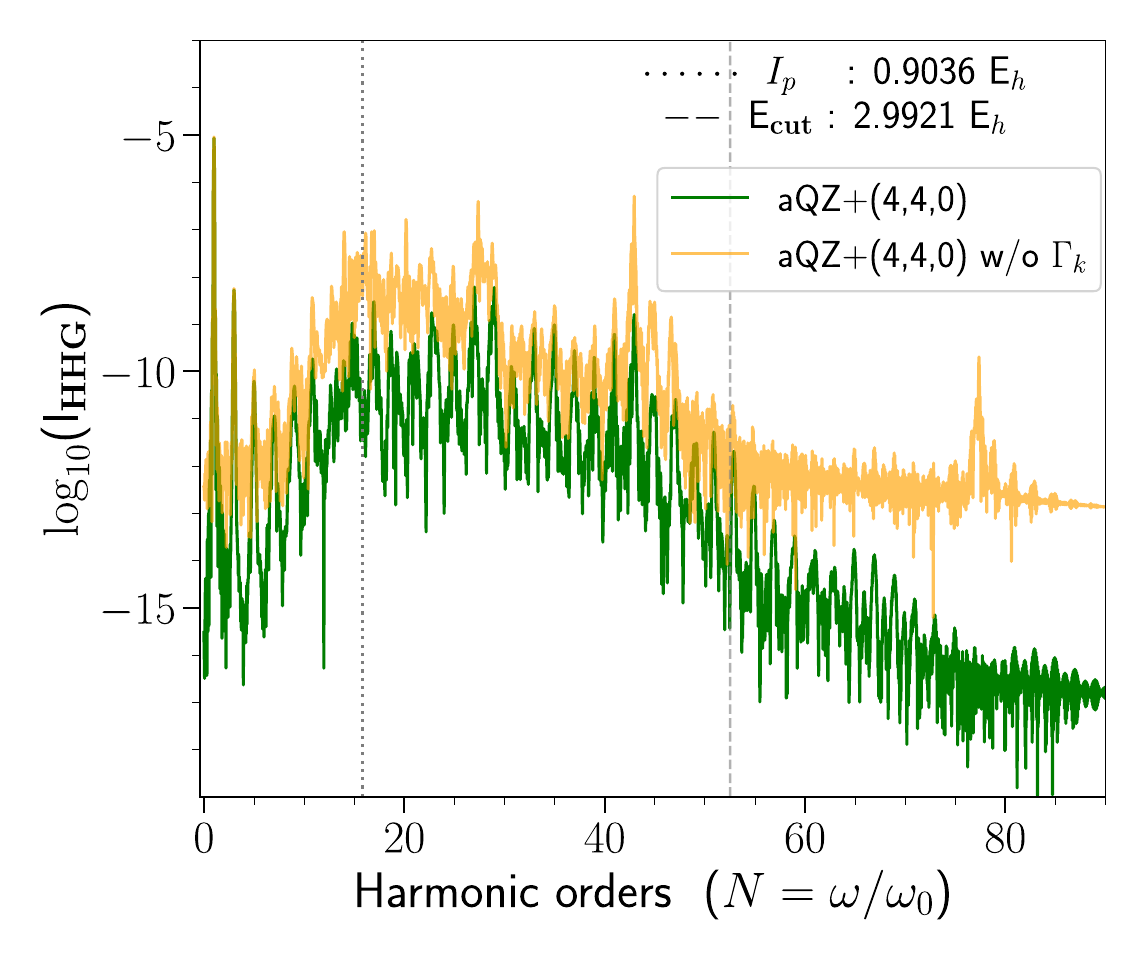}
    \caption{Comparison of HHG spectra of Helium atom generated by a laser pulse with $\text{I}_{0} = 3\times10^{14}$W/cm$^{2}$, calculated using TD-CIS/aQZ+(4,4,0) with and without heuristic lifetimes.}
    \label{fig:he_3E14_ltcomp_hhg}
  \end{figure}

  In Fig. [\ref{fig:he_3E14_ltcomp_hhg}], we compare the HHG spectra calculated using TD-CIS/aQZ+(4,4,0) for a Helium atom, both with and without lifetimes, for a peak laser intensity of I$_{0} = 3 \times 10^{14} \text{ W/cm}^{2}$. It is evident that the addition of finite lifetimes to states above the $I_{p}$ improves the peak-to-peak resolution and gives an accurate cutoff behavior. In particular, it removes the spurious interference effects originating from unphysical recombination events and reduces the overall background signal. This is due to the fact, that heuristic lifetime models account for possible ionization losses by treating the unbound states above $I_{p}$ as non-stationary and thereby eliminate any artifactual contributions to the HHG spectra. Additionally, in our calculations, we employ a modified version of the heuristic lifetime model\cite{coccia2016gaussian} proposed by Luppi \textit{et al.}, which provides a better estimate of the overall ionization rates compared to the original model\cite{klinkusch2009laser}. 
  % \textcolor{blue}{TO DO: Improve the lifetime comparison analysis and compare different models with TD-CIS Fig. [\ref{fig:he_3E14_methodcomp_hhg}]. Specifically, the qualitative aspects: cut-off, features in the plateau, and a possible reason for the high variation of background signal across different methods. These can be discussed based on the levels of approximation/ level of detail of the electronic wavefunction, a similar analysis to what was provided by Head-Gordan and Luppi.}   
  
  \begin{figure}
      \centering
      \includegraphics[width=.50\textwidth]{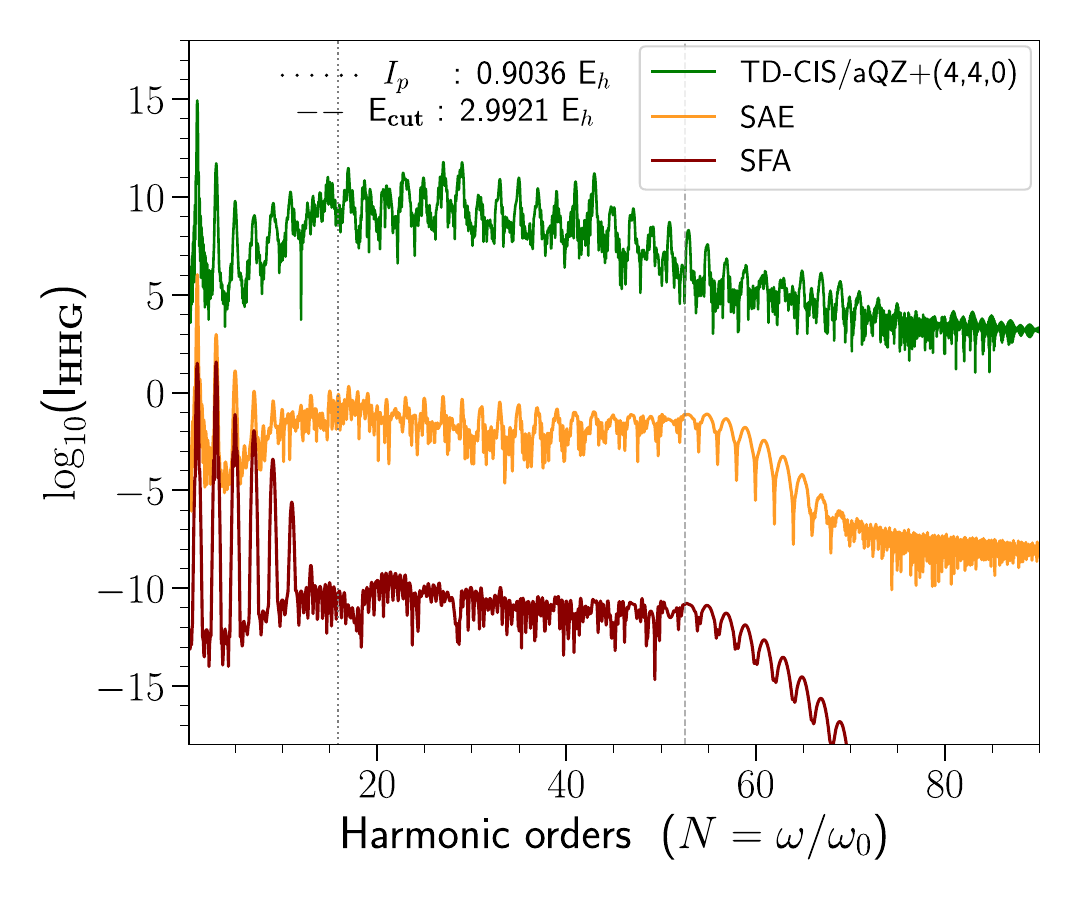}
      \caption{Comparison of HHG spectra of Helium atom generated by a laser pulse with $\text{I}_{0} = 3\times10^{14}$W/cm$^{2}$, calculated using TD-CIS/aQZ+(4,4,0) approach, grid-based numerical method with SAE approximation and Lewenstein's model based on SFA.}
      \label{fig:he_3E14_methodcomp_hhg}
  \end{figure}

  \begin{figure*}
      \centering
      \begin{subfigure}[b]{.49\textwidth}
          \includegraphics[width=\textwidth]{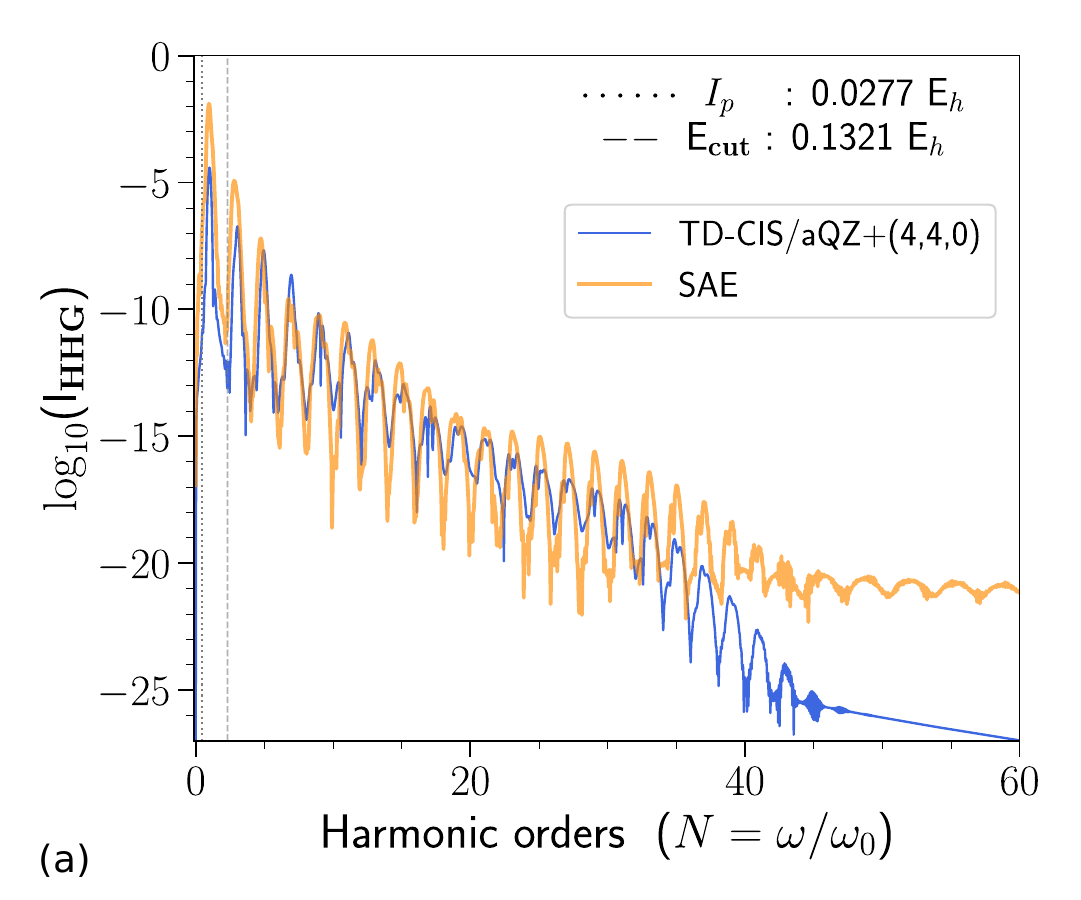}
          \refstepcounter{subfigure}
          \label{fig:h_15E12_sfa_comp_hhg}
      \end{subfigure}
      \hfill
      \begin{subfigure}[b]{.49\textwidth}
          \includegraphics[width=.99\textwidth]{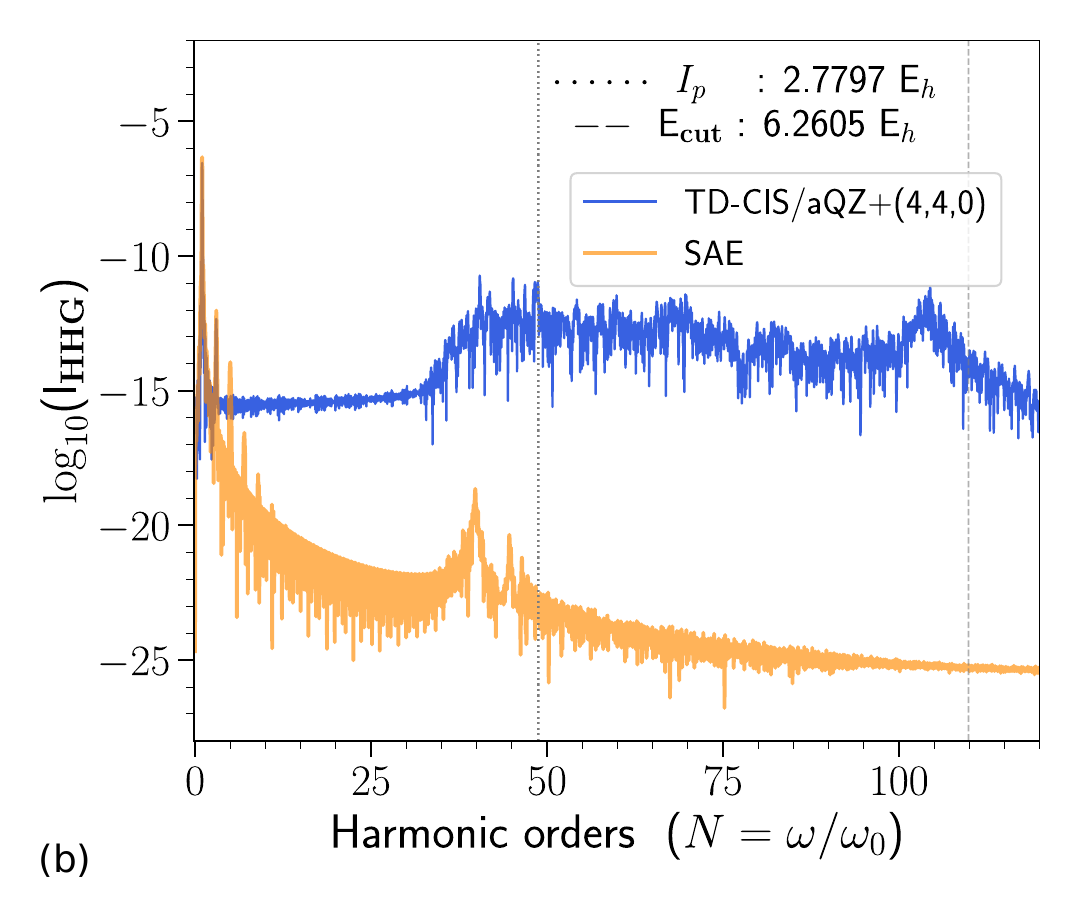}
          \refstepcounter{subfigure}
          \label{fig:li_aqz_nn0_5E14_hhg_comp}
      \end{subfigure}
    \vspace{-2em}  
    \caption{Comparison of HHG spectra of (a) Hydride ion (H$^{-}$) generated by a laser pulse with $\text{I}_{0} = 15\times 10^{12}$ W/cm$^{2}$ and (b) Lithium cation (Li$^{+}$) generated by a laser pulse with $\text{I}_{0} = 5\times 10^{14}$ W/cm$^{2}$, calculated using a numerical grid-based method within SAE approximation  and TD-CIS/aQZ+(4,4,0) with heuristic lifetime.}
    \label{fig:ions_aqz_nn0_hhg_comp}
  \end{figure*}

  \subsection{Effect of doubles on HHG spectra}
  
  \begin{figure}
    \centering
    \begin{subfigure}[b]{0.47\textwidth}  
        \includegraphics[width=\textwidth]{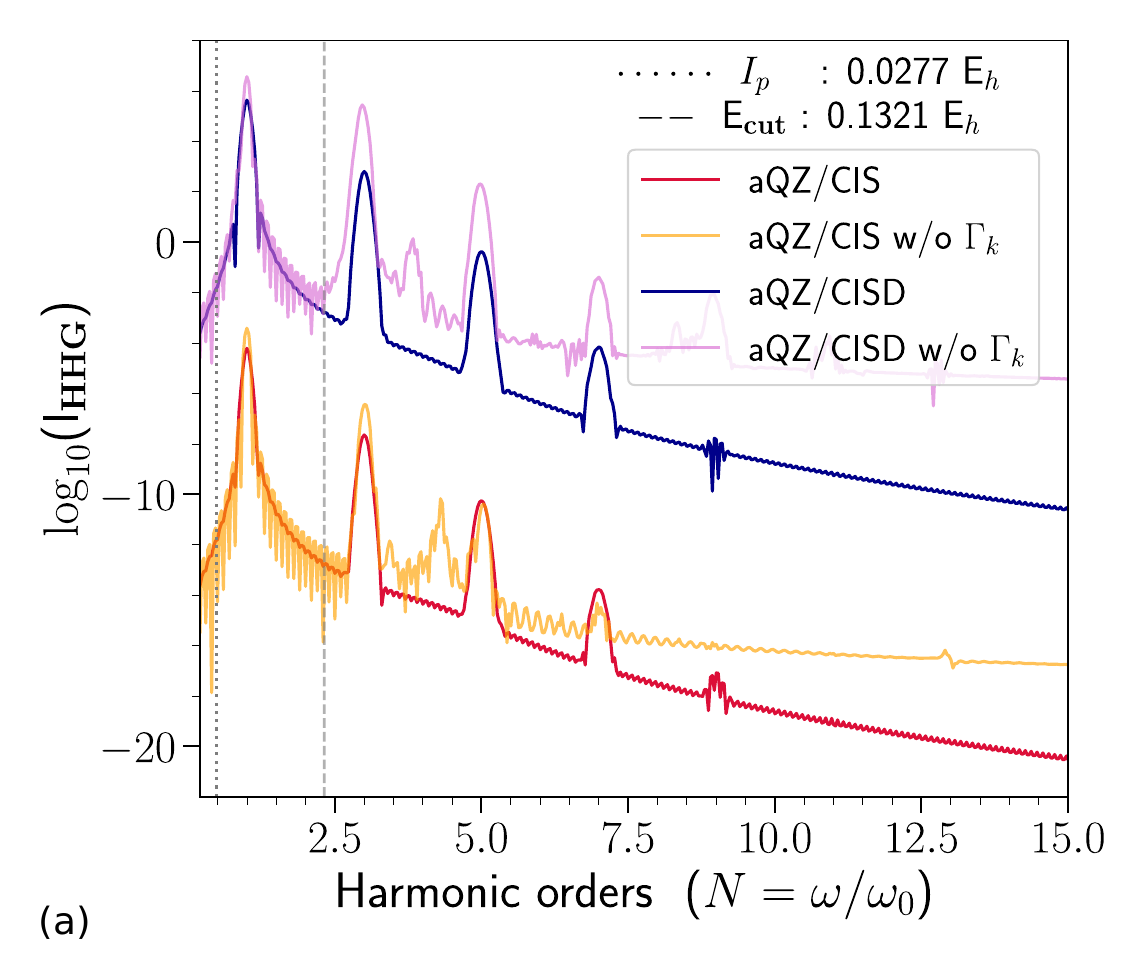}
        \refstepcounter{subfigure}
        \label{fig:h_fci_15E12_hhg_comp}      
    \end{subfigure}
    \begin{subfigure}[b]{0.47\textwidth}
        \includegraphics[width=\textwidth]{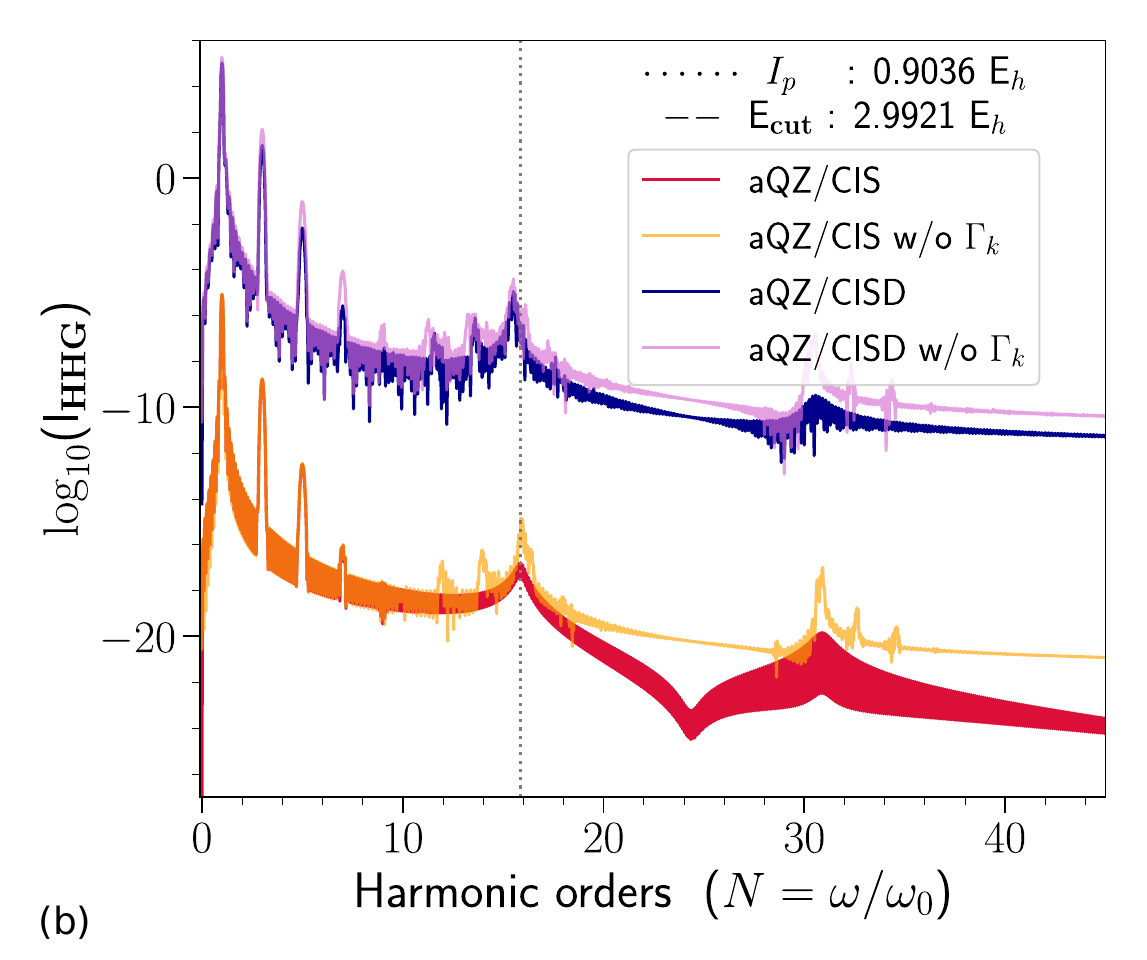}
        \refstepcounter{subfigure}
        \label{fig:he_fci_3E14_hhg_comp}      
    \end{subfigure}
    \begin{subfigure}[b]{0.47\textwidth}
        \includegraphics[width=\textwidth]{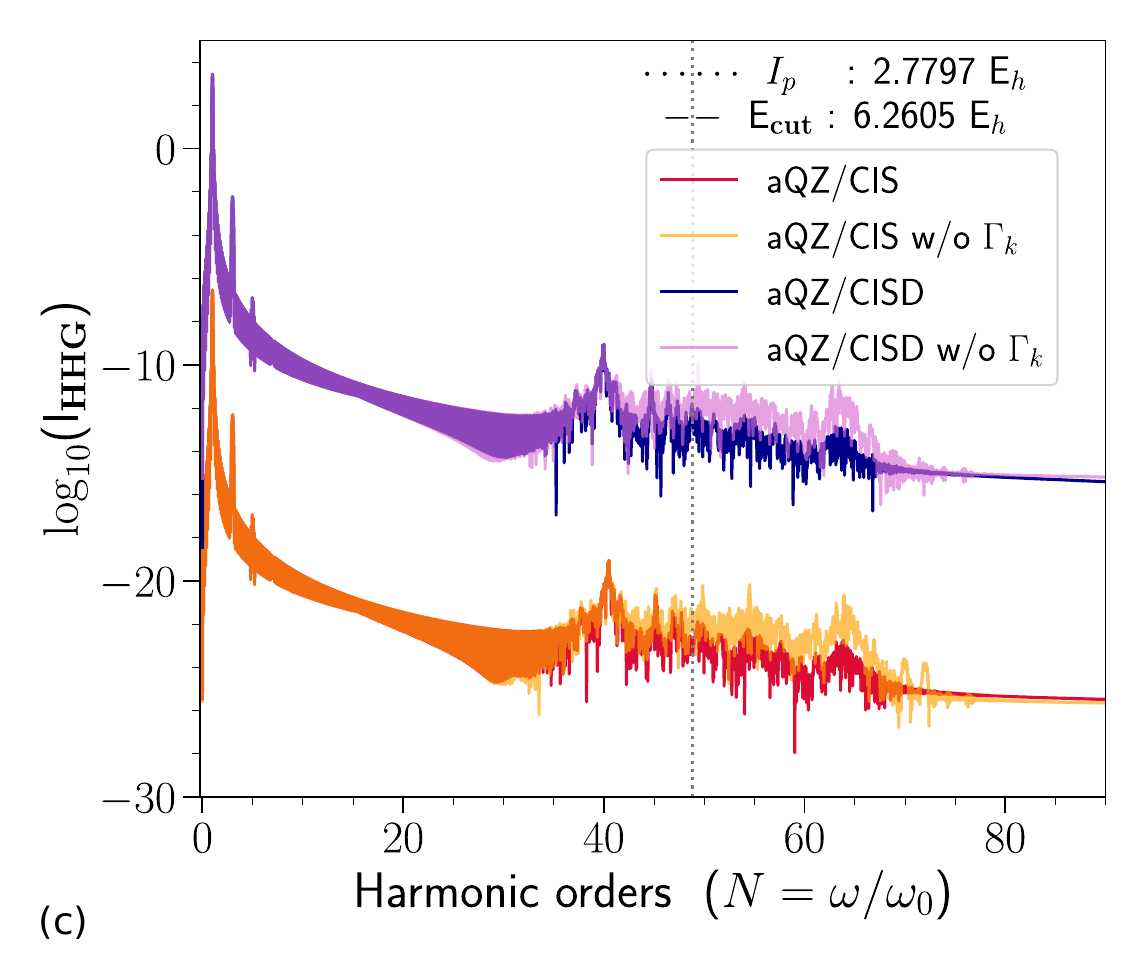}
        \refstepcounter{subfigure}
        \label{fig:li_fci_5E14_hhg_comp}      
    \end{subfigure}
    \caption{Comparision of HHG spectra of (a) Hydride anion (H$^{-}$), (b) Helium atom (He) and (c) Lithium cation (Li$^{+}$), calculated using TD-CIS/aQZ and TD-CISD/aQZ. For clarity, the spectra obtained with TD-CIS are upshifted by +10.}
    \label{fig:fci_hhg_comp}
  \end{figure}
       
 To evaluate the broader applicability of our method, we decided to study the HHG spectra of H$^{-}$ and Li$^{+}$, two ionic systems that are isoelectronic to Helium. Previously, H$^{-}$ has been studied using model calculations as it presents an interesting case of harmonic generation from a non-Coulombic potential\cite{becker1994modeling,kuchiev1999quantum}. On the other hand, HHG from ionized alkali metals and ionized plasmas have been studied for enhanced harmonic efficiency and extended cutoff\cite{kubodera1993high,popmintchev2015ultraviolet}. It is worth noting that the HHG spectra of anions and neutral atoms have been comparatively investigated in the literature to examine the influence of the Coulomb interaction between the active electron and the residual core electrons on HHG rates\cite{ostrovsky2005high}. In the context of TDCI, this would be equivalent to studying the effect of electron correlations on the HHG by using TD-CISD. In Fig. [\ref{fig:fci_hhg_comp}], we compare the HHG spectra calculated using TD-CIS and TD-CISD for H$^{-}$, He, and Li$^{+}$ at distinct peak laser intensities. We found that the results of the correlated TD-CISD calculation did not significantly differ from those of the correlated TD-CIS calculation. This reinforces the notion that HHG is effectively a one-electron process. 
    
 \section{Conclusions} \label{sec:conclusions}
 To conclude, we have investigated Luppi \textit{et al}'s idea of augmenting Kaufmann functions to Dunning basis sets to prepare hybrid Gaussian-continuum basis sets. Our scheme provides a simple way to systematically construct energy-optimized aXZ+($N$,$l_{max}$,c) basis sets for strong-field electron dynamics calculations. We have shown them to be well-conditioned for calculating higher harmonic spectra, free from any numerical instabilities that were reported earlier\cite{luppi2013role,coccia2016gaussian}.

  \section*{Supplementary Information}
  \begin{itemize}
    \item SI-1: Supplementary material
  \end{itemize}
  
  \section*{Data Availability}
  All the data as well as Python scripts and Jupyter notebooks (used for simulations, analysis, and plotting) related to this study are available on the public repository {\small \href{https://github.com/vijaymocherla/si_hhg_gaussian_basissets}{https://github.com/vijaymocherla/si\_hhg\_gaussian\_basissets}}. 
  Any other relevant information would be made available from the corresponding author upon reasonable request.

  \section*{Acknowledgments}
  We acknowledge the support of the Department of Atomic Energy, Government of India, under Project Identification No.~RTI~4007. 
  
\bibliographystyle{apsrev4-2}
\bibliography{references}
\end{document}

\title{Supplementary Material: Variational augmentation of Gaussian continuum basis sets for calculating atomic higher harmonic generation spectra}

\author{Sai Vijay Bhaskar Mocherla}
\affiliation{Tata Institute of Fundamental Research Hyderabad, Hyderabad 500046, India}

\author{Raghunathan Ramakrishnan}
\email{ramakrishnan@tifrh.res.in}
\affiliation{Tata Institute of Fundamental Research Hyderabad, Hyderabad 500046, India}

\maketitle
\section{Spin-adapted configuration state functions}
\begin{eqnarray}
    |^{1}\Phi_{a}^{r} \rangle &=& \frac{1}{\sqrt{2}} ( |\Phi_{a}^{r} \rangle + |\Phi_{\bar{a}}^{\bar{r}}\rangle ) \\ 
    |^{1}\Phi_{aa}^{rr} \rangle &=& |\Phi_{a\bar{a}}^{r\bar{r}} \rangle \\
    |^{1}\Phi_{aa}^{rs} \rangle &=& \frac{1}{\sqrt{2}} ( |\Phi_{a\bar{a}}^{r\bar{s}} \rangle + |\Phi_{a\bar{a}}^{s\bar{r}} \rangle ) \\  
    |^{1}\Phi_{ab}^{rr} \rangle &=& \frac{1}{\sqrt{2}} ( |\Phi_{a\bar{b}}^{r\bar{r}}\rangle + |\Phi_{b\bar{a}}^{r\bar{r}}\rangle )  \\
    |^\text{A}\Phi_{ab}^{rs} \rangle &=& \frac{1}{\sqrt{12}} ( 2|\Phi_{ab}^{rs}\rangle + 2|\Phi_{\bar{a}\bar{b}}^{\bar{r}\bar{s}}\rangle - |\Phi_{\bar{a}b}^{\bar{s}r}\rangle - |\Phi_{a\bar{b}}^{s\bar{r}}\rangle + |\Phi_{\bar{a}b}^{\bar{r}s}\rangle + |\Phi_{a\bar{b}}^{r\bar{s}}\rangle )\\
    |^\text{B}\Phi_{ab}^{rs} \rangle &=& \frac{1}{2} (|\Phi_{\bar{a}b}^{\bar{s}r}\rangle + |\Phi_{a\bar{b}}^{s\bar{r}}\rangle + |\Phi_{\bar{a}b}^{\bar{r}s}\rangle + |\Phi_{a\bar{b}}^{r\bar{s}}\rangle   ) \
\end{eqnarray}

\section{Laser paramters}
\begin{figure}
    \centering
    \includegraphics[width=.5\textwidth]{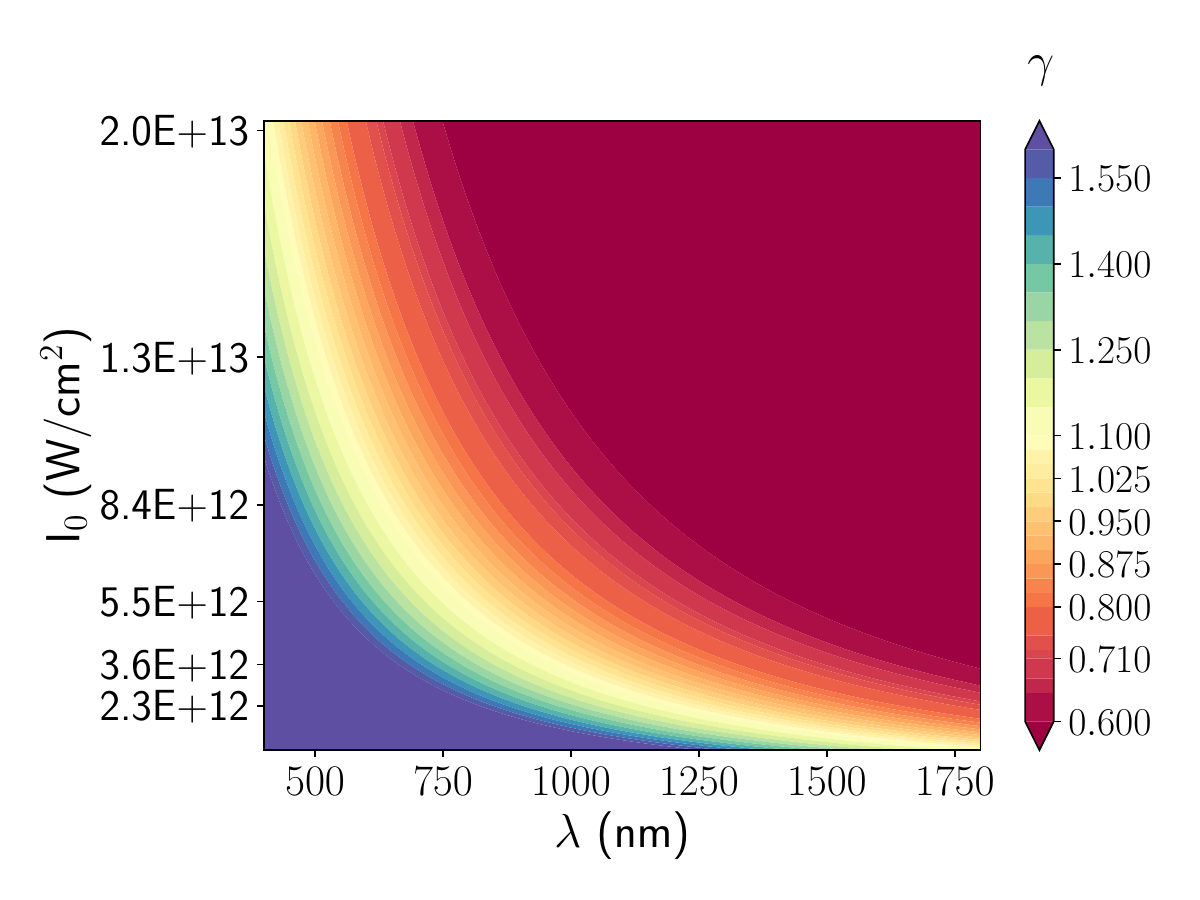}
    \vspace{-2em}
    \caption{Keldysh parameter plot for H$^{-}$ anion.}
    \label{fig:h_minus_keldysh}
\end{figure}

\begin{figure}
   \centering
   \includegraphics[width=.5\textwidth]{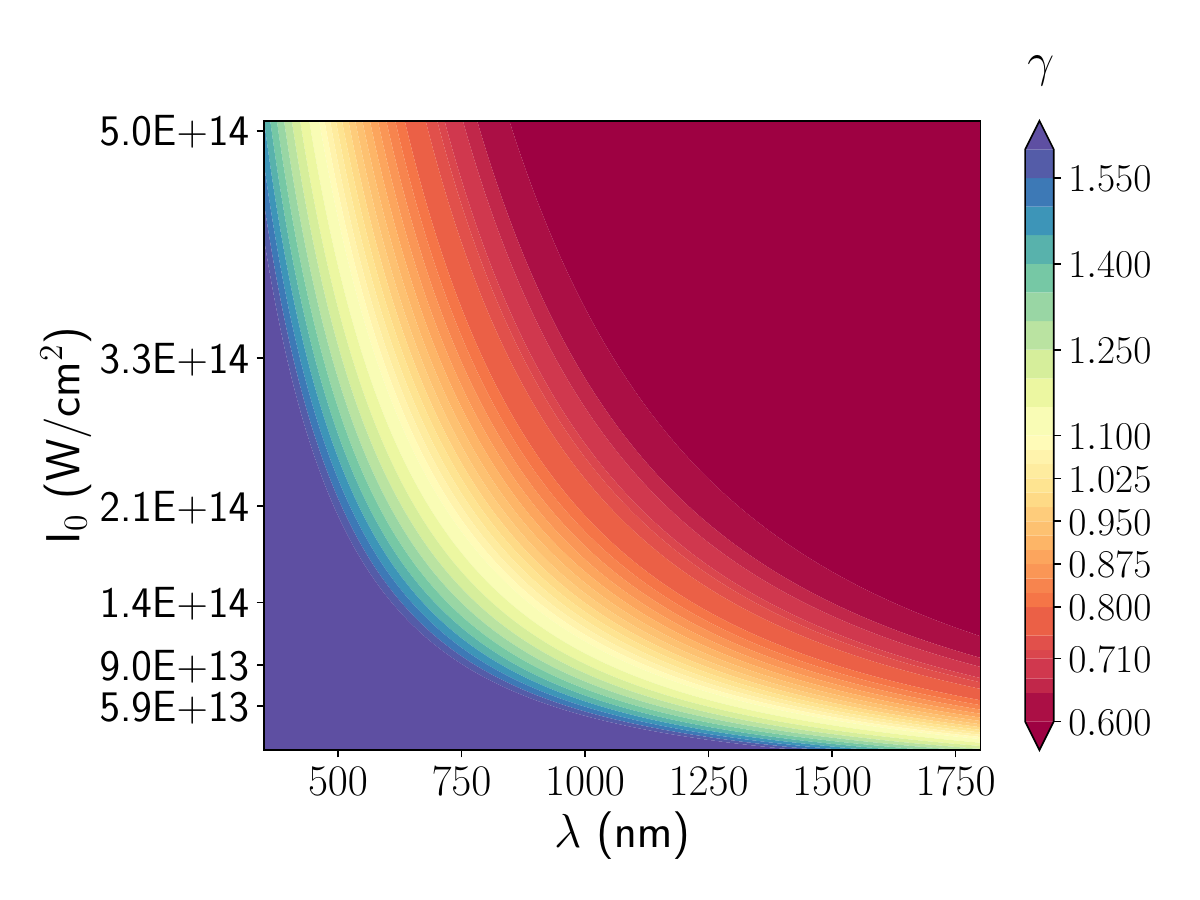}
   \vspace{-2em}
   \caption{Keldysh parameter plot for He atom.}
   \label{fig:he_keldysh}
\end{figure}

\begin{figure}
    \centering
    \includegraphics[width=.5\textwidth]{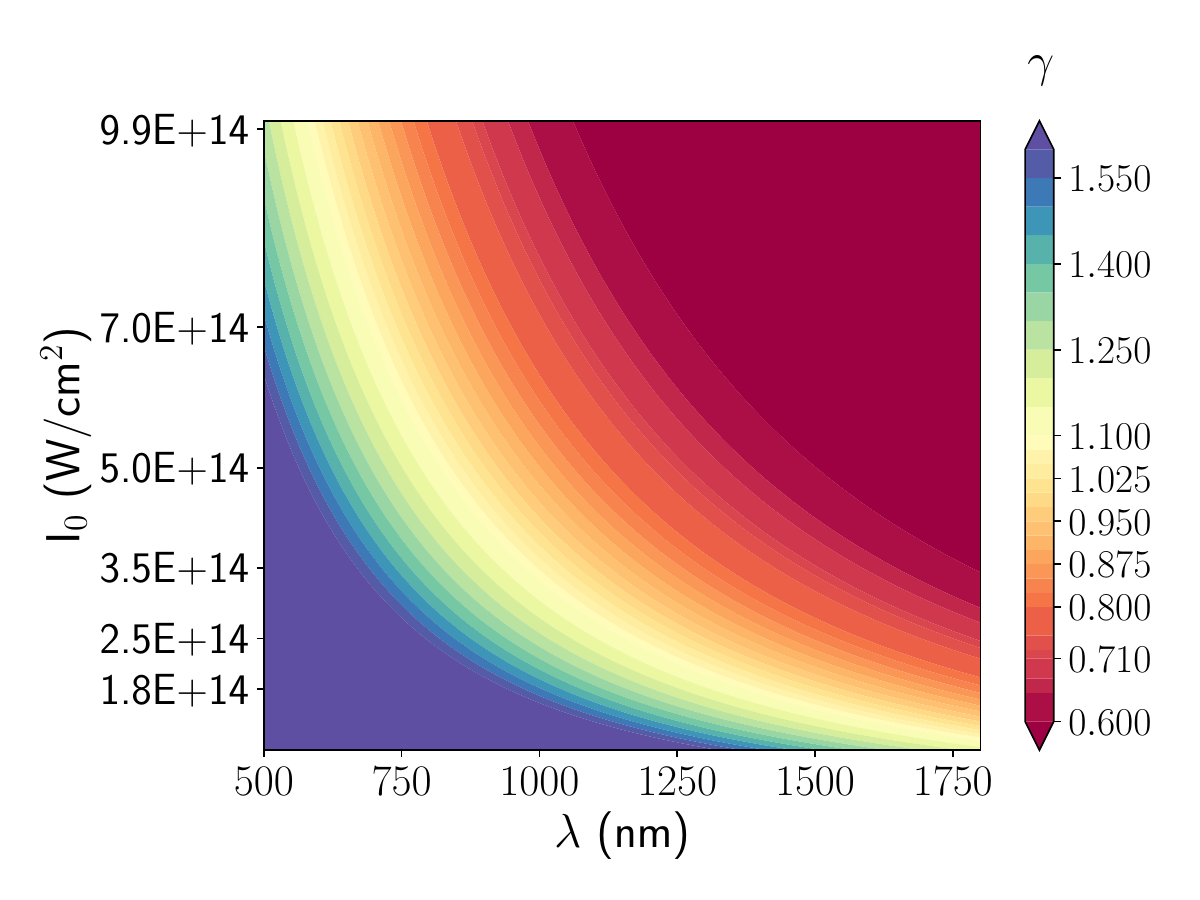}
    \vspace{-2em}
    \caption{Keldysh parameter plot for Li$^{+}$ cation.}
    \label{fig:li_plus_keldysh}
  \end{figure}

Figures [\ref{fig:h_minus_keldysh},\ref{fig:he_keldysh},\ref{fig:li_plus_keldysh}] show different regimes of ionization for the Helium atom as characterized by the Keldysh 
parameter.

\section{Preparation of Hybrid Basis sets}

\begin{figure}[h]
    \includegraphics[width=.45\textwidth]{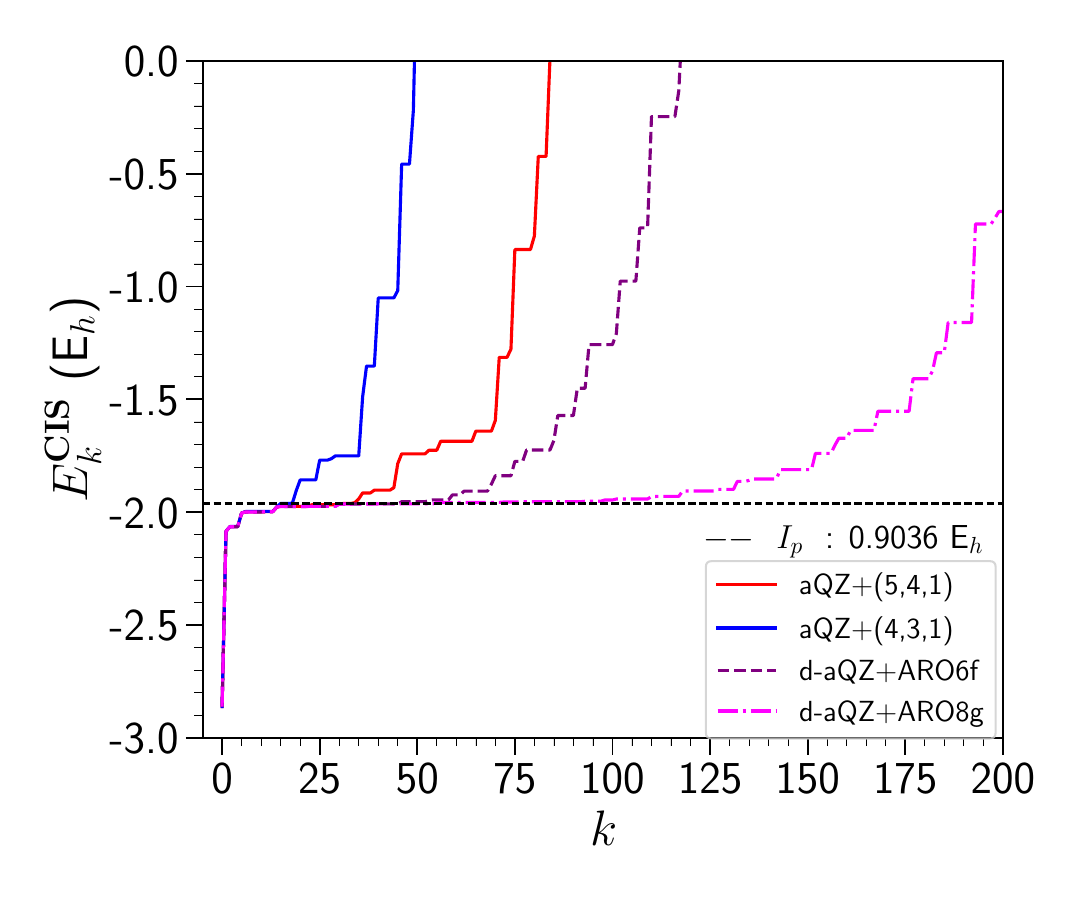}
    \includegraphics[width=.45\textwidth]{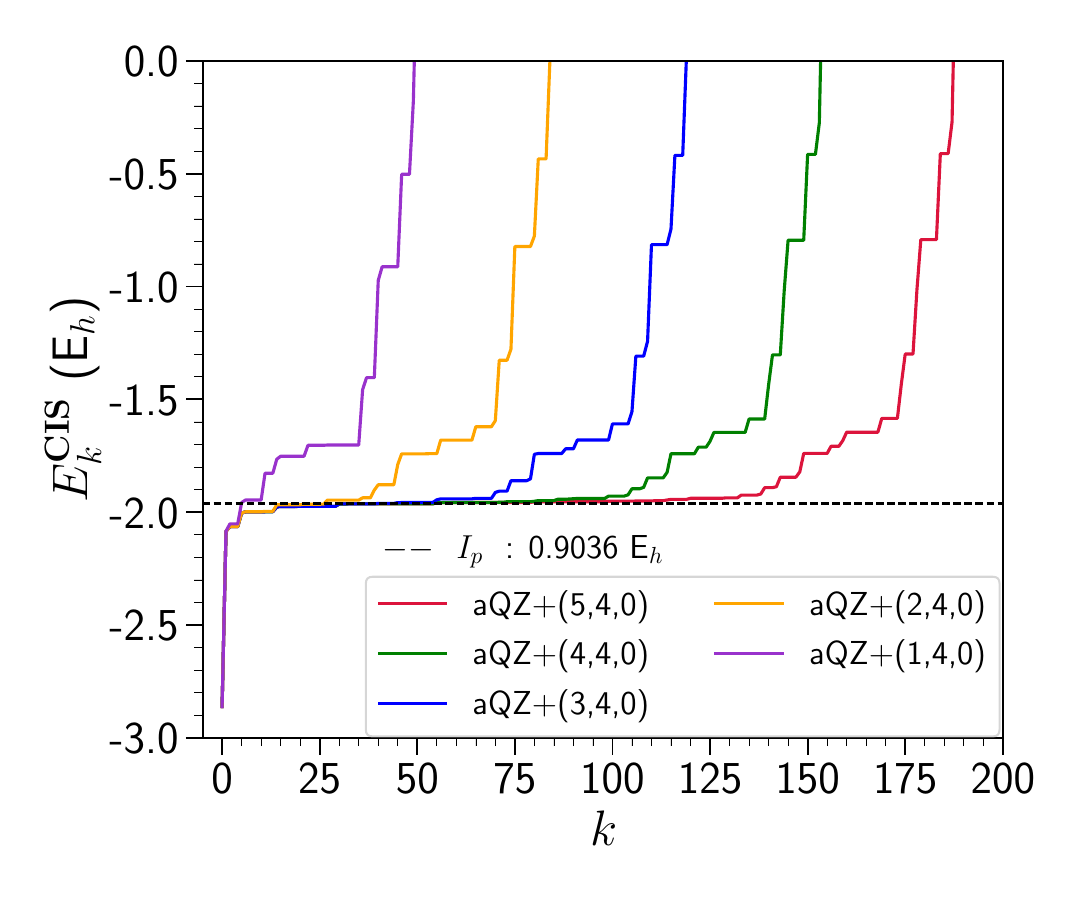}
    \caption{Energy distribution of CIS states of Helium atom calculated with aQZ+($N$,$l$,1) and aQZ+($N$,4,0) basis sets.}
    \label{fig:he_nl1_n40_ecis_comp}
\end{figure}

\renewcommand{\arraystretch}{1.8}
\setlength{\tabcolsep}{6pt}
\begin{table*}[h!]
% \begin{adjustbox}{max width = \textwidth}
  \centering
  \caption{The lowest Gaussian exponents ($\alpha_{\text{min}}$ in bohr$^{-2}$) of each type Kaufmann function from the aQZ+(N,4,0) energy optimized basis sets for Helium atom, along with their corresponding half-width at half-maximum (HWHM in bohr) representing their average radial spread.}
  \label{table:he_gauss_exp_n40}
  \begin{tabular}{cccccc}
    \hline\hline
    \multirow{2}{*}{Basis sets} & \multicolumn{5}{c}{$\alpha_\text{min}$   (HWHM $\approx \sqrt{\ln(2)/\alpha_\text{min}}$ in bohr)}\\\cline{2-6}
    ~ & S & P & D & F & G \\ \hline
    aQZ+(1,4,0) & 1.05$\times 10^{-1}$ (2.57) & 1.24$\times 10^{-1}$ (2.36) & 8.85$\times 10^{-2}$ (2.80) & 9.17$\times 10^{-2}$ (2.75) & 7.65$\times 10^{-2}$ (3.01)\\
    aQZ+(2,4,0) & 1.82$\times 10^{-2}$ (6.17) & 1.66$\times 10^{-2}$ (6.46) & 1.52$\times 10^{-2}$ (6.76) & 1.39$\times 10^{-2}$ (7.05) & 1.28$\times 10^{-2}$ (7.35)\\
    aQZ+(3,4,0) & 4.70$\times 10^{-3}$ (12.15) & 4.41$\times 10^{-3}$ (12.54) & 4.15$\times 10^{-3}$ (12.93) & 3.91$\times 10^{-3}$ (13.32) & 3.69$\times 10^{-3}$ (13.71)\\
    aQZ+(4,4,0) & 1.72$\times 10^{-3}$ (20.08) & 1.64$\times 10^{-3}$ (20.57) & 1.56$\times 10^{-3}$ (21.06) & 1.49$\times 10^{-3}$ (21.55) & 1.43$\times 10^{-3}$ (22.03)\\
    aQZ+(5,4,0) & 7.72$\times 10^{-4}$ (29.97) & 7.42$\times 10^{-4}$ (30.56) & 7.14$\times 10^{-4}$ (31.15) & 6.88$\times 10^{-4}$ (31.74) & 6.63$\times 10^{-4}$ (32.32)\\\hline
    \hline     
  \end{tabular}
% \end{adjustbox}
\end{table*}

\begin{figure*}[h!]
    \includegraphics[width=\textwidth]{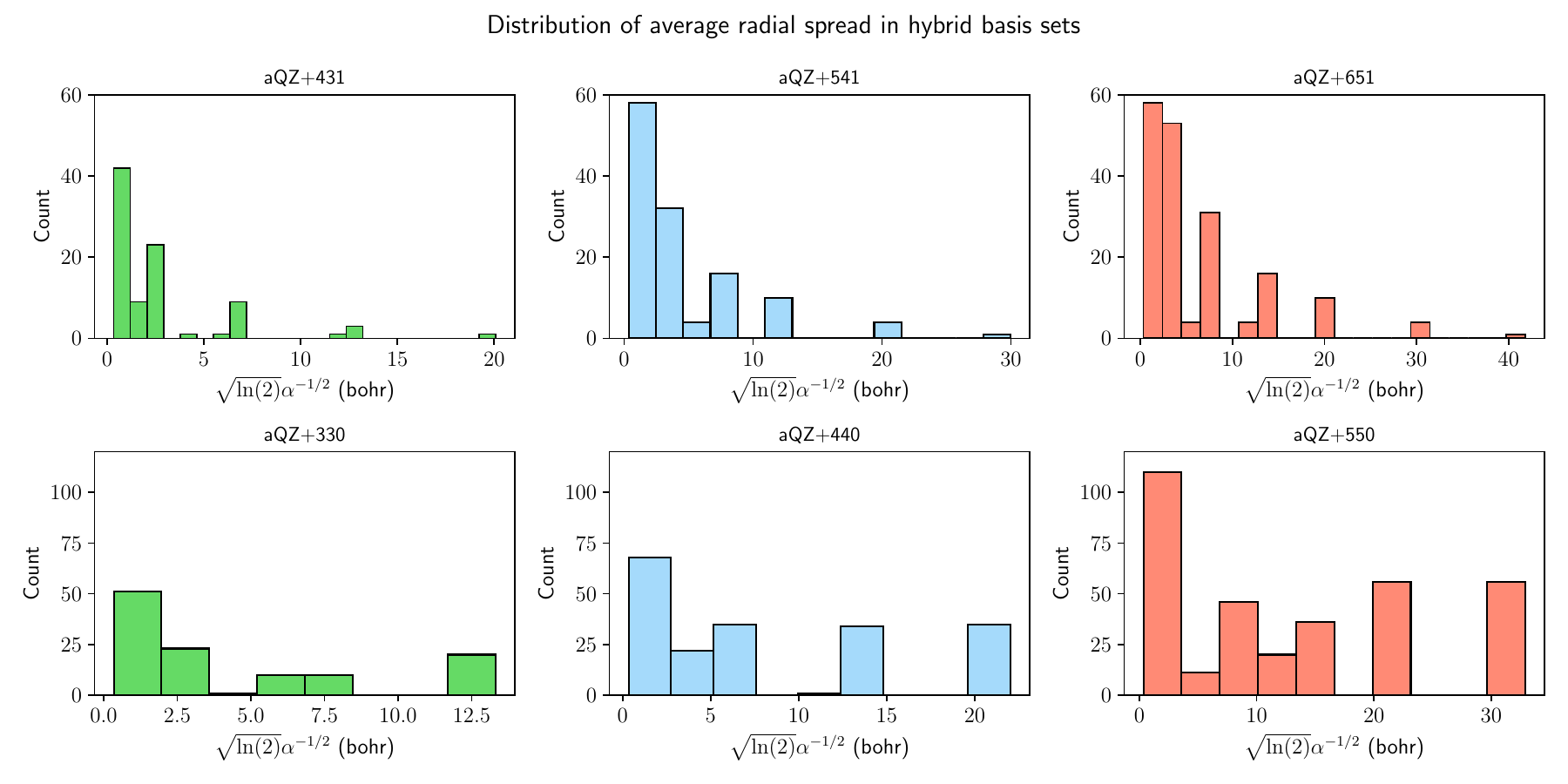}
    \caption{Distributions of average radial spread ($\sqrt{\ln{2}}\alpha^{-1/2}$) of basis functions in different aQZ+(N,4,0) basis sets.}
    \label{fig:he_aqz_nlc_hist}
  \end{figure*}

\begin{figure*}[H]
    \includegraphics[width=\textwidth]{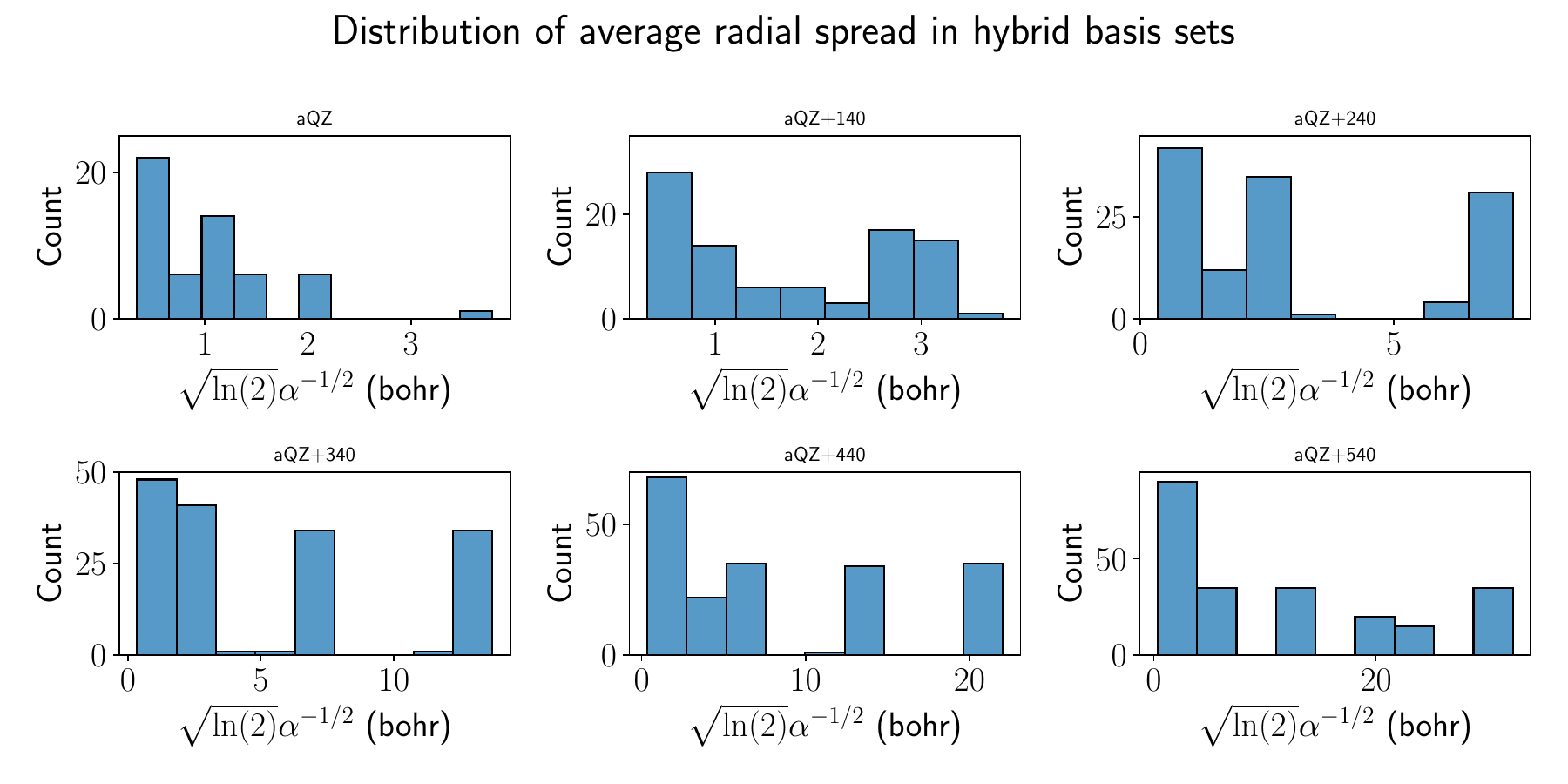}
    \caption{Distributions of average radial spread ($\sqrt{\ln{2}}\alpha^{-1/2}$) of basis functions in different aQZ+(N,4,0) basis sets.}
    \label{fig:he_aqz_n40_hist}
\end{figure*}
    
\section{HHG Plots}
\begin{figure}[h]
    \includegraphics[width=.49\textwidth]{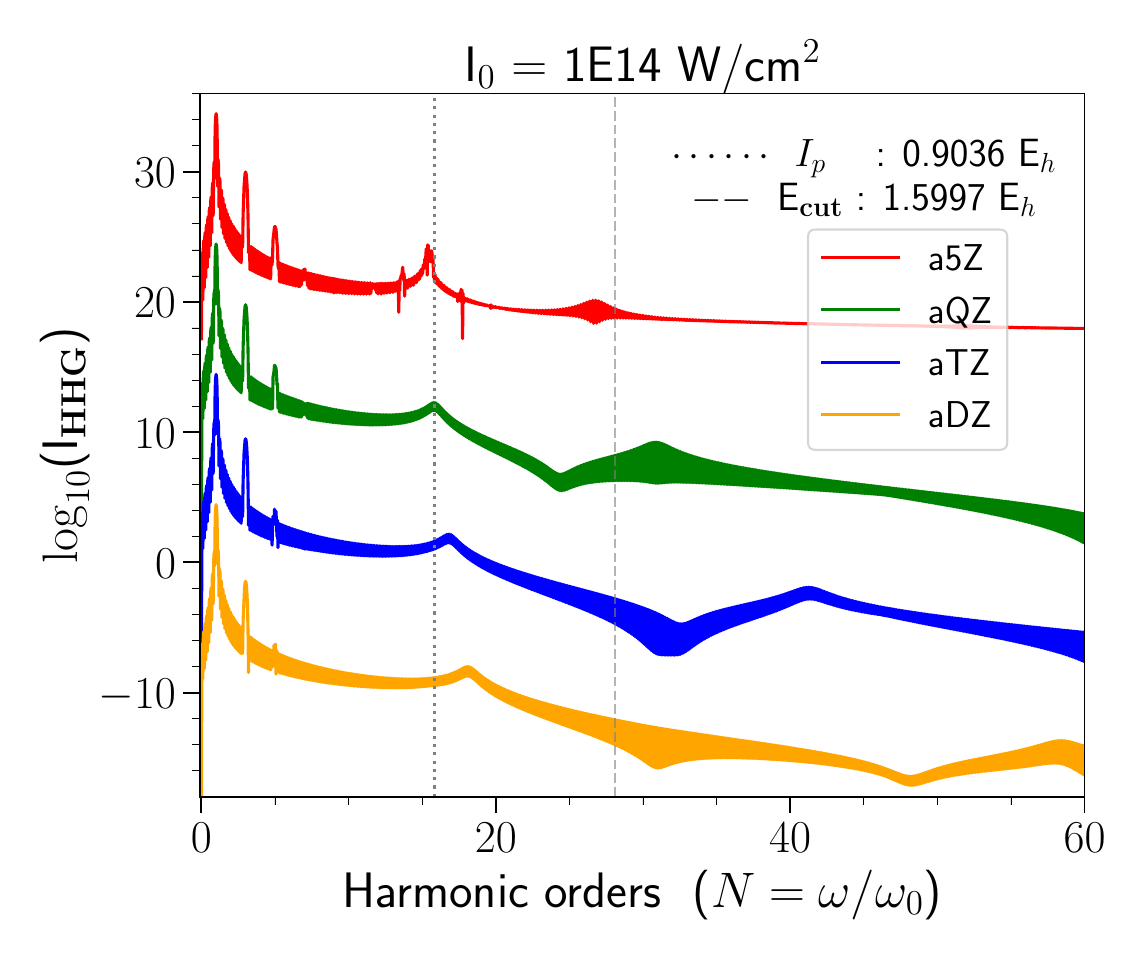}
    \includegraphics[width=.49\textwidth]{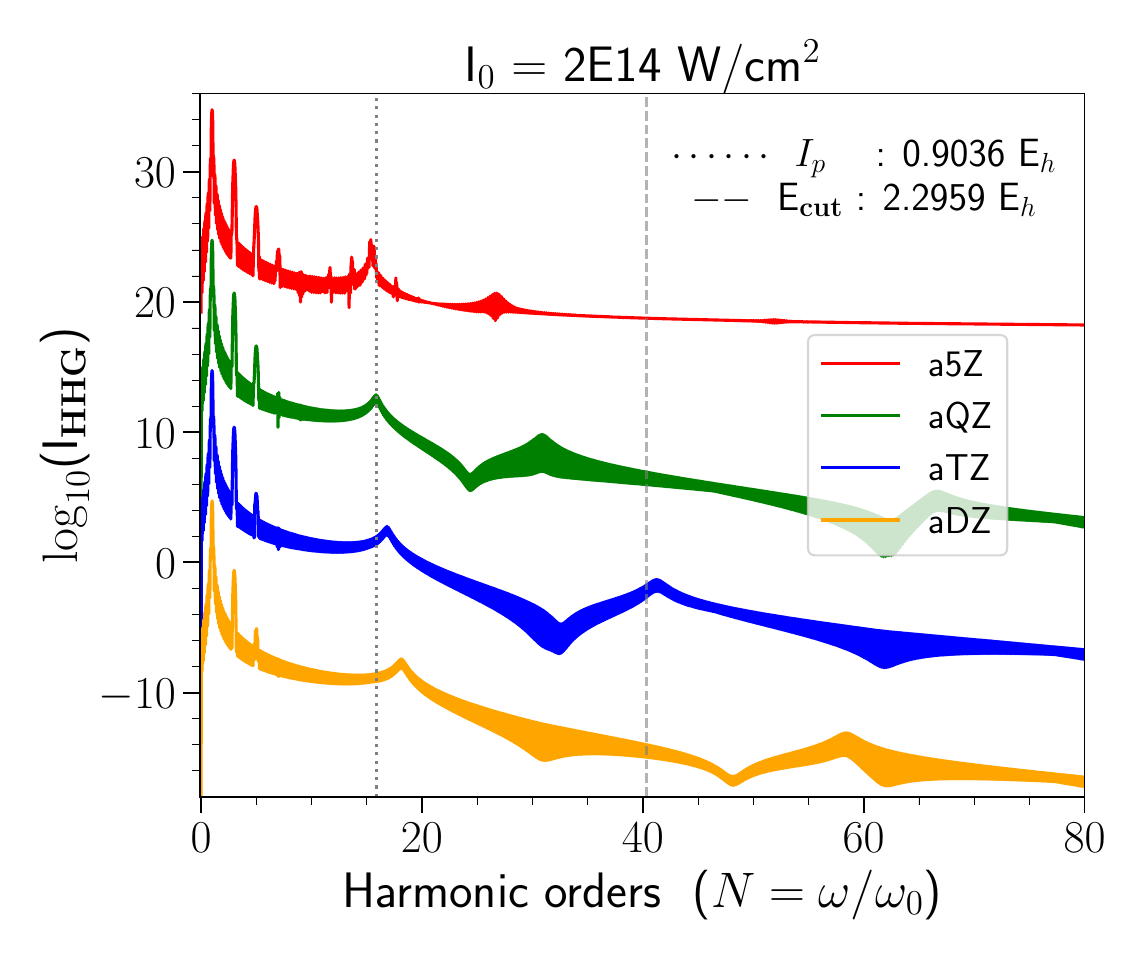}
    \includegraphics[width=.49\textwidth]{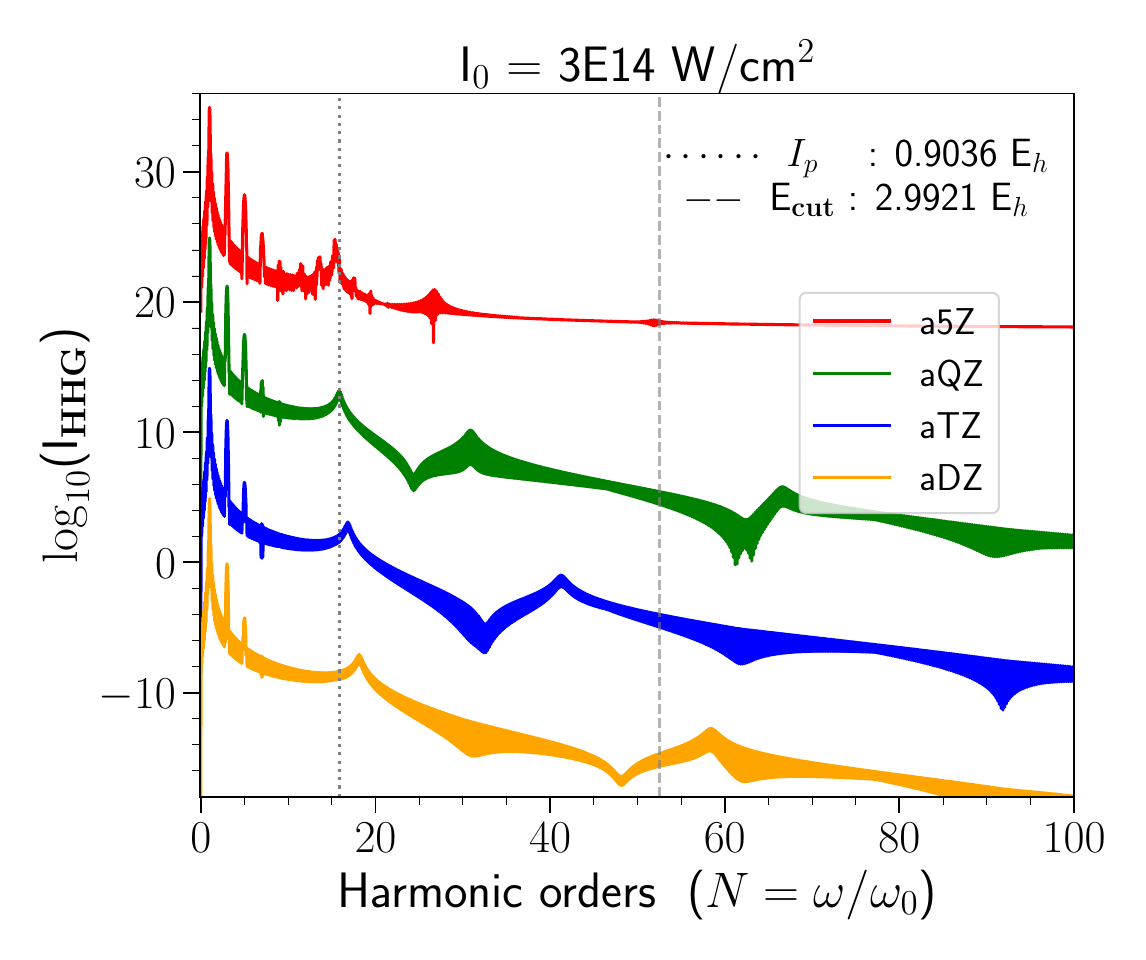}
    \includegraphics[width=.49\textwidth]{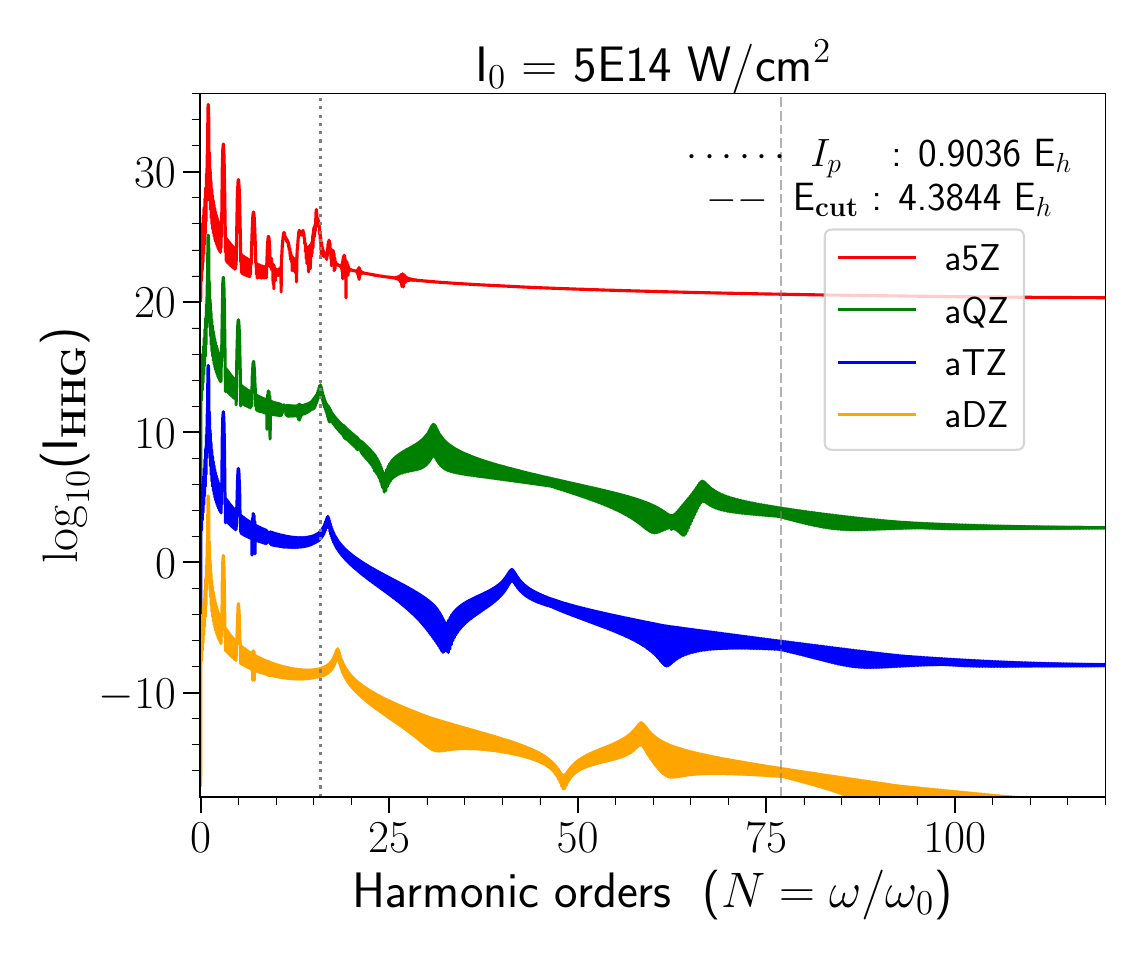}

    \caption{Comparison of HHG spectra of Helium atom generated by driving laser pulses of peak intensities $\text{I}_{0} = \{1,2,3,5\}\times 10^{14}$ W/cm$^{2}$, calculated with different a$X$Z basis sets using TD-CIS. For clarity, the spectra have been upshifted in multiples of +10 from aDZ to a5Z.}
    \label{fig:he_axz_hhg_comp}
\end{figure}

\begin{figure}[h]
    \includegraphics[width=.49\textwidth]{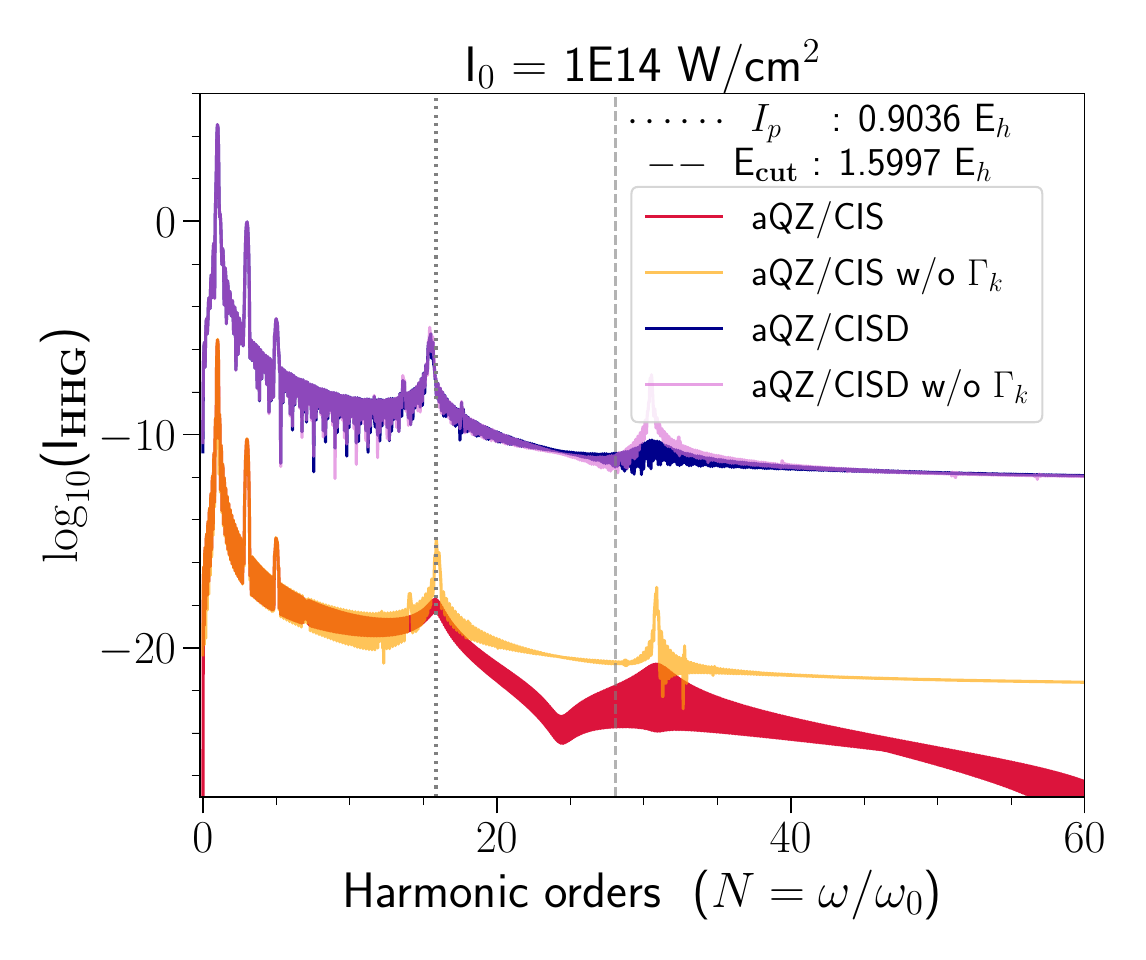}
    \includegraphics[width=.49\textwidth]{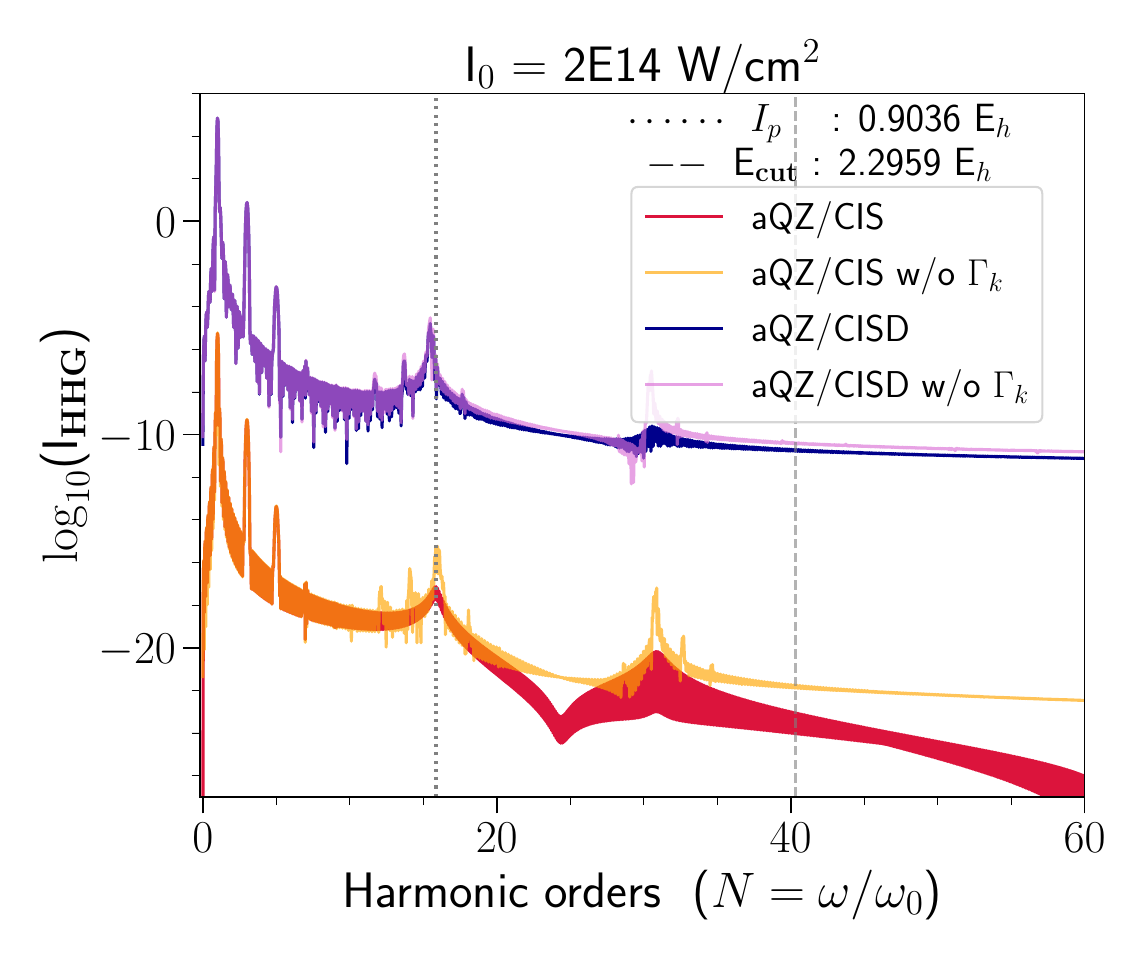}
    \includegraphics[width=.49\textwidth]{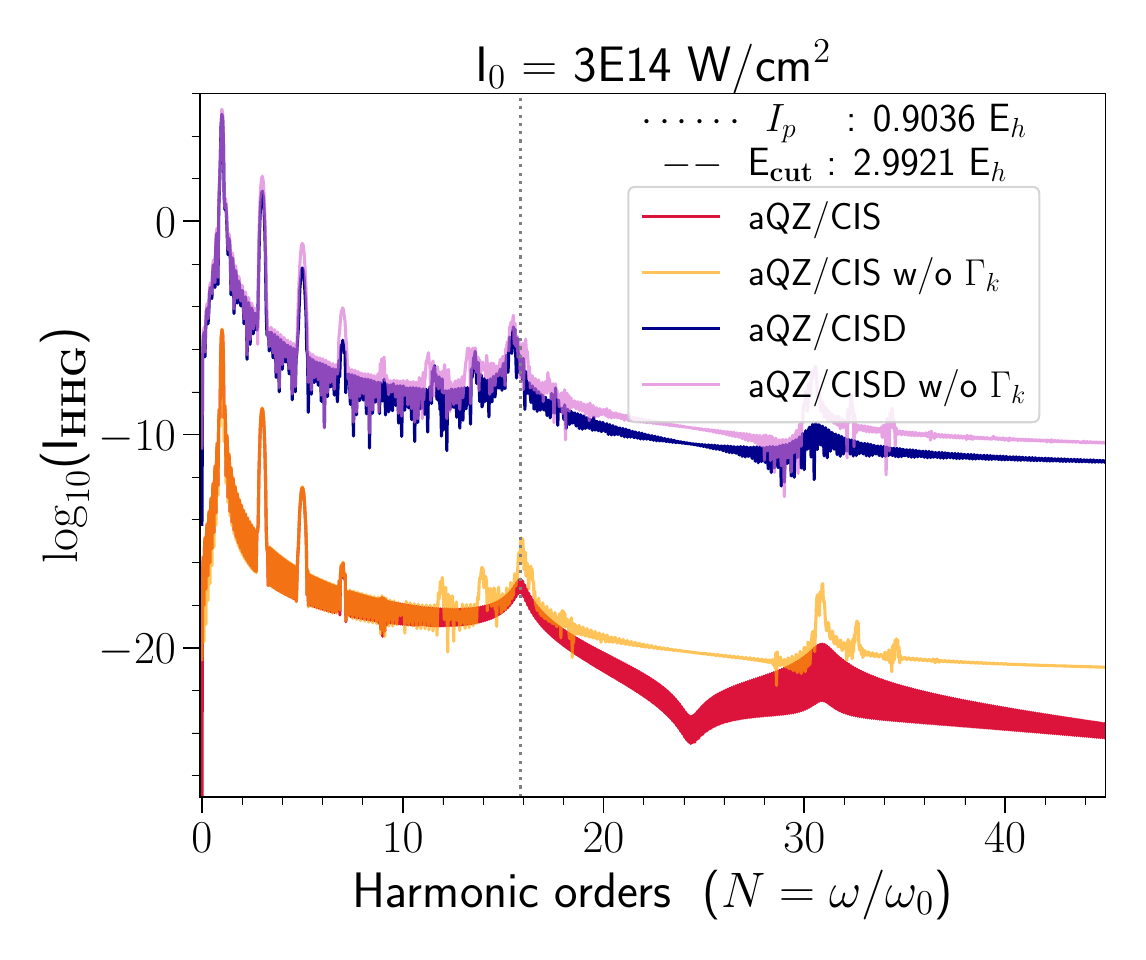}
    \includegraphics[width=.49\textwidth]{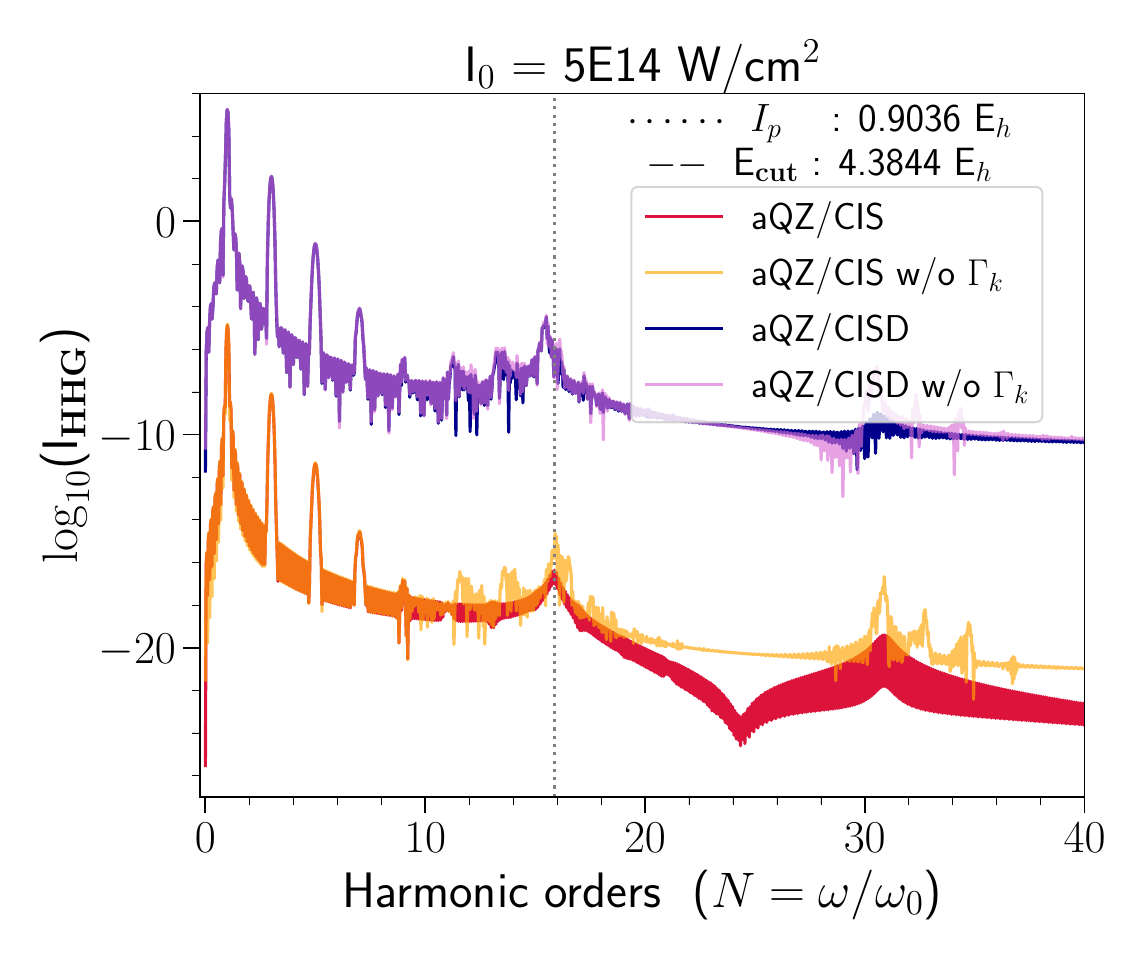}
    \caption{Comparison of HHG spectra of Helium atom generated by driving laser pulses of peak intensities $\text{I}_{0} = \{1,2,3,5\}\times 10^{14}$ W/cm$^{2}$, calculated using TD-CIS/aQZ and TD-CISD/aQZ. For clarity, the spectra have been upshifted in multiples of +10 from aDZ to a5Z.}
    \label{fig:he_fci_hhg_comp}
\end{figure}

\begin{figure}[h]
    \includegraphics[width=.49\textwidth]{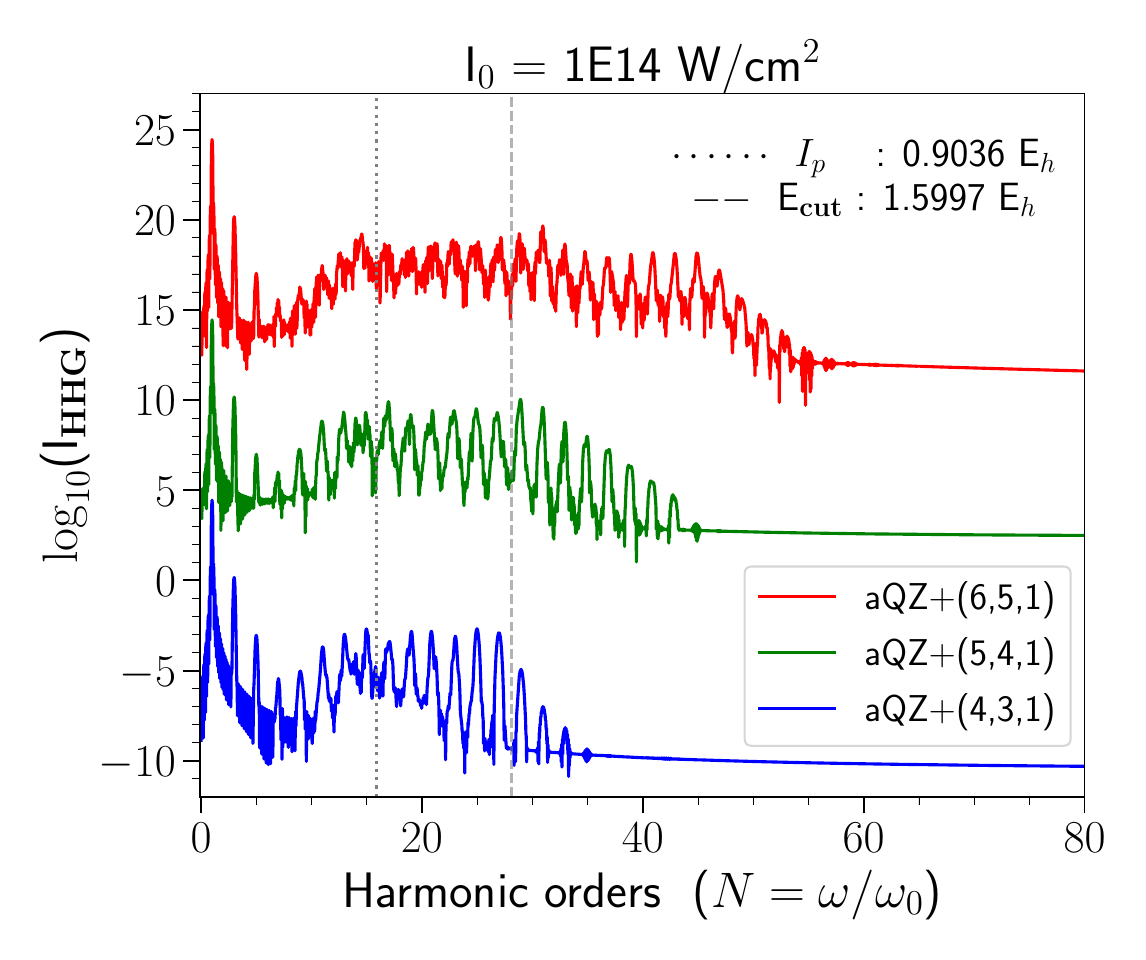}
    \includegraphics[width=.49\textwidth]{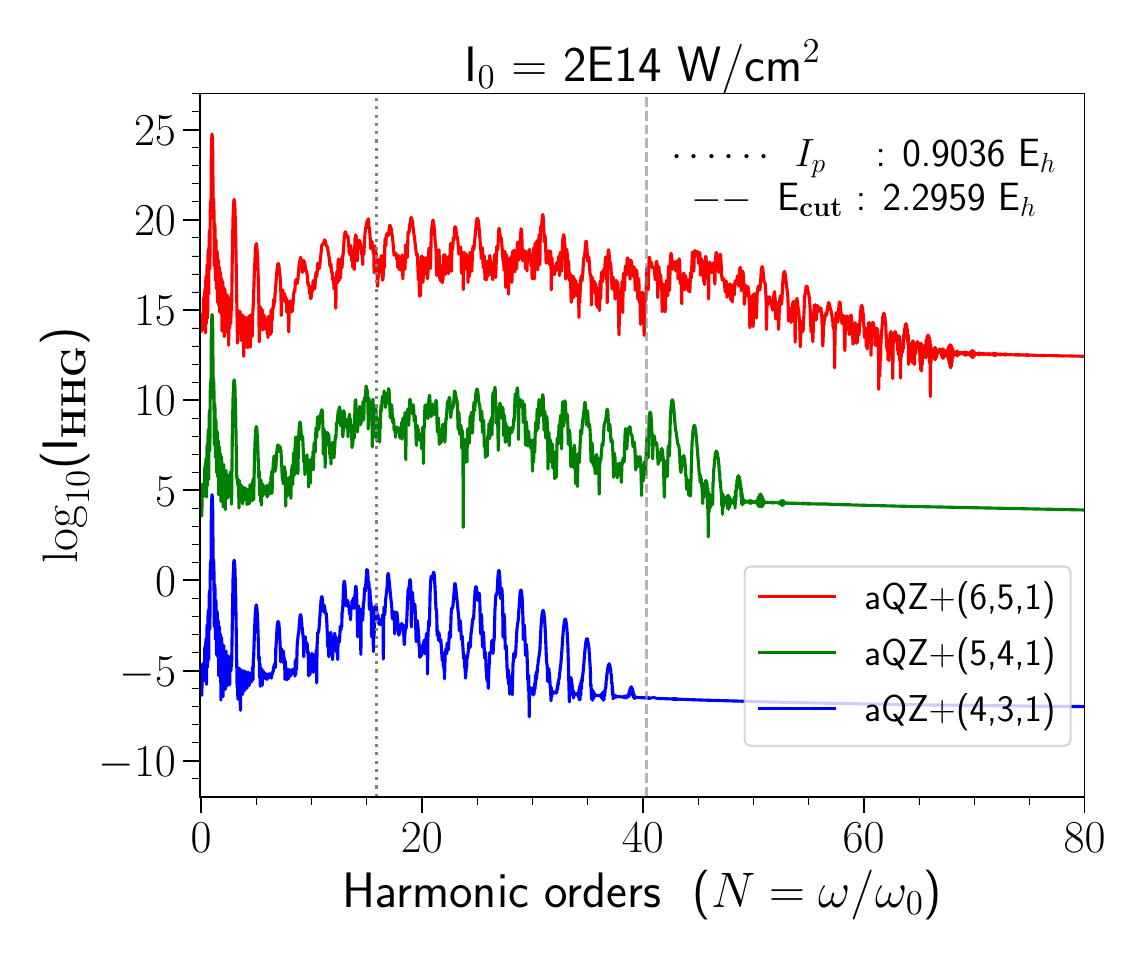}
    \includegraphics[width=.49\textwidth]{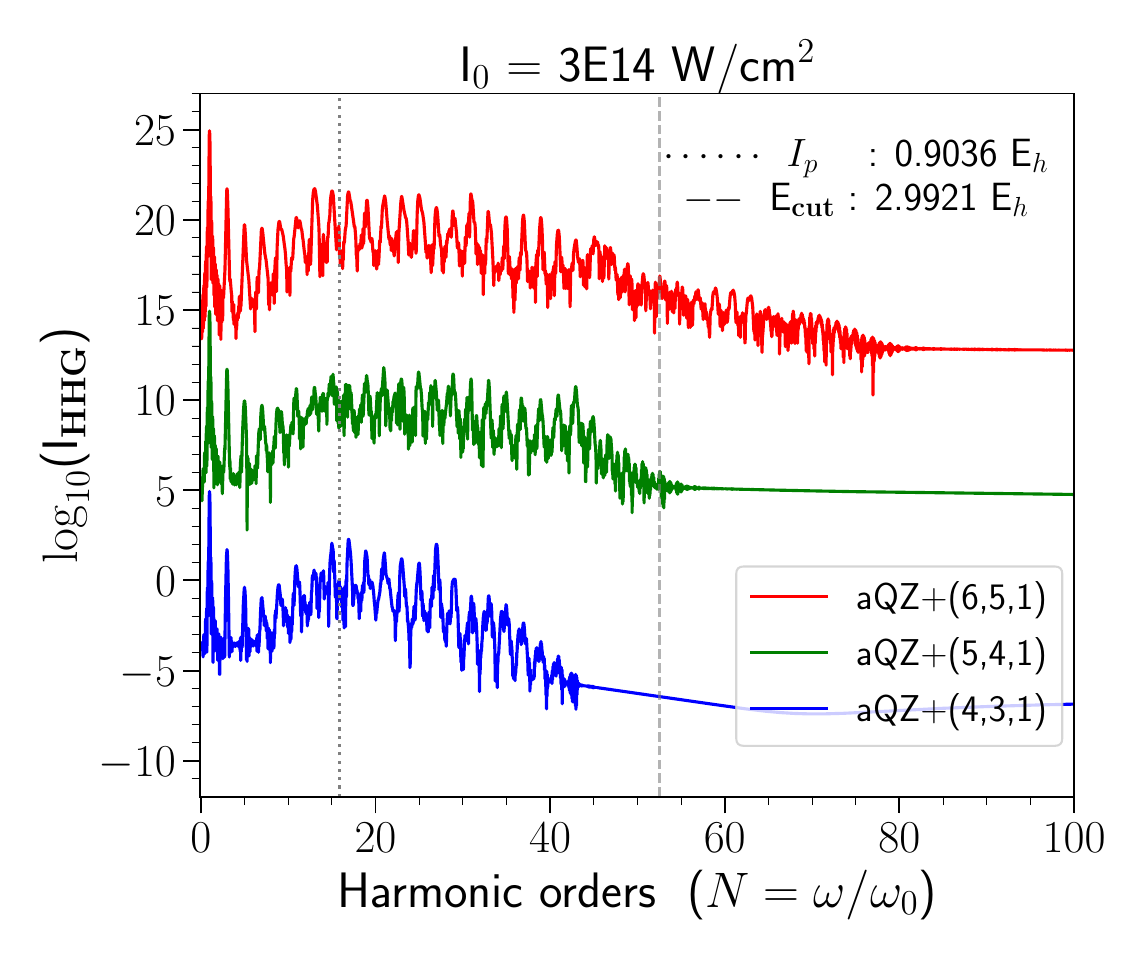}
    \includegraphics[width=.49\textwidth]{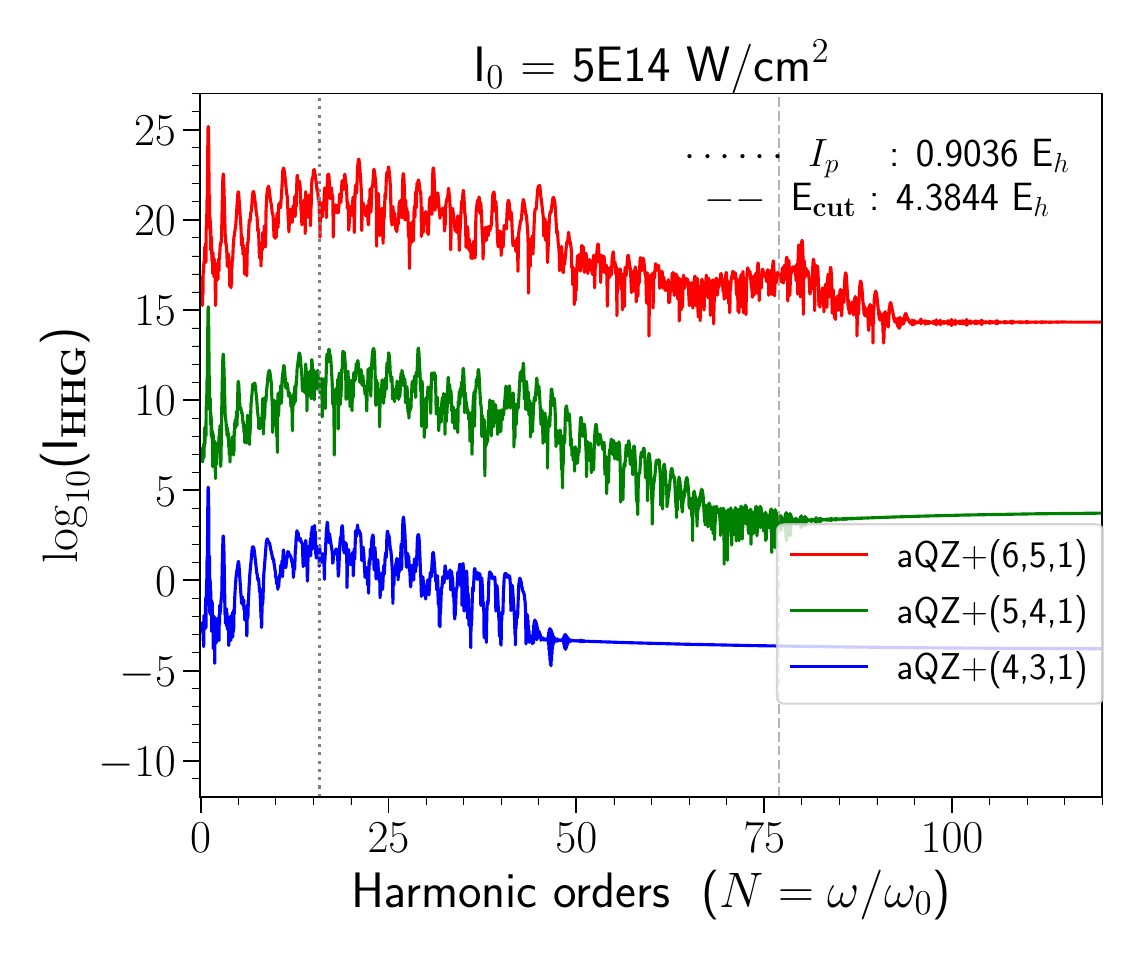}
    \caption{Comparison of HHG spectra of Helium atom for $\text{I}_{0} = \{1,2,3,5\}\times10^{14}$W/cm$^{2}$, calculated using TD-CIS with differnt aQZ+($N$,$l_{max}$,c) basis sets. For clarity, the spectra have been upshifted in multiples of +10 from aDZ to a5Z.} 
    \label{fig:he_aqz_nl1_hhg_comp}
\end{figure}

\begin{figure}[h]
    \includegraphics[width=.49\textwidth]{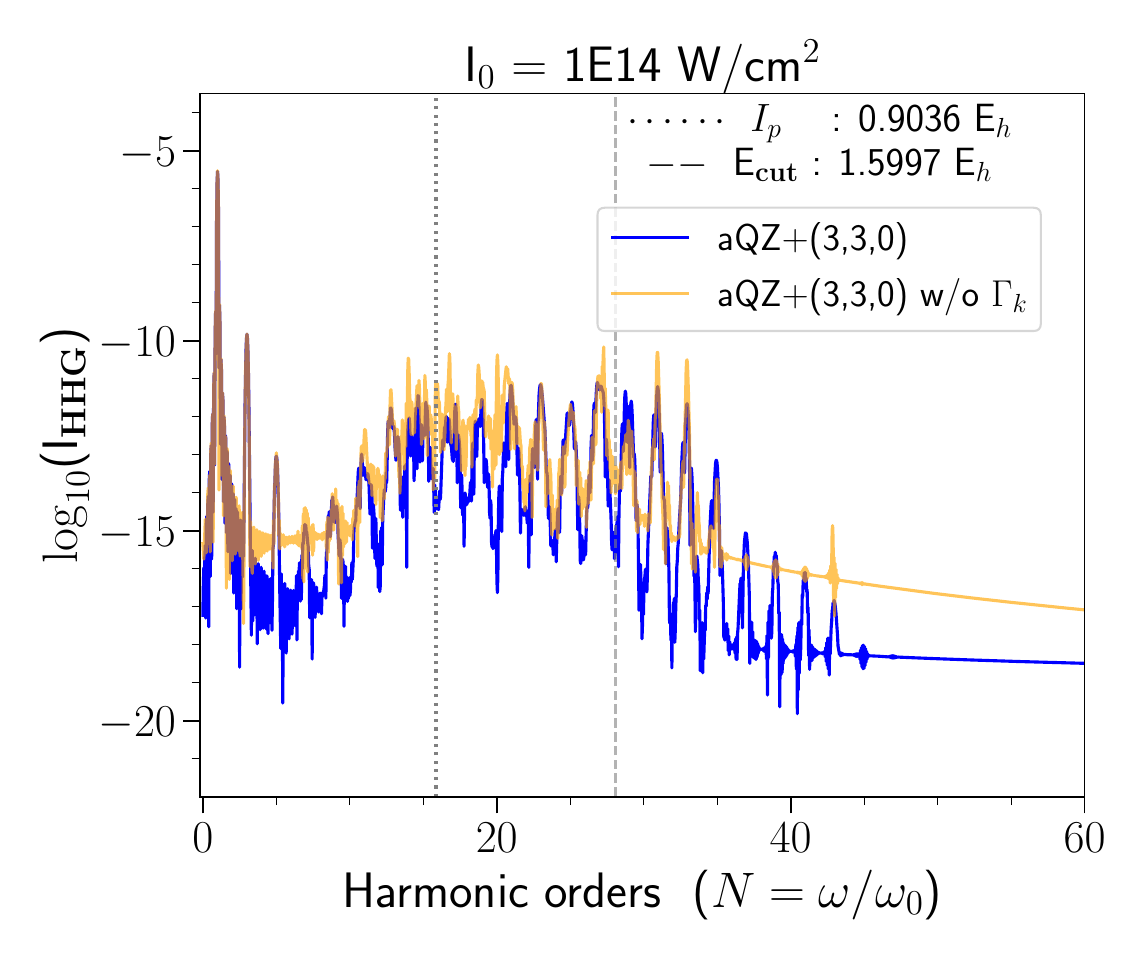}
    \includegraphics[width=.49\textwidth]{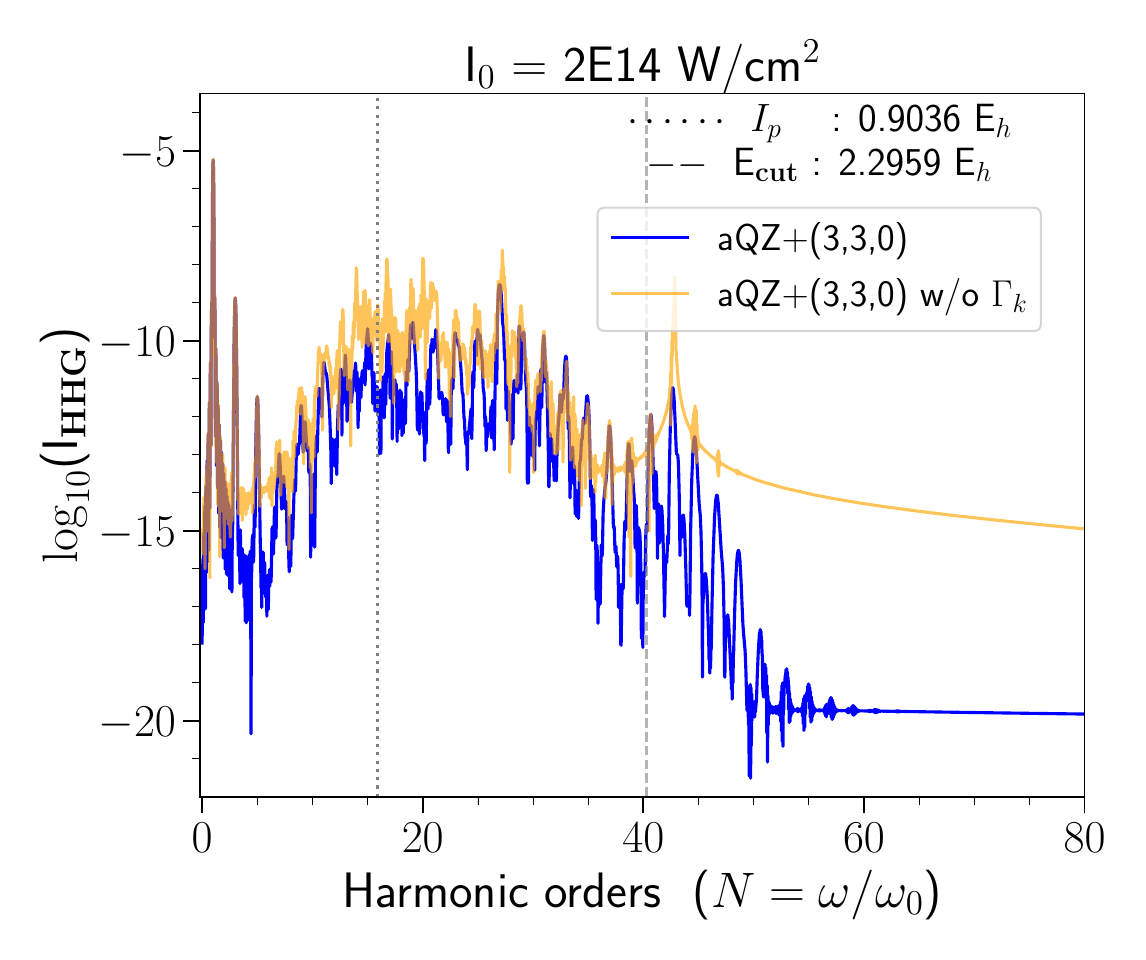}
    \includegraphics[width=.49\textwidth]{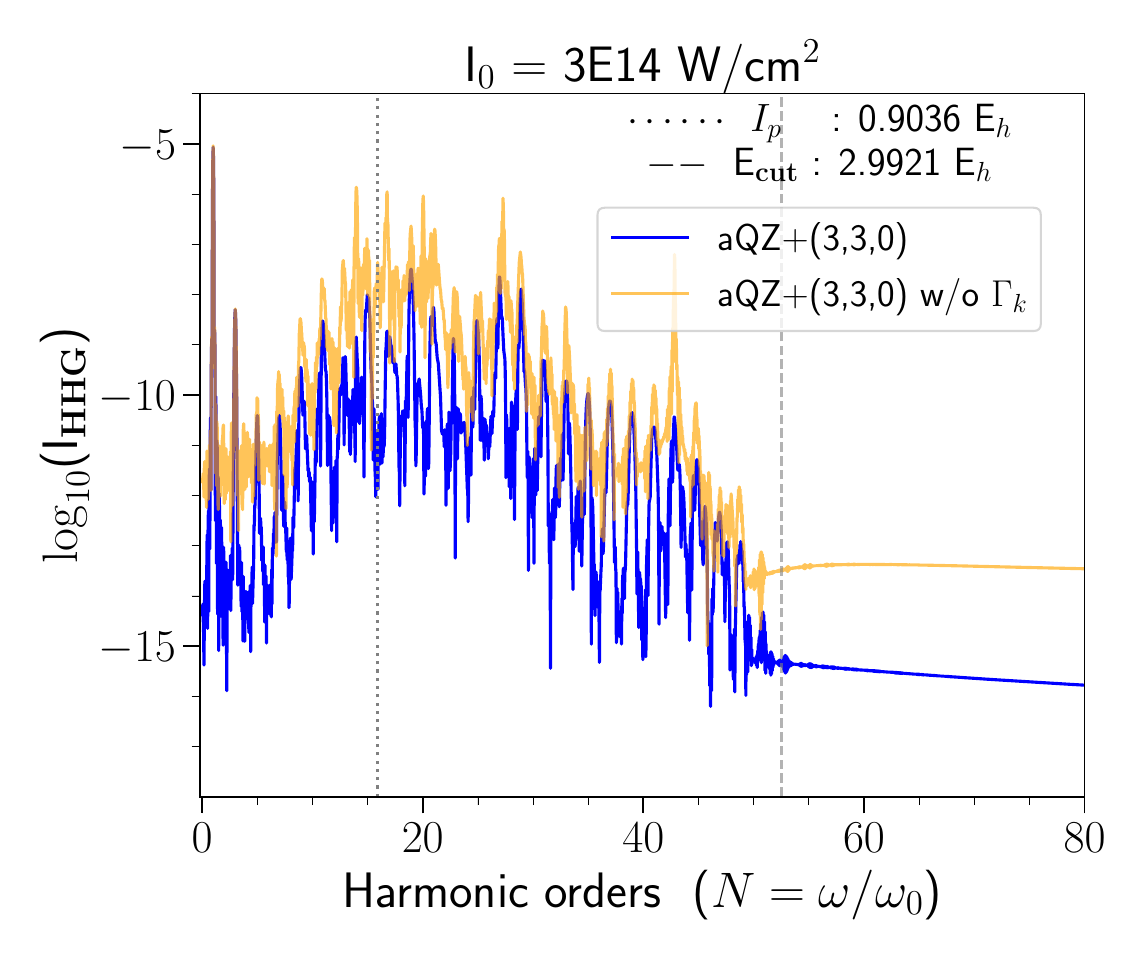}
    \includegraphics[width=.49\textwidth]{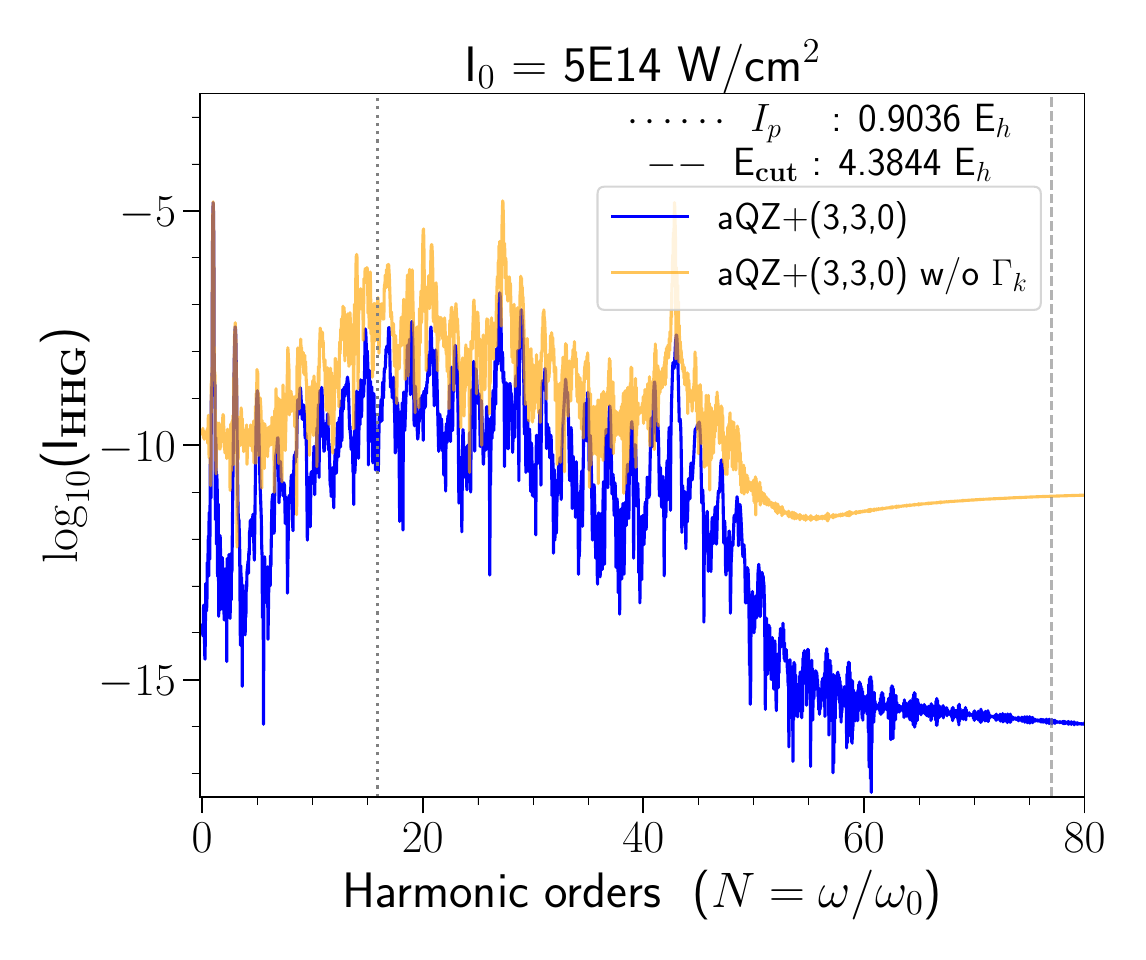}
    \caption{Comparison of HHG spectra of Helium atom generated by driving laser pulses of peak intensities $\text{I}_{0} = \{1,2,3,5\}\times 10^{14}$ W/cm$^{2}$, calculated with and without heuristic lifetimes using TD-CIS/aQZ+(3,3,0).}
    \label{fig:he_aqz_330_ltcomp_hhg_comp}
\end{figure}

\begin{figure}[h]
    \includegraphics[width=.49\textwidth]{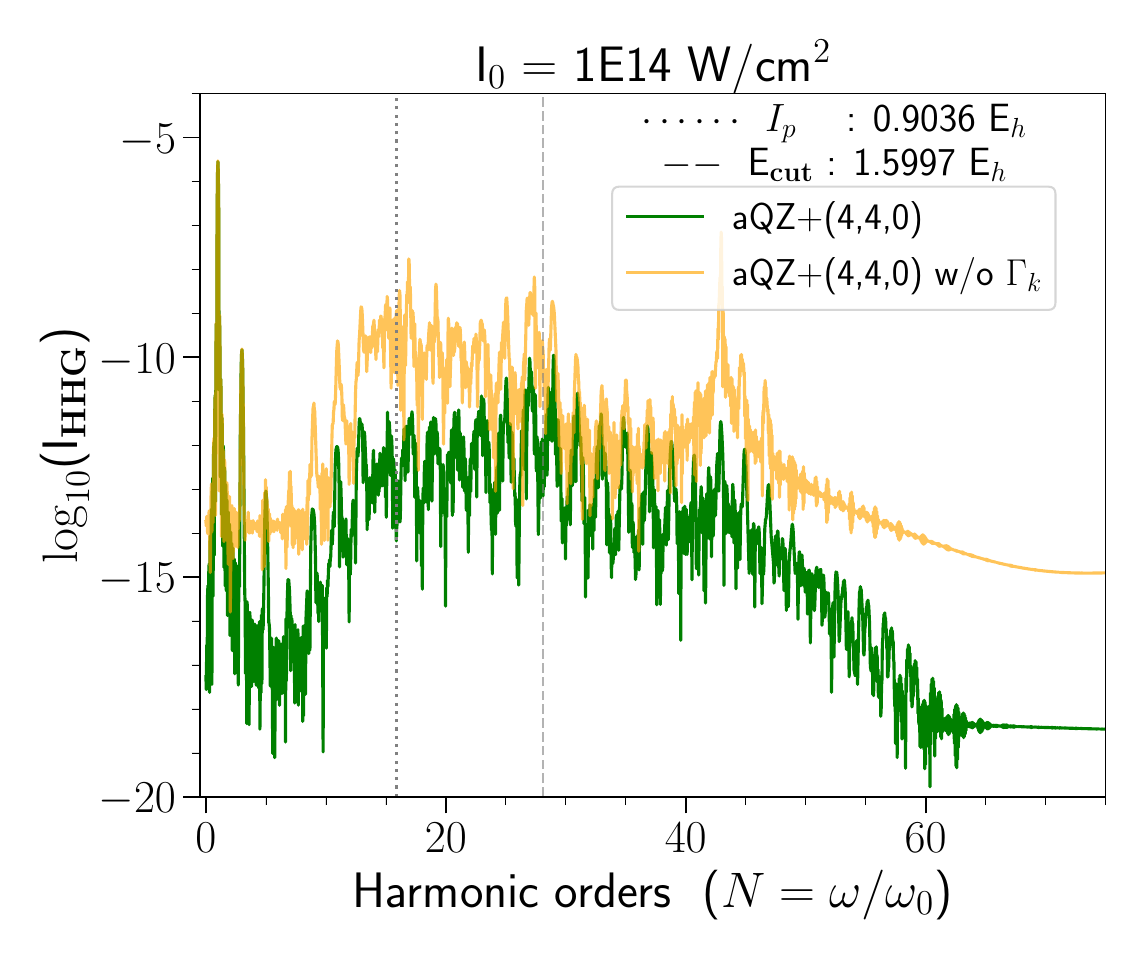}
    \includegraphics[width=.49\textwidth]{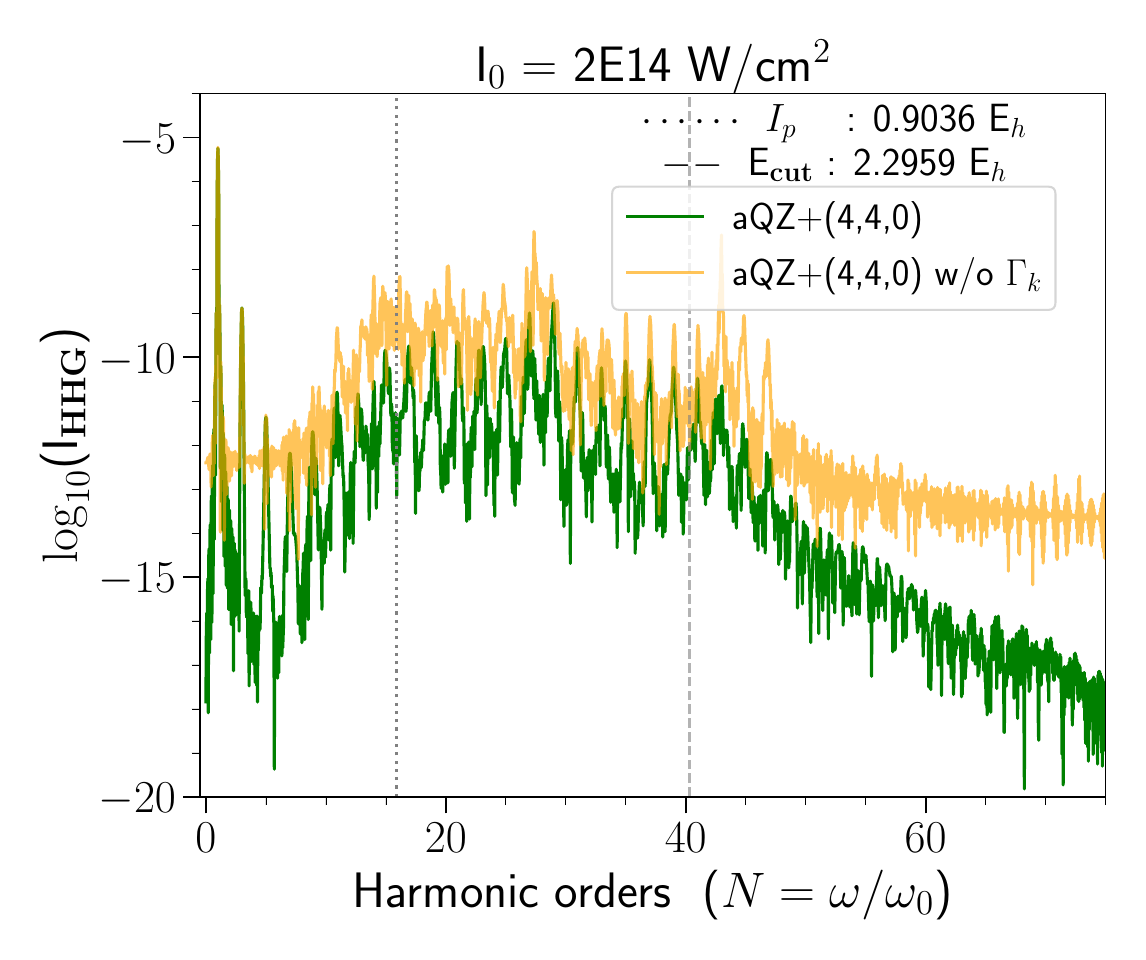}
    \includegraphics[width=.49\textwidth]{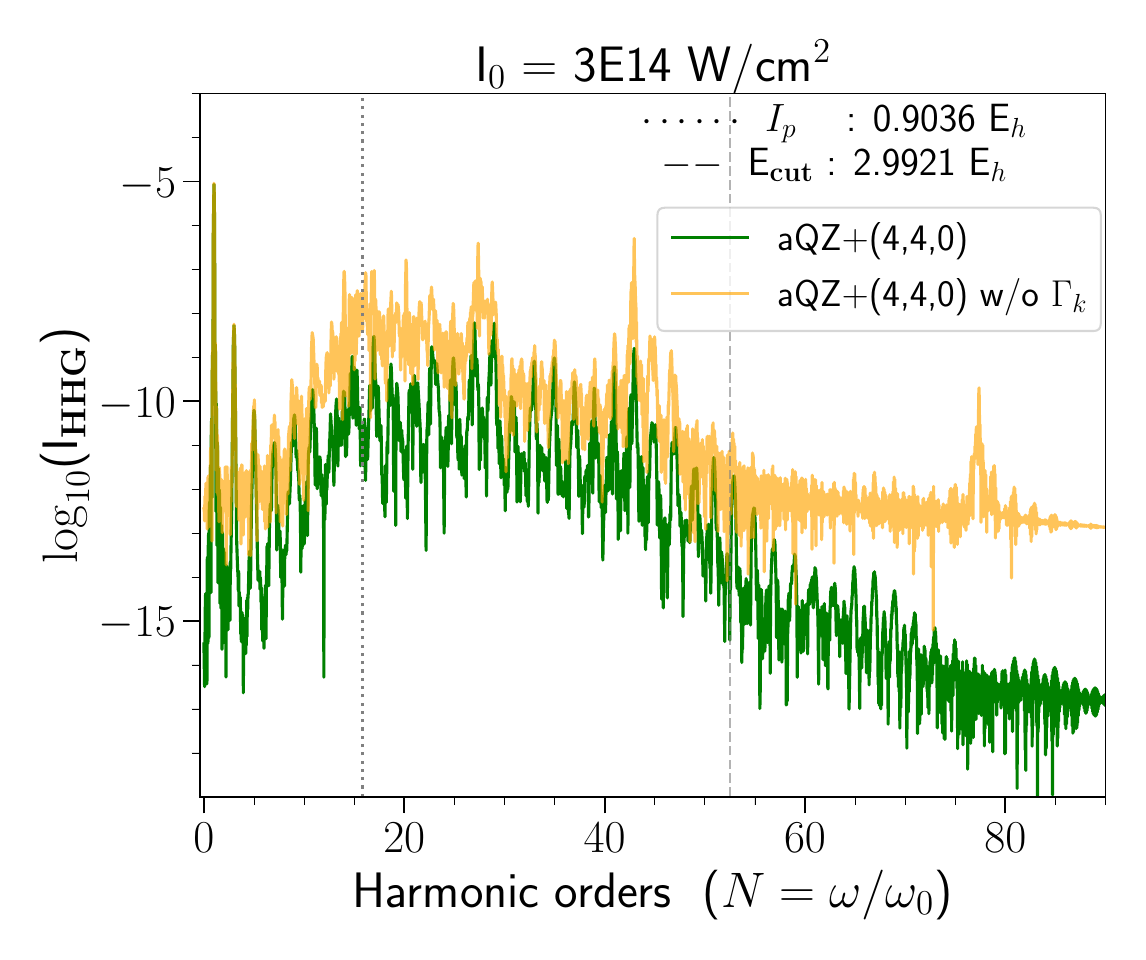}
    \includegraphics[width=.49\textwidth]{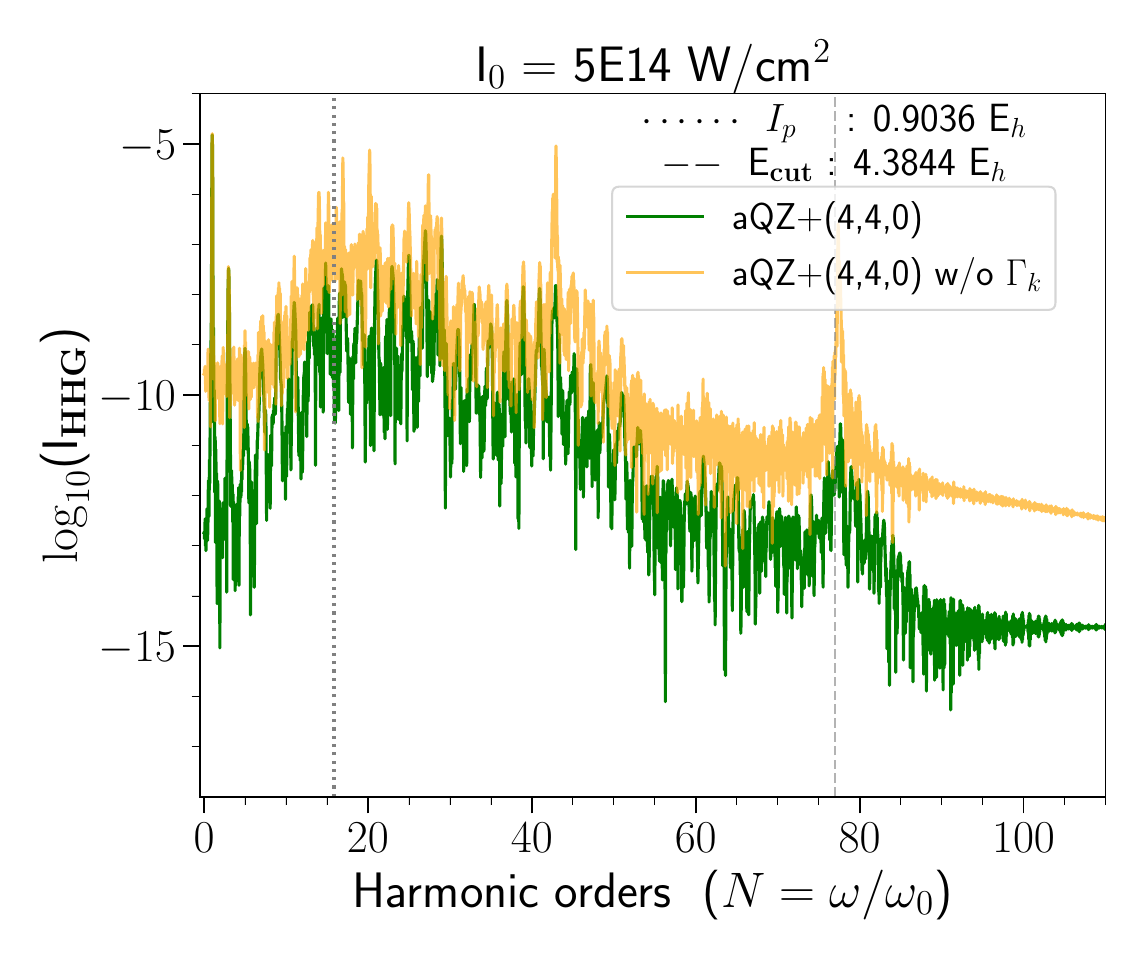}
    \caption{Comparison of HHG spectra of Helium atom generated by driving laser pulses of peak intensities $\text{I}_{0} = \{1,2,3,5\}\times 10^{14}$ W/cm$^{2}$, calculated with and without heuristic lifetimes using TD-CIS/aQZ+(4,4,0).}
    \label{fig:he_aqz_440_ltcomp_hhg_comp}
\end{figure}

\begin{figure}[h]
    \includegraphics[width=.49\textwidth]{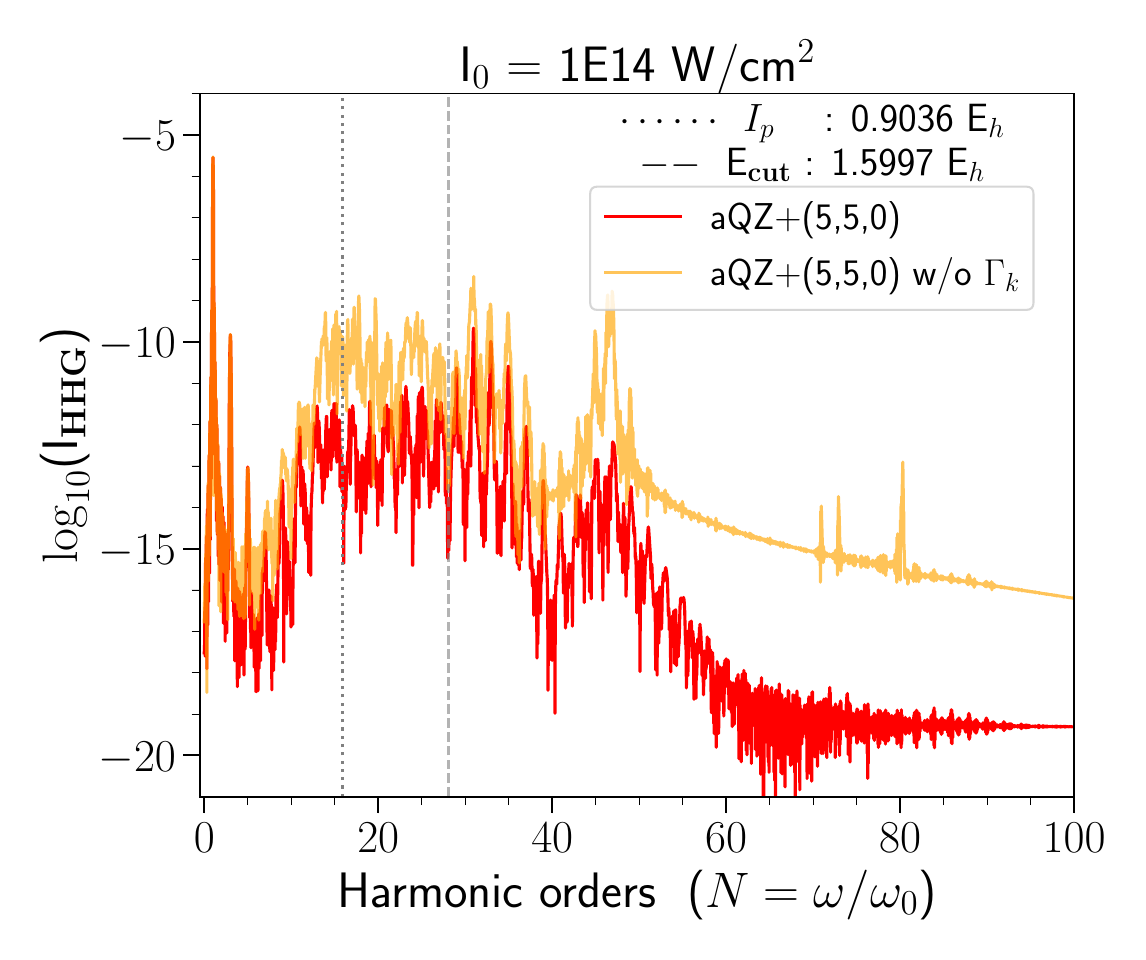}
    \includegraphics[width=.49\textwidth]{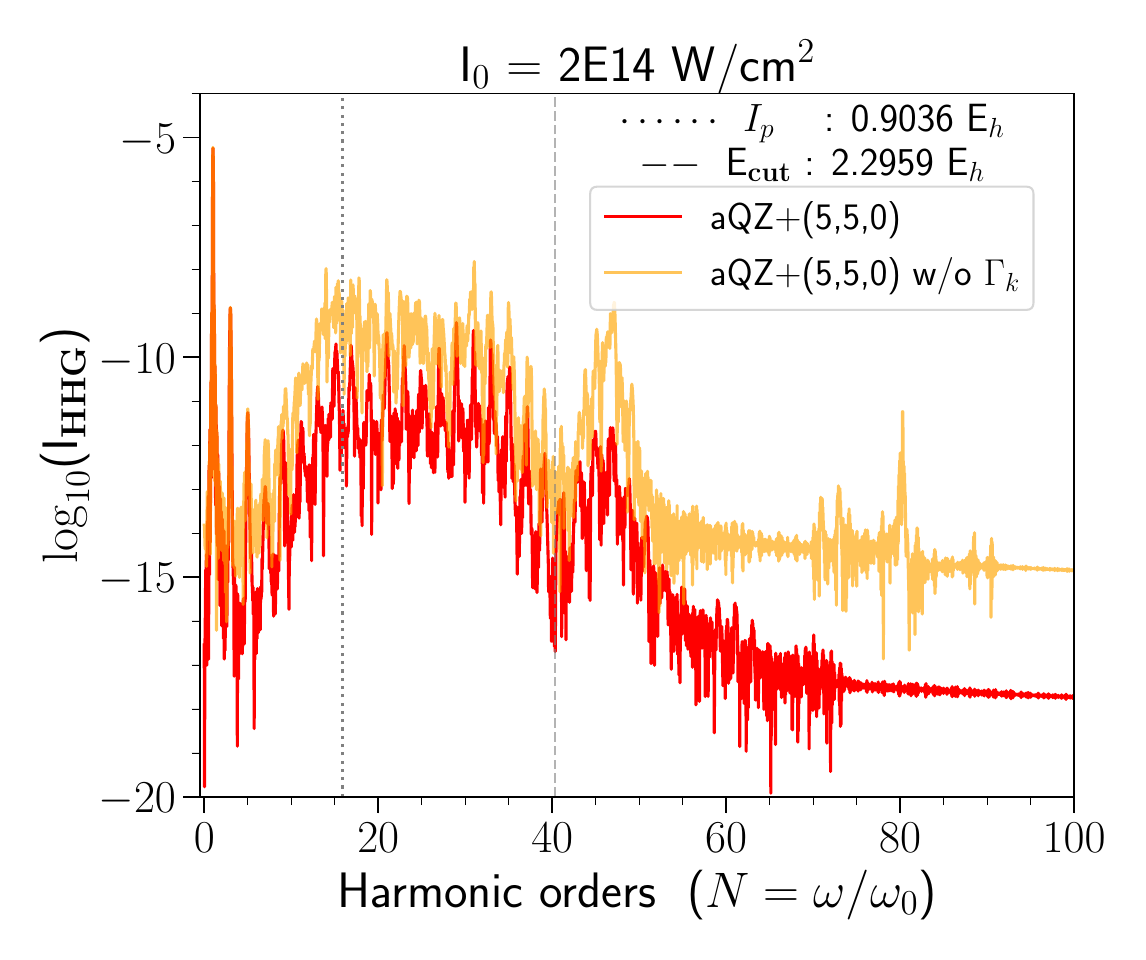}
    \includegraphics[width=.49\textwidth]{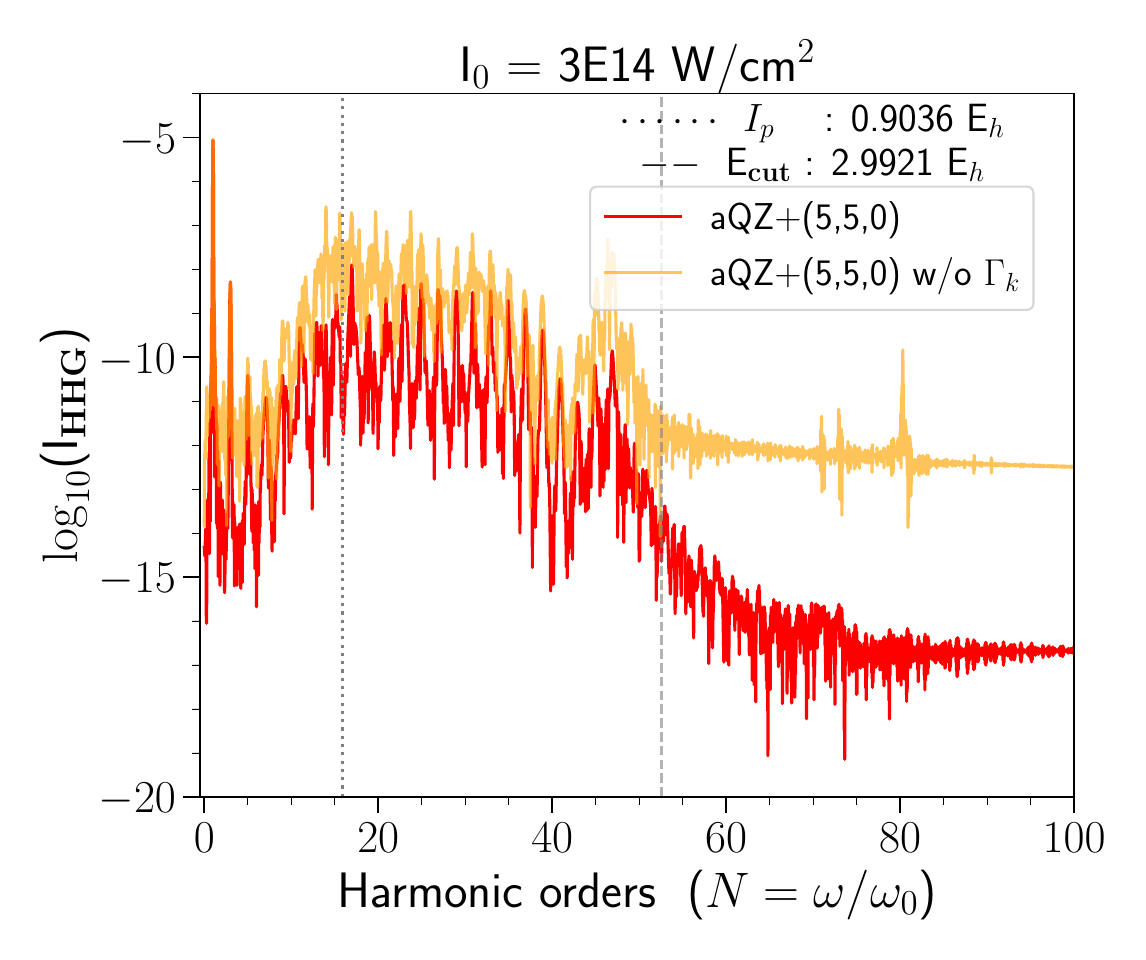}
    \includegraphics[width=.49\textwidth]{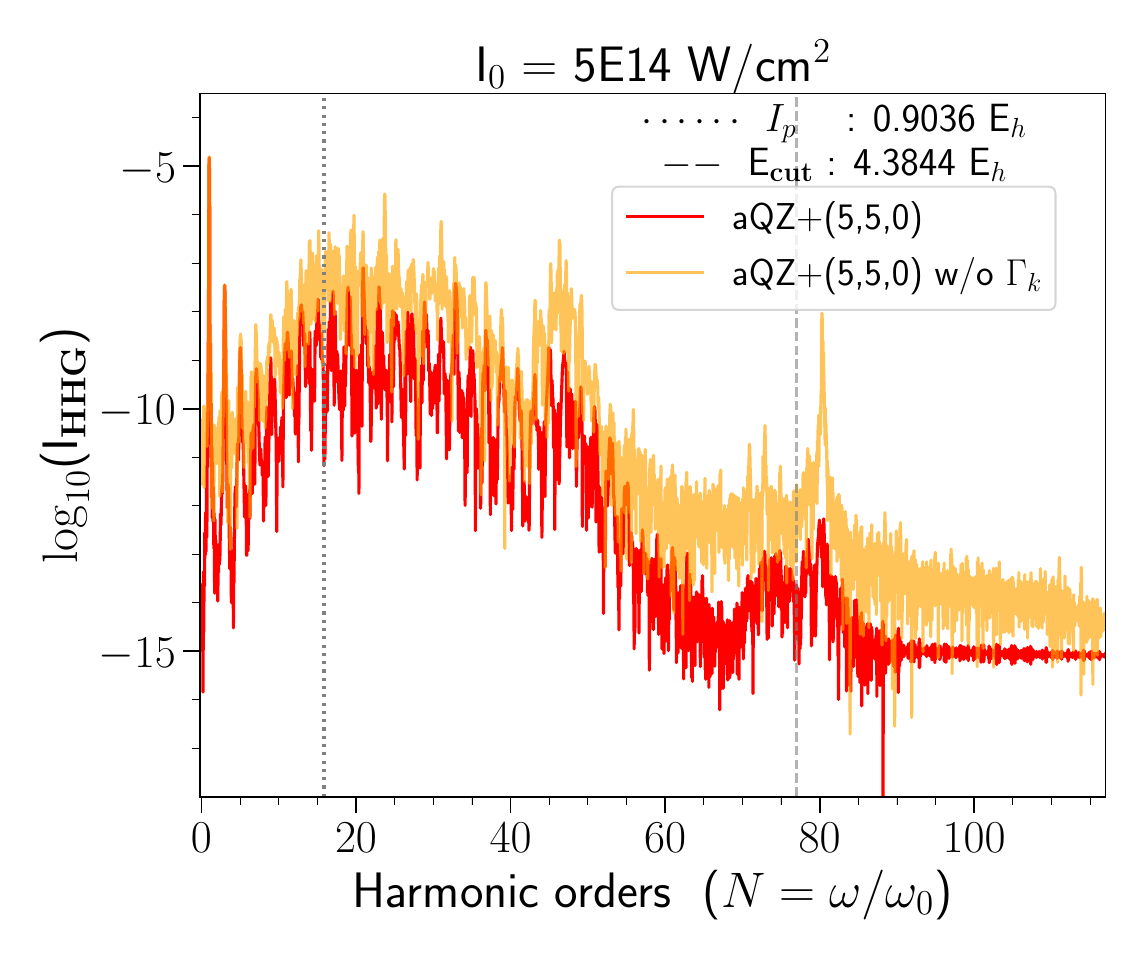}
    \caption{Comparison of HHG spectra of Helium atom generated by driving laser pulses of peak intensities $\text{I}_{0} = \{1,2,3,5\}\times 10^{14}$ W/cm$^{2}$, calculated with and without heuristic lifetimes using TD-CIS/aQZ+(5,5,0).}
    \label{fig:he_aqz_550_ltcomp_hhg_comp}
\end{figure}
\subsection{Comparison of harmonic spectra generated by different pulse envelopes}
\renewcommand{\arraystretch}{1.5}
\begin{table}
  \caption{Envelope functions for cosine-squared ($\cos^{2}$), Gaussian, and trapezoidal pulses.}
  \label{table:envelopes}
  \begin{tabular}{cl}\hline
    \hline \\
    $\cos^{2}$ &
    $ f(t) = \begin{cases}
         \cos^{2}(\frac{\pi}{2\sigma} (t - t_{p})) & |t- t_{p}| \leq \sigma,\\
          0 & \text{otherwise} ,
      \end{cases}$ 
    \\ ~ \\
    Gaussian & 
    $ f(t) = \begin{cases}
         \exp\left[-4\ln2 \left(\frac{t-t_{p}}{\sigma}\right)^{2} \right] & |t- t_{p}| \leq \sigma,\\
          0 & \text{otherwise} ,
      \end{cases}  $ 
    \\ ~ \\
    Trapezoidal &
    $ f(t) = 
      \begin{cases}
        \frac{\sigma+(t-t_{p})}{0.6 \sigma} &  -\sigma \leq (t-t_{p}) < -0.4\sigma,\\
         1.0 & -0.4\sigma \leq (t-t_{p}) \leq 0.4\sigma,\\
         \frac{\sigma-(t-t_{p})}{0.6 \sigma}  &  0.4\sigma < (t-t_{p}) \leq \sigma,\\
         0 & \text{otherwise} ,
      \end{cases}$ 
      \\ ~ \\
      \hline
      \hline
  \end{tabular}
\end{table}

\begin{figure}[h]
    \includegraphics[width=\textwidth]{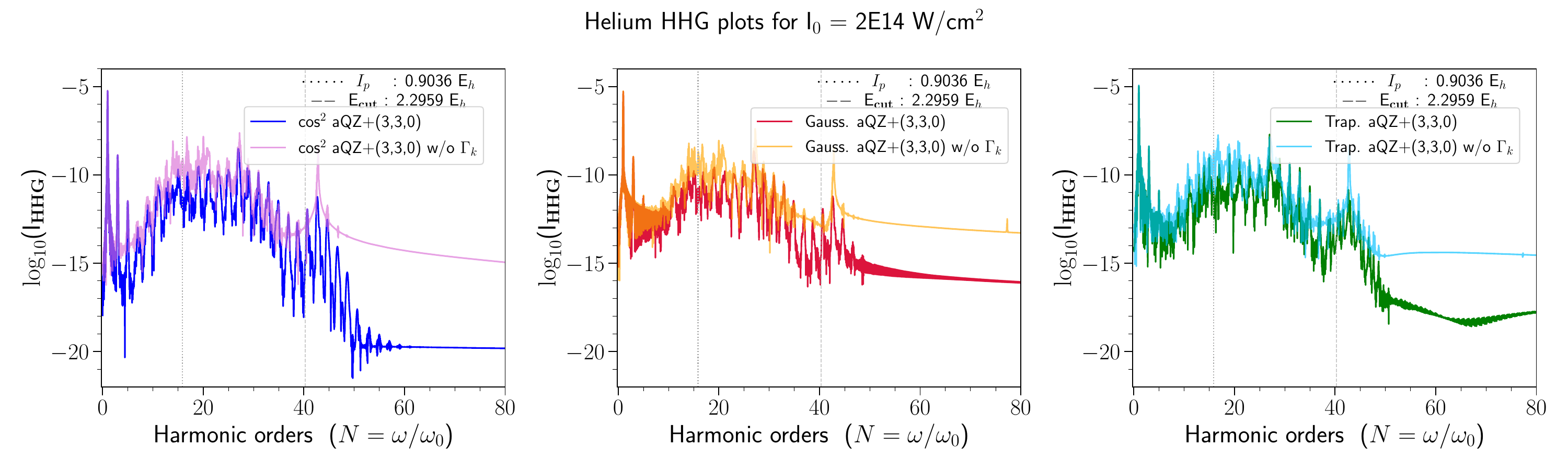}
    \caption{Comparison of HHG spectra of Helium atom generated by driving pulses with different envelop shapes and a peak intensity $\text{I}_{0} = 2\times 10^{14}$ W/cm$^{2}$, calculated using TD-CIS/aQZ+(3,3,0).}
    \label{fig:he_aqz_330_2E14_pulse_hhg_comp}
\end{figure}

\begin{figure}[h]
    \includegraphics[width=\textwidth]{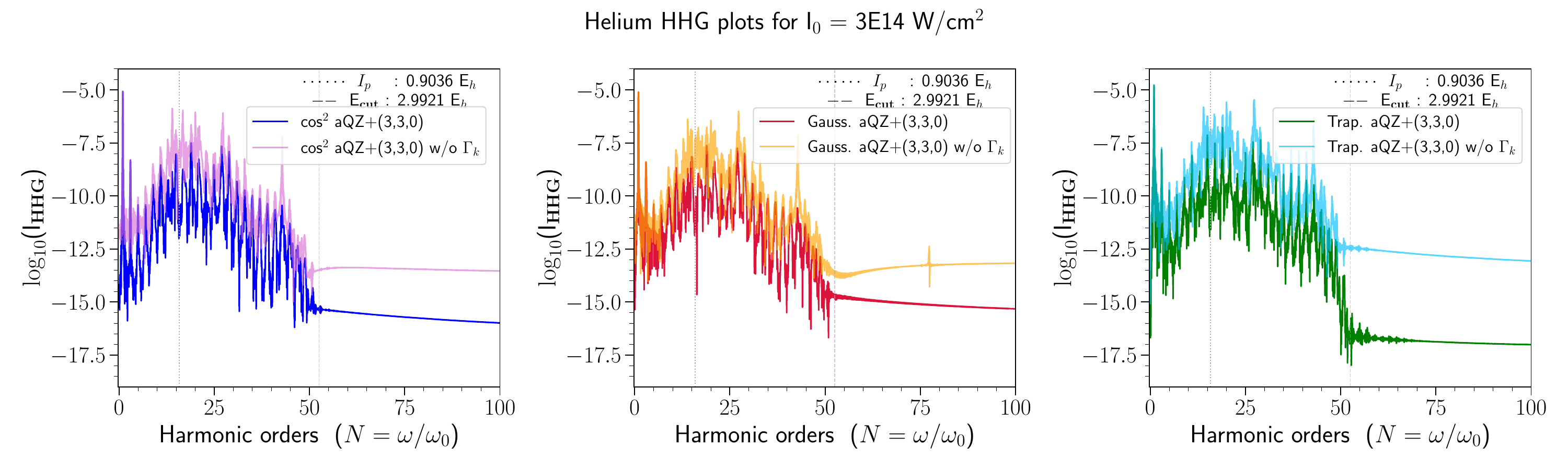}
    \caption{Comparison of HHG spectra of Helium atom generated by driving pulses with different envelop shapes and a peak intensity $\text{I}_{0} = 3\times 10^{14}$ W/cm$^{2}$, calculated using TD-CIS/aQZ+(3,3,0).}
    \label{fig:he_aqz_330_3E14_pulse_hhg_comp}
\end{figure}

\begin{figure}[h]
    \includegraphics[width=\textwidth]{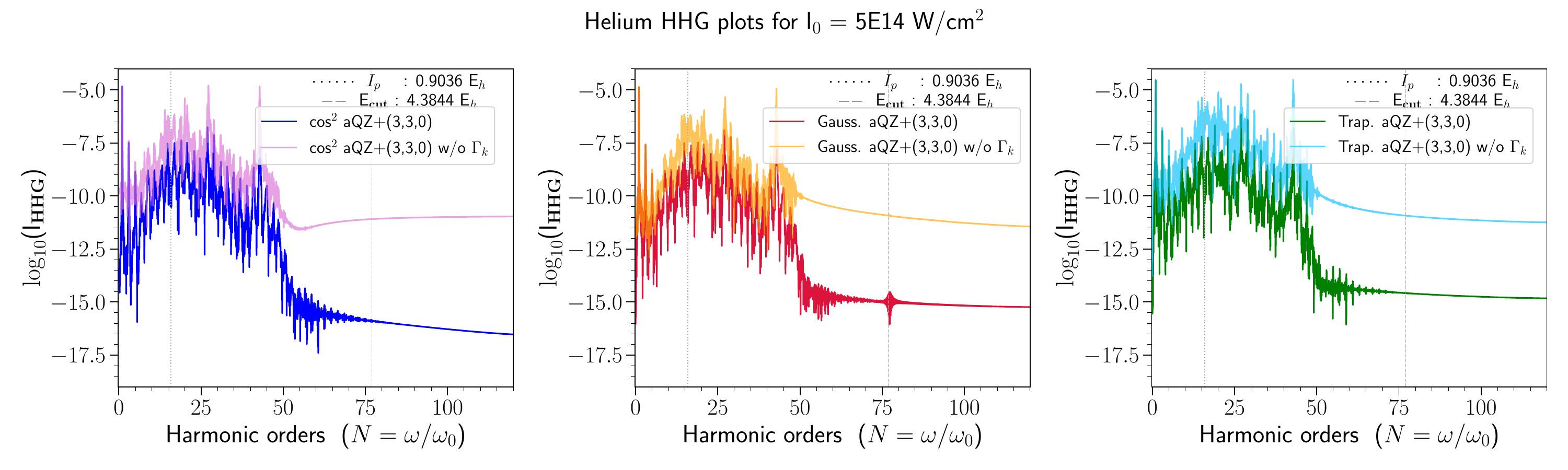}
    \caption{Comparison of HHG spectra of Helium atom generated by driving pulses with different envelop shapes and a peak intensity $\text{I}_{0} = 5\times 10^{14}$ W/cm$^{2}$, calculated using TD-CIS/aQZ+(3,3,0).}
    \label{fig:he_aqz_330_5E14_pulse_hhg_comp}
\end{figure}

\begin{figure}[h]
    \includegraphics[width=\textwidth]{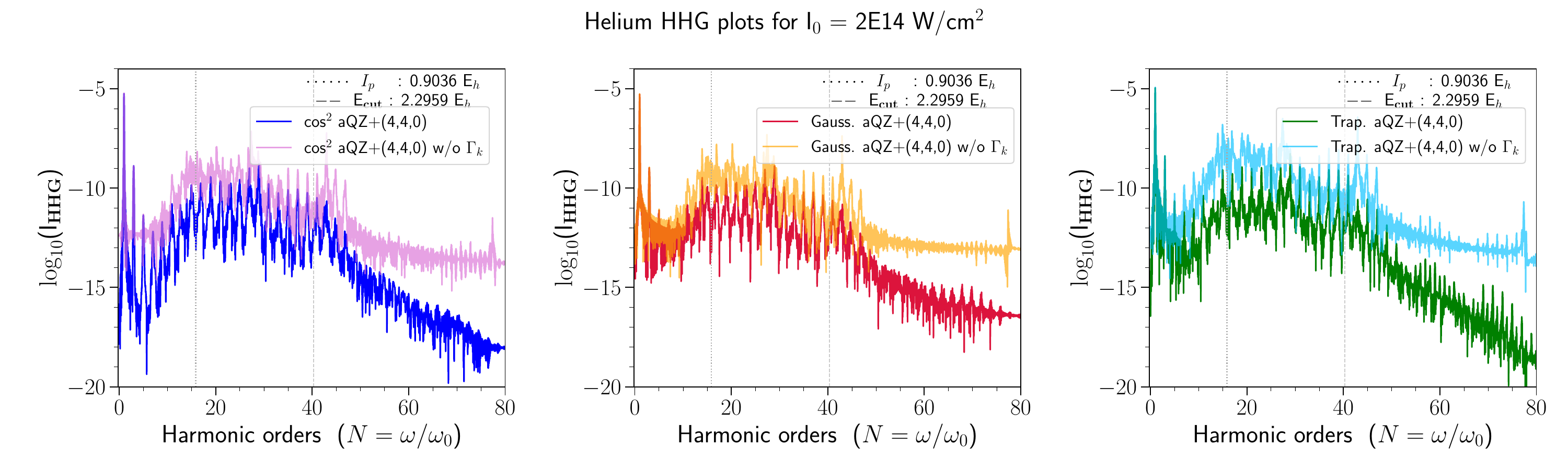}
    \caption{Comparison of HHG spectra of Helium atom generated by driving pulses with different envelop shapes and a peak intensity $\text{I}_{0} = 2\times 10^{14}$ W/cm$^{2}$, calculated using TD-CIS/aQZ+(4,4,0).}
    \label{fig:he_aqz_440_2E14_pulse_hhg_comp}
\end{figure}

\begin{figure}[h]
    \includegraphics[width=\textwidth]{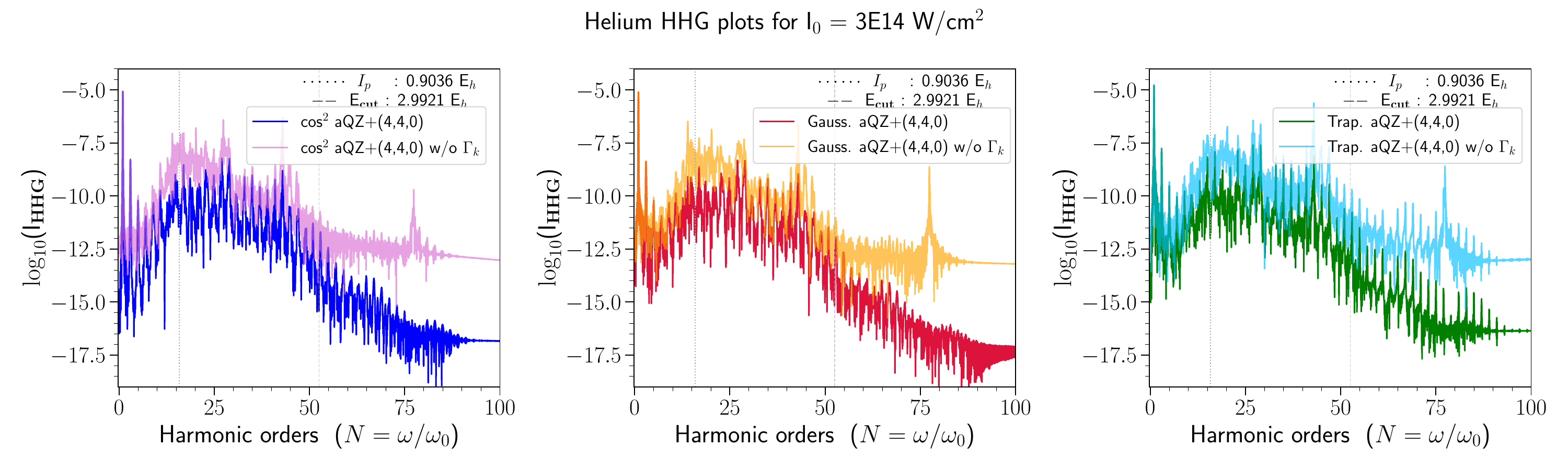}
    \caption{CComparison of HHG spectra of Helium atom generated by driving pulse with different envelop shapes and a peak intensity $\text{I}_{0} = 3\times 10^{14}$ W/cm$^{2}$, calculated using TD-CIS/aQZ+(4,4,0).}
    \label{fig:he_aqz_440_3E14_pulse_hhg_comp}
\end{figure}

\begin{figure}[h]
    \includegraphics[width=\textwidth]{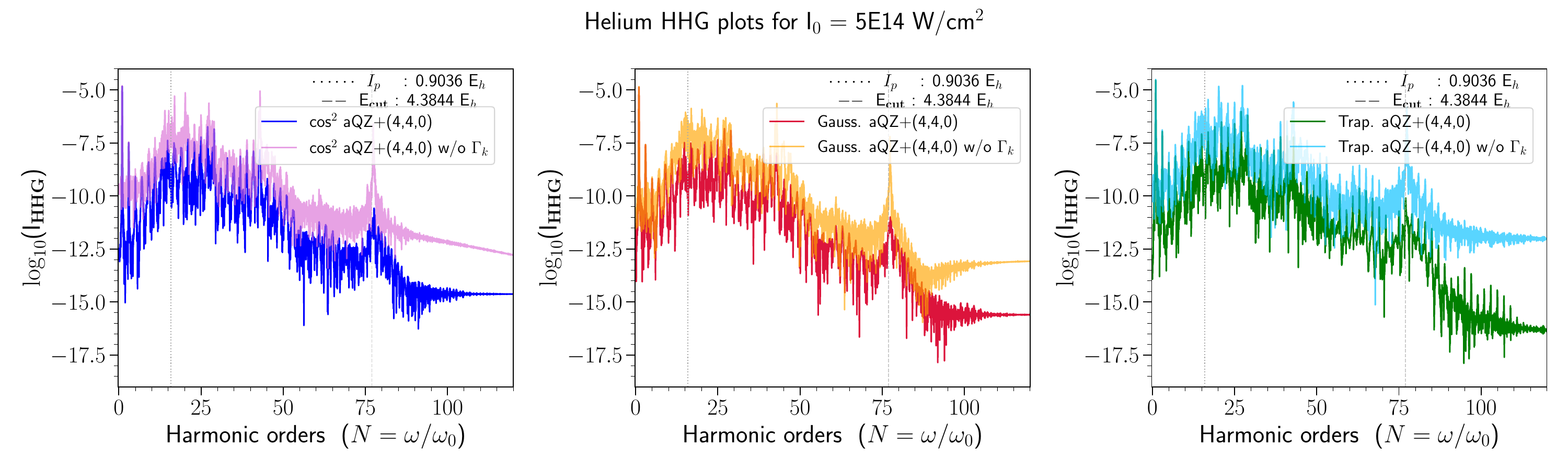}
    \caption{Comparison of HHG spectra of Helium atom generated by driving pulses with different envelop shapes and a peak intensity $\text{I}_{0} = 5\times 10^{14}$ W/cm$^{2}$, calculated using TD-CIS/aQZ+(4,4,0).}
    \label{fig:he_aqz_440_5E14_pulse_hhg_comp}
\end{figure}

\begin{figure}[h]
    \includegraphics[width=\textwidth]{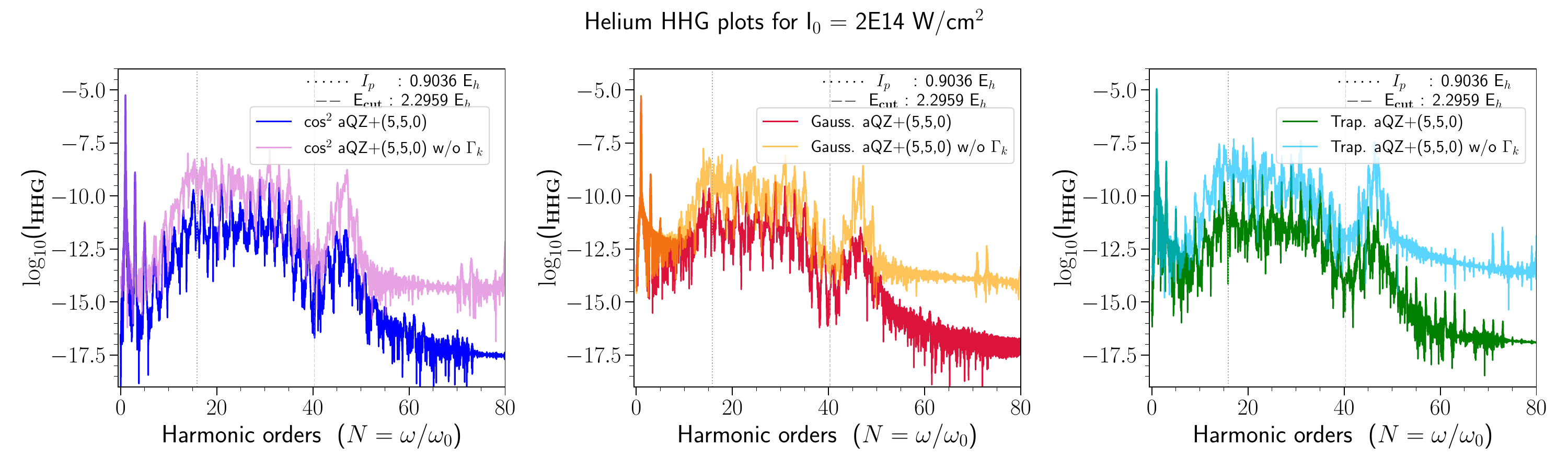}
    \caption{Comparison of HHG spectra of Helium atom generated by driving with different envelop shapes and a peak intensity $\text{I}_{0} = 2\times 10^{14}$ W/cm$^{2}$lopes, calculated using TD-CIS/aQZ+(5,5,0).}
    \label{fig:he_aqz_550_2E14_pulse_hhg_comp}
\end{figure}

\begin{figure}[h]
    \includegraphics[width=\textwidth]{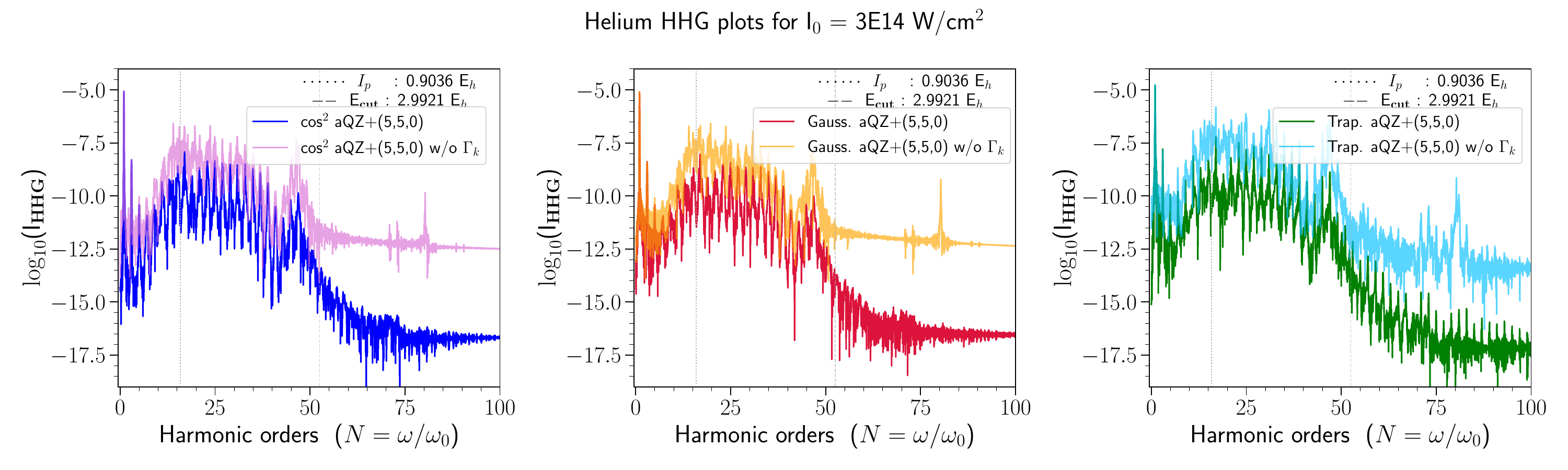}
    \caption{Comparison of HHG spectra of Helium atom generated by driving pulses with different envelop shapes and a peak intensity $\text{I}_{0} = 3\times 10^{14}$ W/cm$^{2}$, calculated using TD-CIS/aQZ+(5,5,0).}
    \label{fig:he_aqz_550_3E14_pulse_hhg_comp}
\end{figure}

\begin{figure}[h]
    \includegraphics[width=\textwidth]{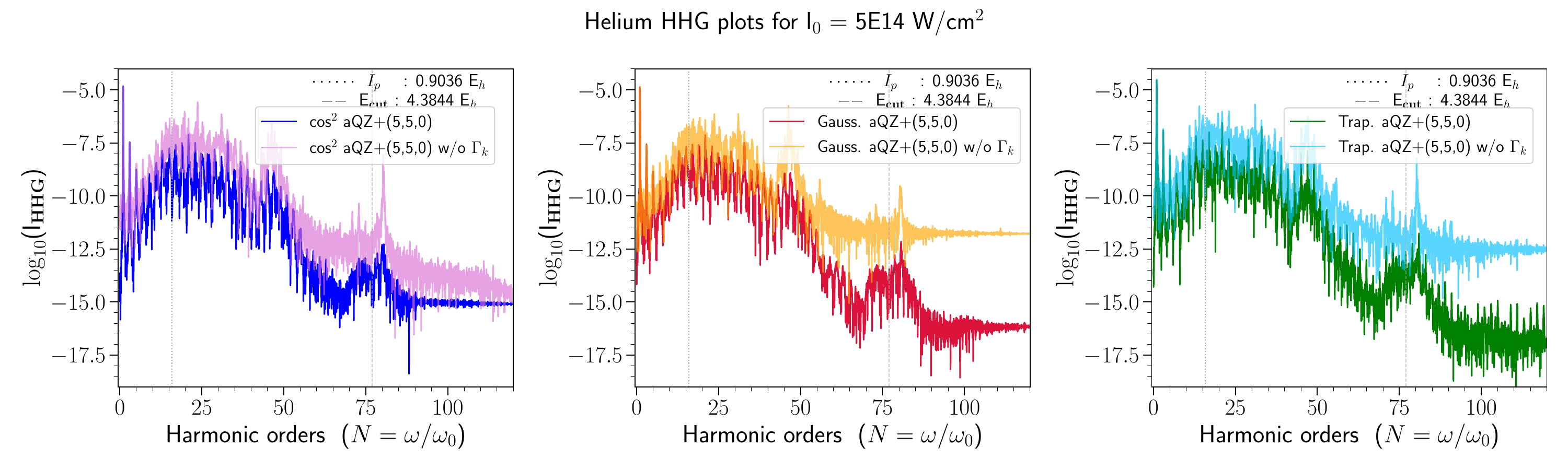}
    \caption{Comparison of HHG spectra of Helium atom generated by driving pulses with different envelop shapes and a peak intensity $\text{I}_{0} = 5\times 10^{14}$ W/cm$^{2}$, calculated using TD-CIS/aQZ+(5,5,0).}
    \label{fig:he_aqz_550_5E14_pulse_hhg_comp}
\end{figure}

\begin{figure}[h]
    \includegraphics[width=.49\textwidth]{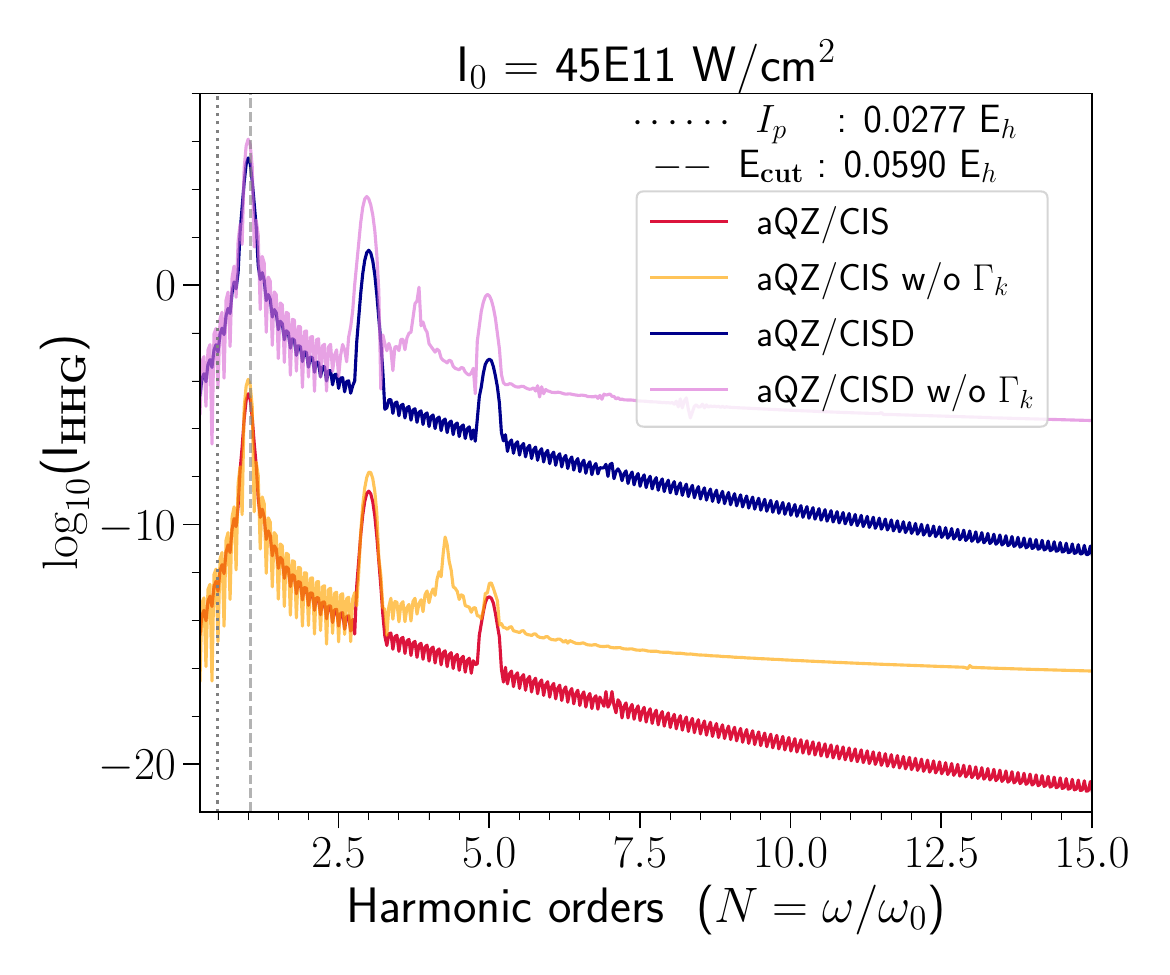}
    \includegraphics[width=.49\textwidth]{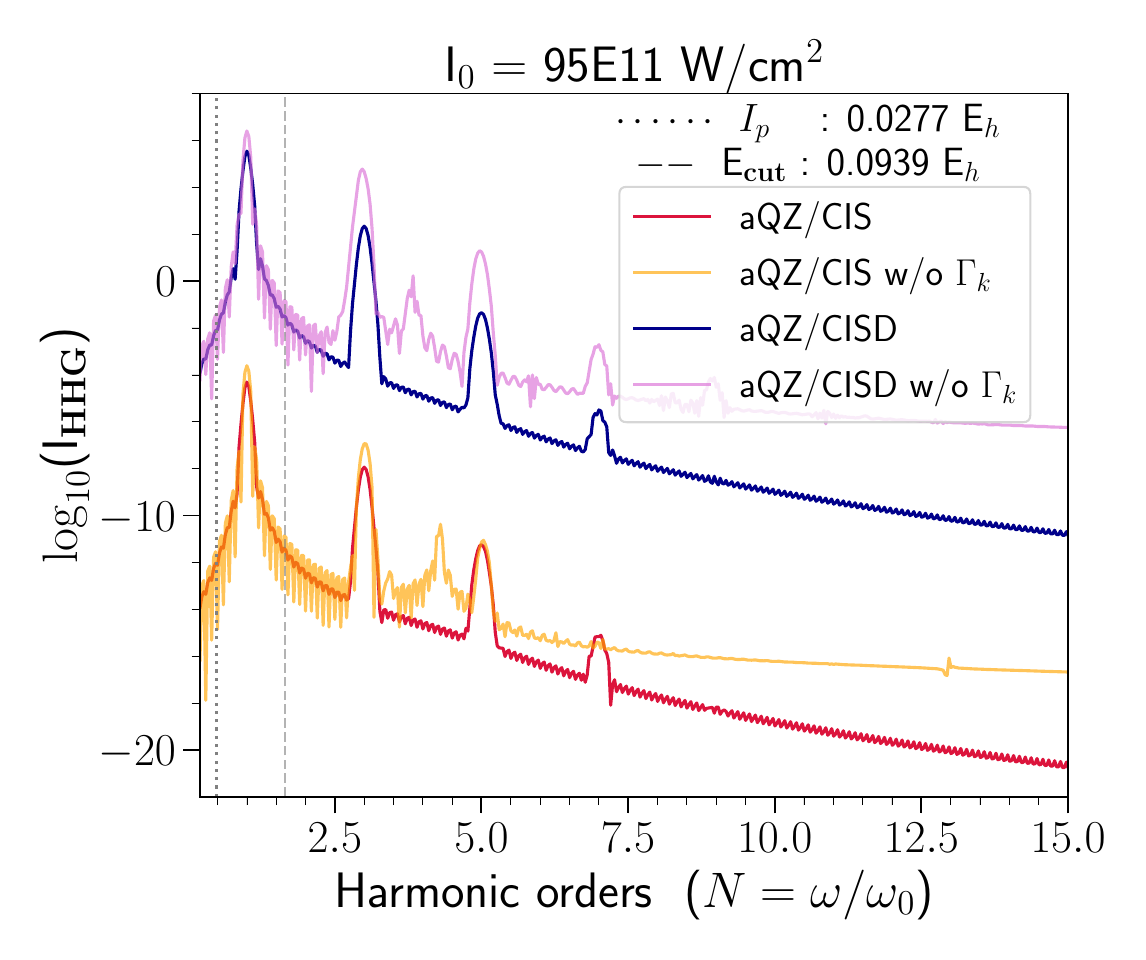}
    \includegraphics[width=.49\textwidth]{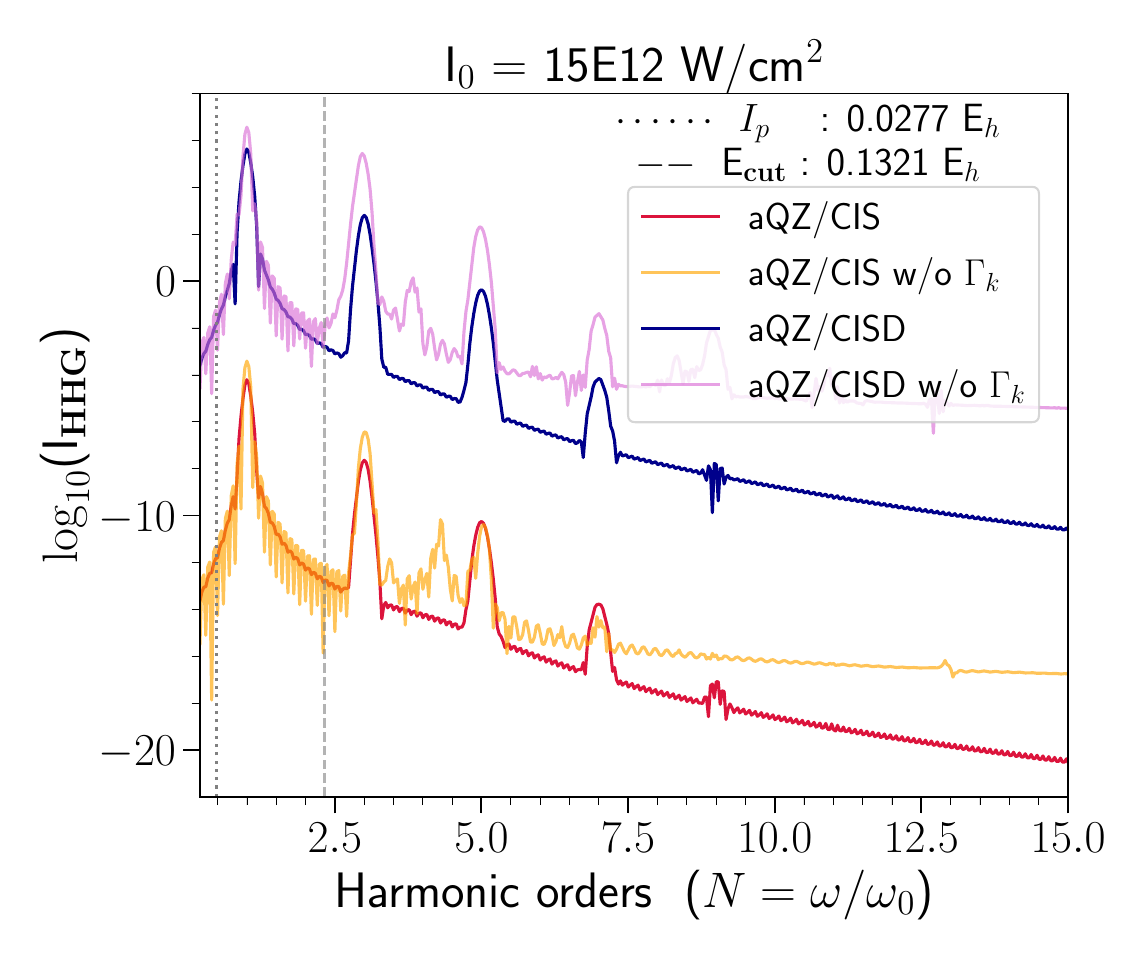}
    \caption{Comparison of HHG spectra of Hydride anion (H$^{-}$) generated by driving laser pulses of peak intensities $\text{I}_{0} = \{4.5,9.5,15\}\times 10^{12}$ W/cm$^{2}$, calculated using TD-CIS/aQZ and TD-CISD/aQZ. For clarity, the spectra obtained with TD-CIS are upshifted by +10.}
    \label{fig:h_minus_fci_hhg_comp}
\end{figure}

\begin{figure}[h]
    \includegraphics[width=.49\textwidth]{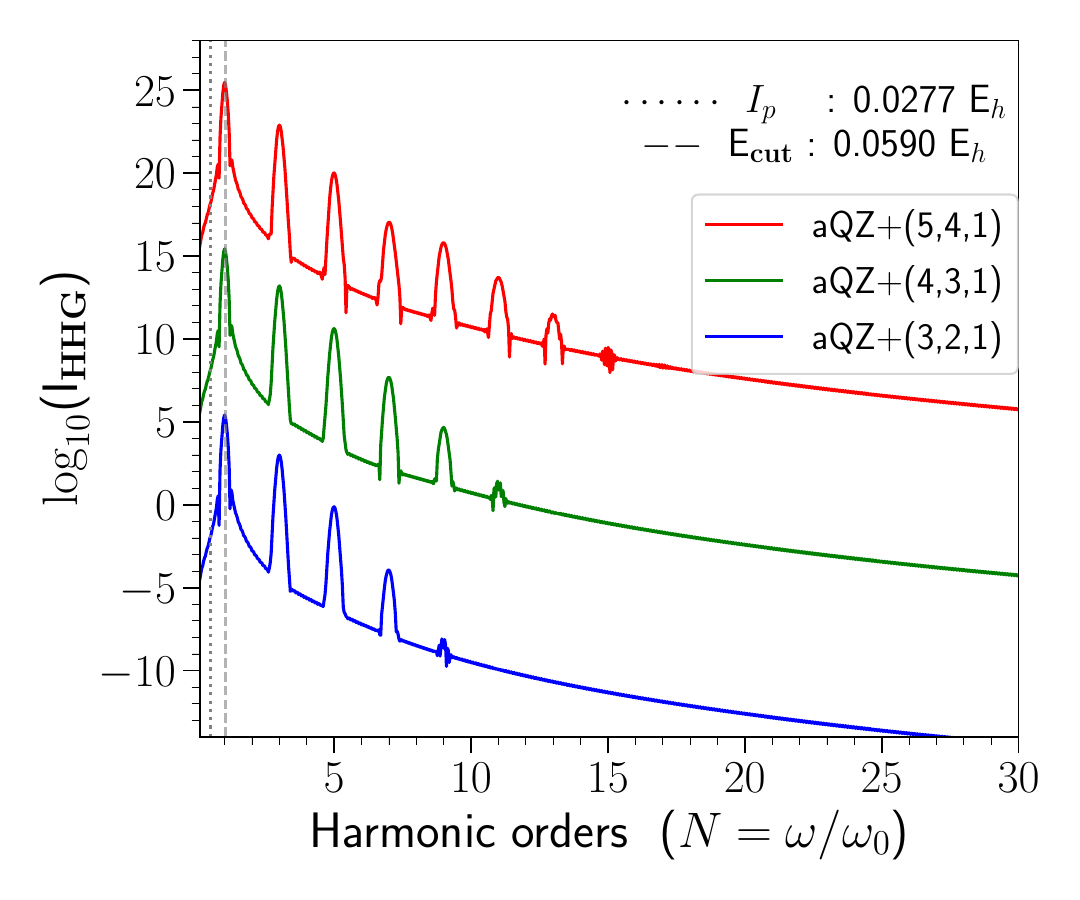}
    \includegraphics[width=.49\textwidth]{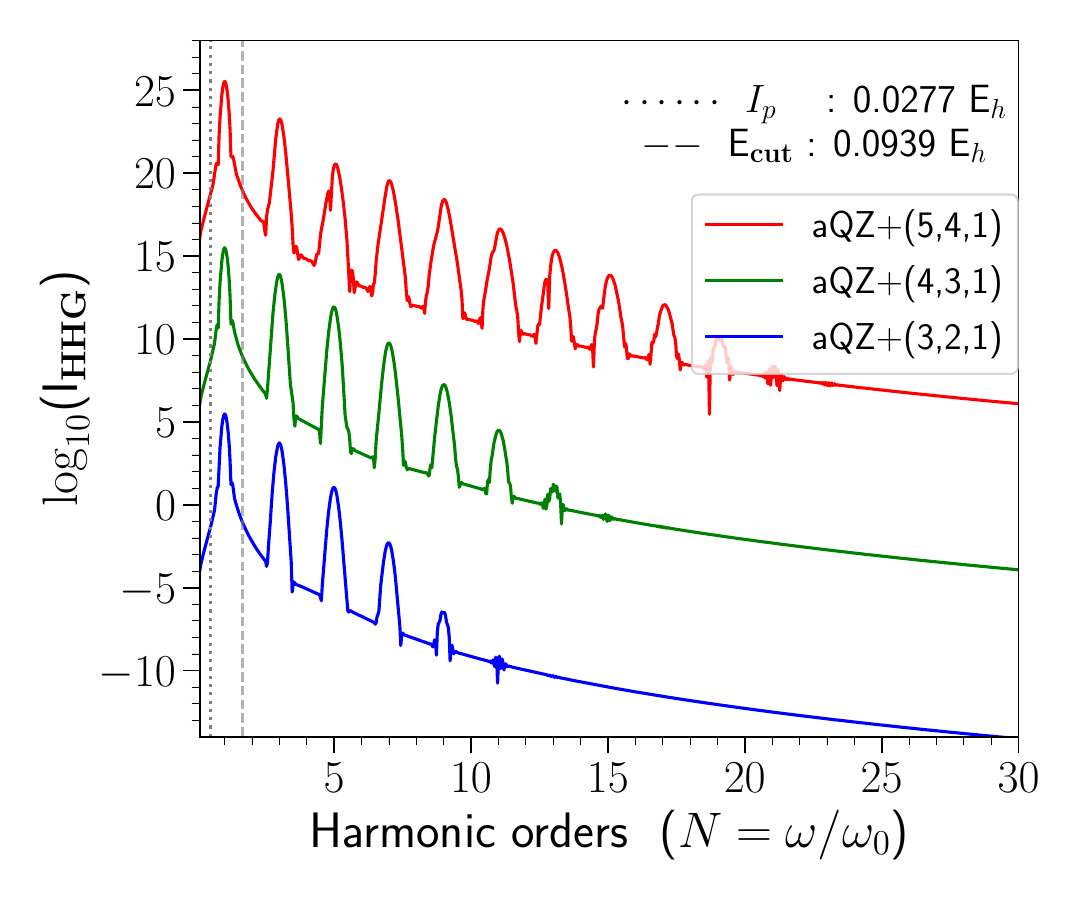}
    \includegraphics[width=.49\textwidth]{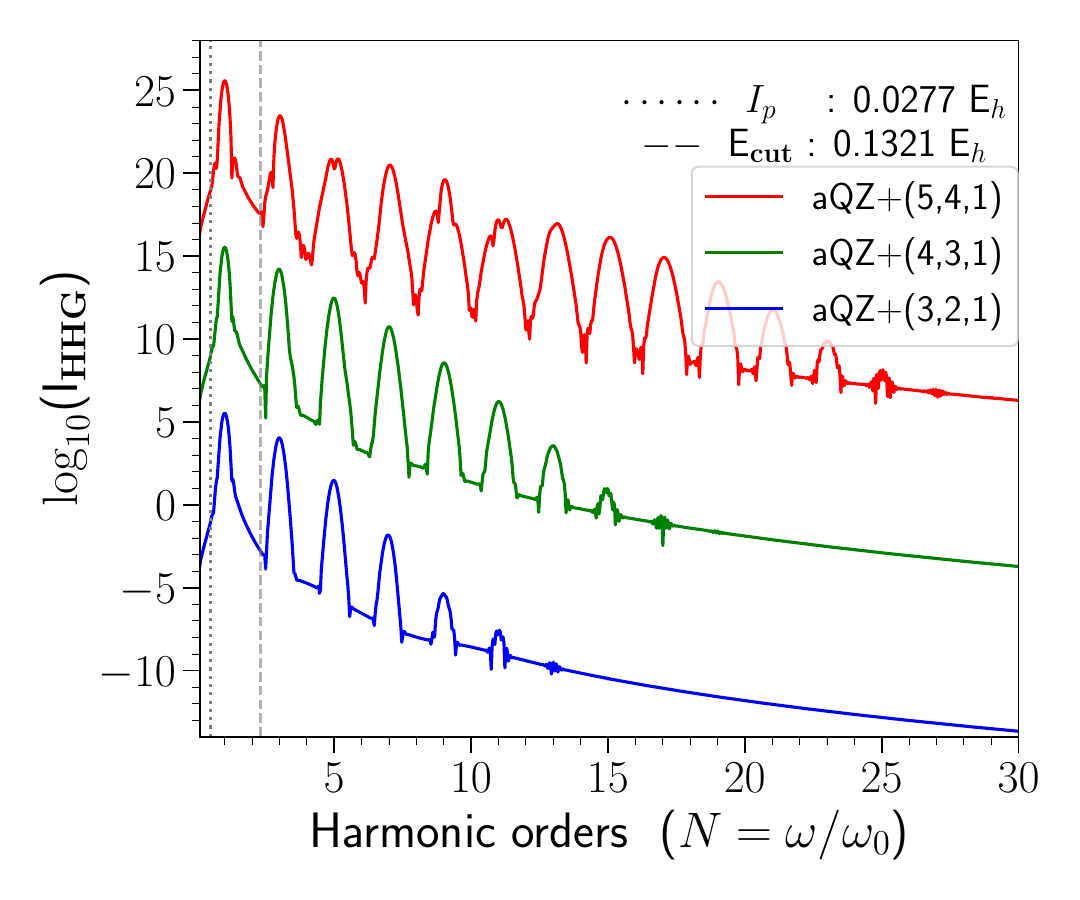}
    \caption{Comparison of HHG spectra of Hydride anion (H$^{-}$) generated by driving laser pulses of peak intensities $\text{I}_{0} = \{4.5,9.5,15\}\times 10^{12}$ W/cm$^{2}$, calculated using TD-CIS with different aQZ+(N,$l$,1) basis sets. Here, spectra calculated with different basis sets are upshifted in multiples of +10 for clarity.}
    \label{fig:h_minus_nl1_hhg_comp}
\end{figure}

\begin{figure}[h]
    \includegraphics[width=.49\textwidth]{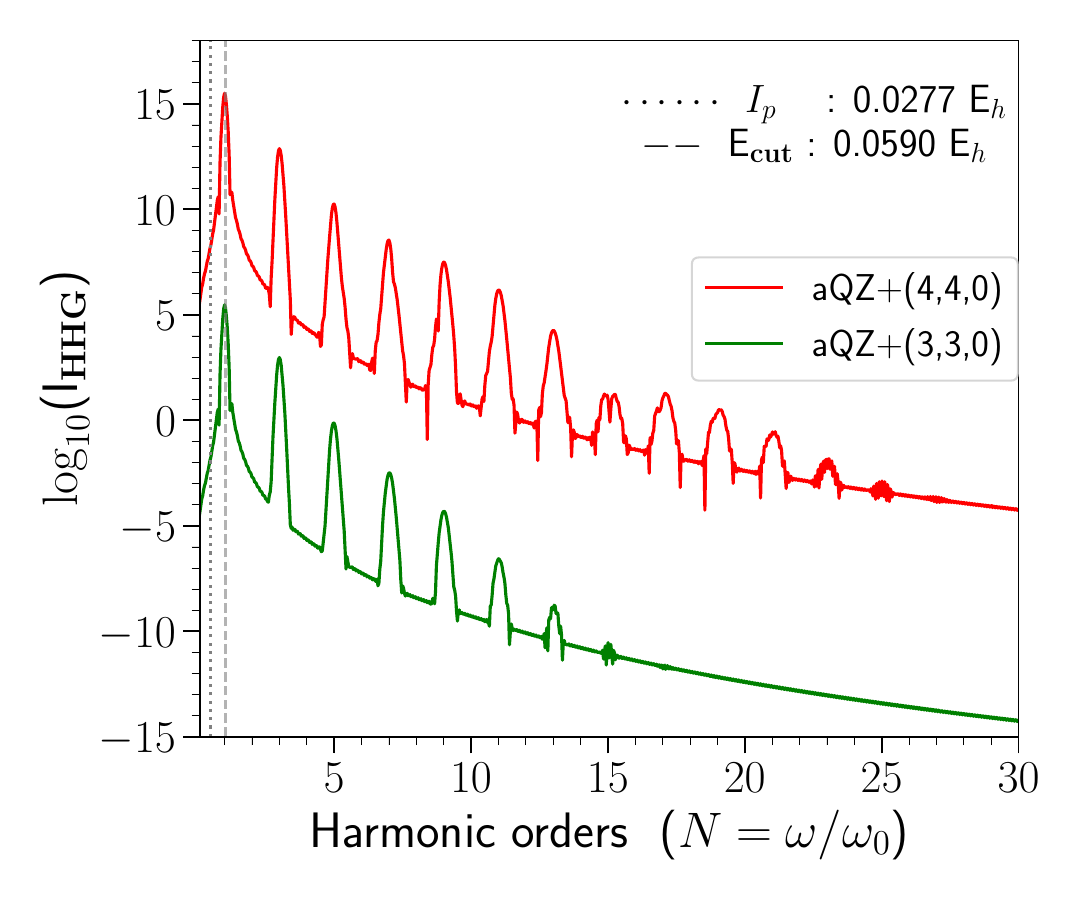}
    \includegraphics[width=.49\textwidth]{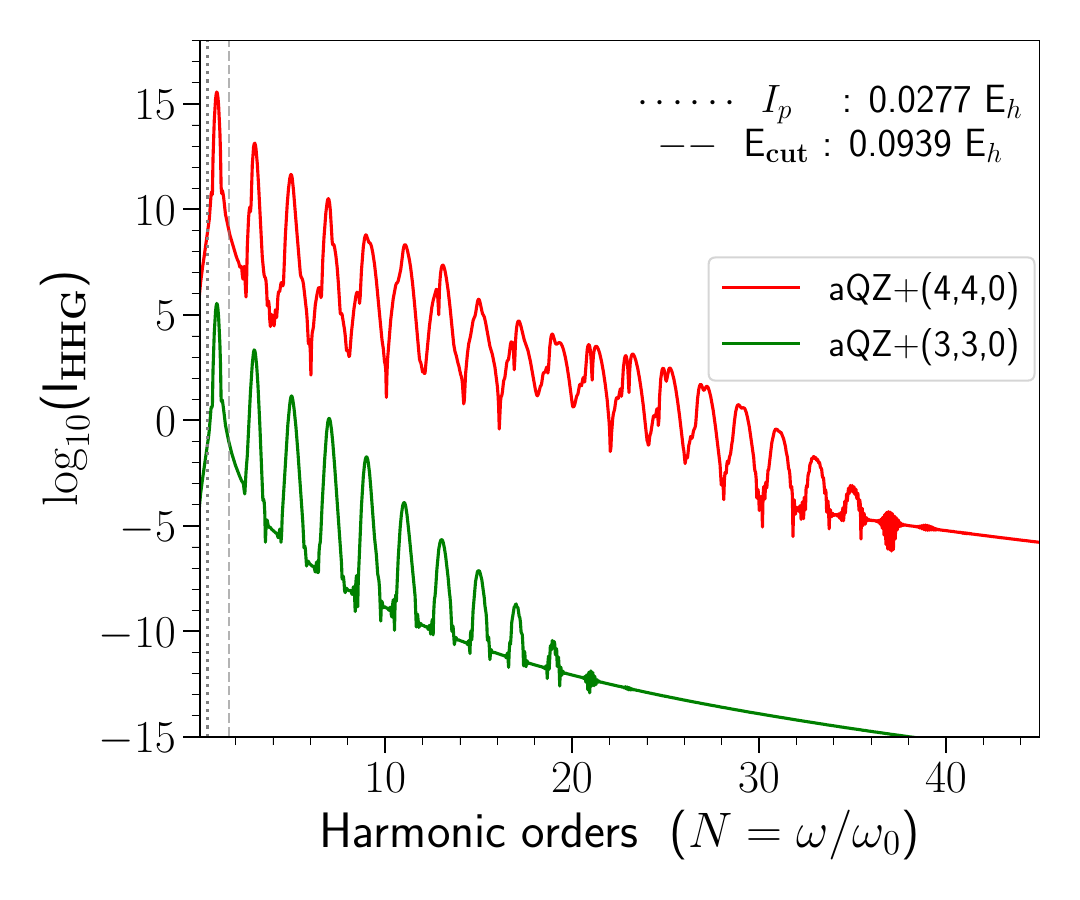}
    \includegraphics[width=.49\textwidth]{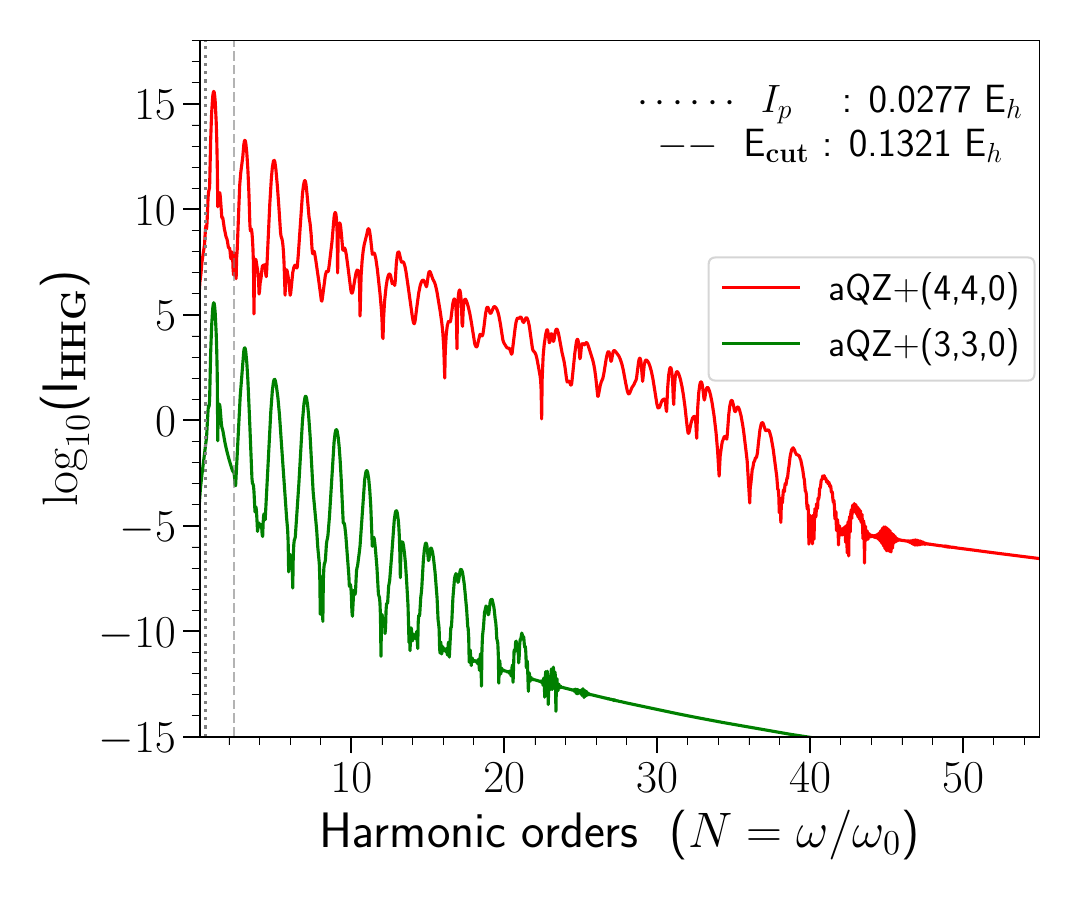}
    \caption{Comparison of HHG spectra of Hydride anion (H$^{-}$) generated by driving laser pulses of peak intensities $\text{I}_{0} = \{4.5,9.5,15\}\times 10^{12}$ W/cm$^{2}$, calculated using TD-CIS with different aQZ+(N,$l$,0) basis sets. Here, spectra calculated with different basis sets are upshifted in multiples of +10 for clarity.}
    \label{fig:h_minus_nn0_hhg_comp}
\end{figure}
  
  \begin{figure}[h]
    \includegraphics[width=.49\textwidth]{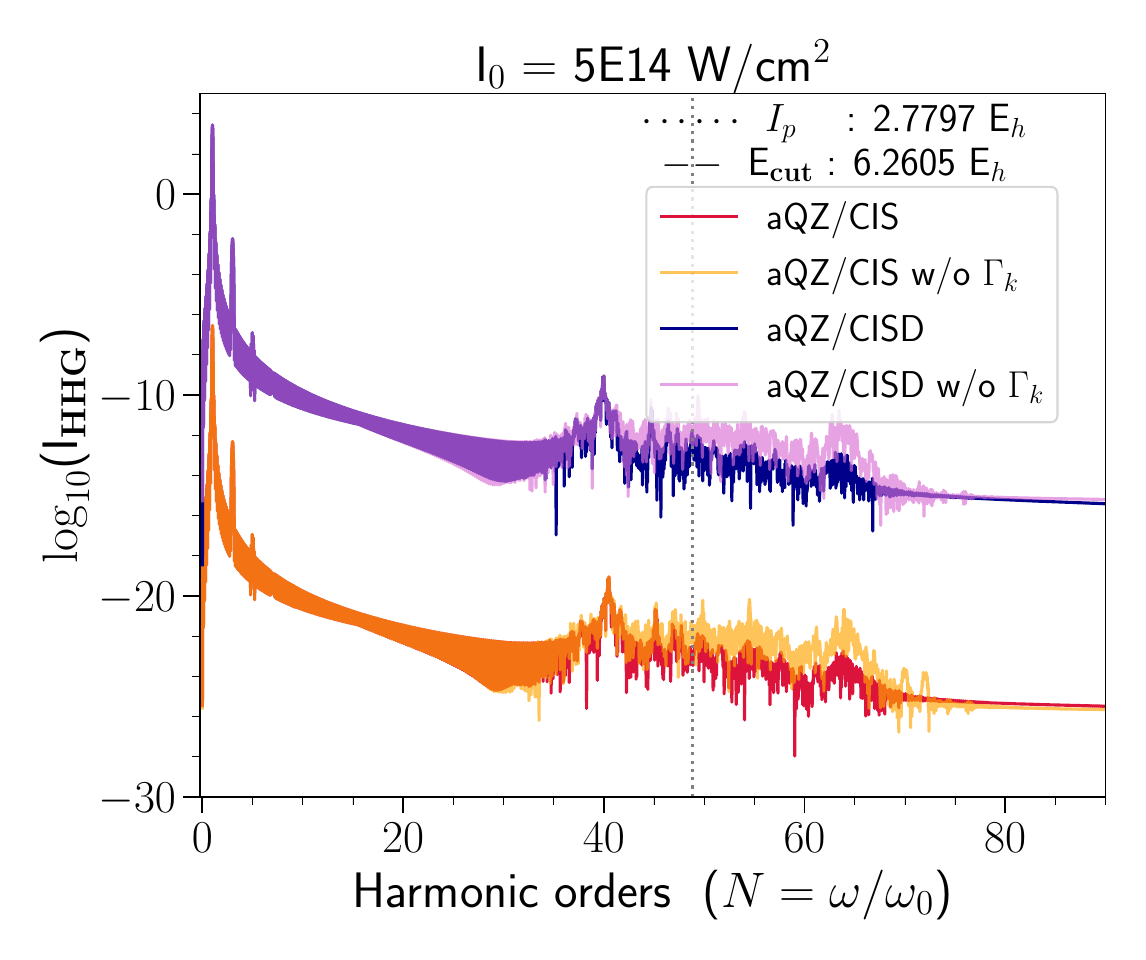}
    \includegraphics[width=.49\textwidth]{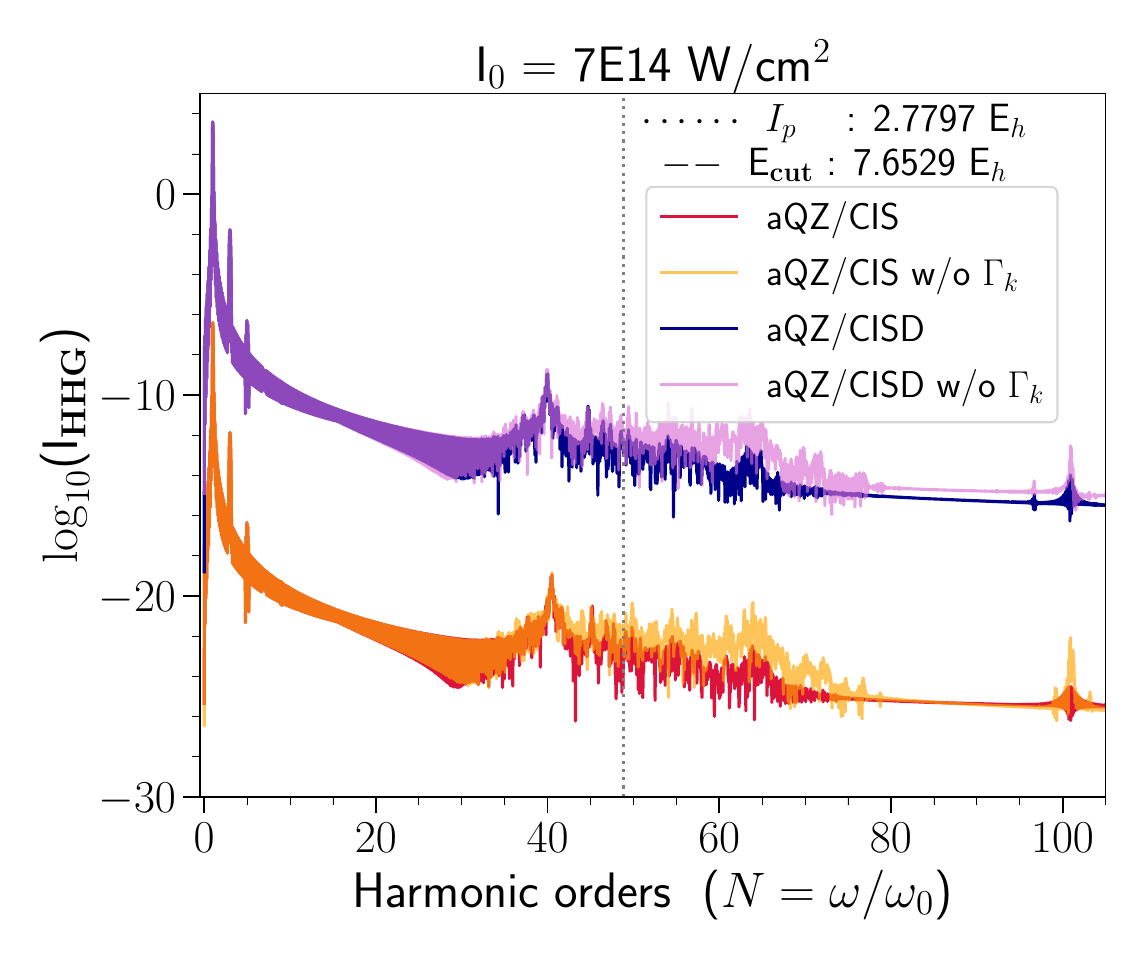}
    \includegraphics[width=.49\textwidth]{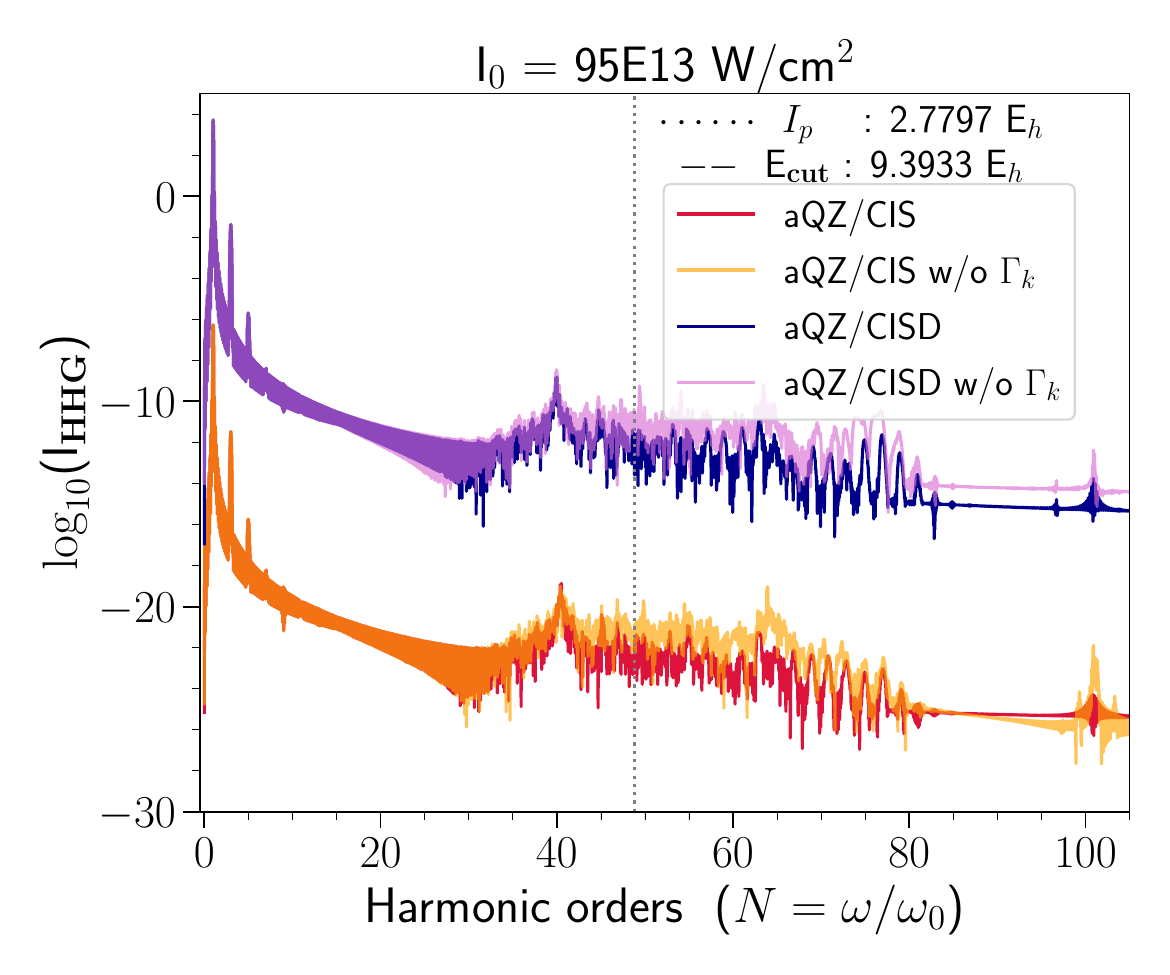}
    \caption{Comparison of HHG spectra of Lithium cation (Li$^{+}$) generated by driving laser pulses of peak intensities $\text{I}_{0} = \{5,7,9.5\}\times 10^{14}$ W/cm$^{2}$, calculated using TD-CIS/aQZ and TD-CISD/aQZ. The spectra obtained with TD-CIS are upshifted by +10, 
    for the sake of clarity.}
    \label{fig:li_plus_fci_hhg_comp}
\end{figure}

\begin{figure}[h]
    \includegraphics[width=.49\textwidth]{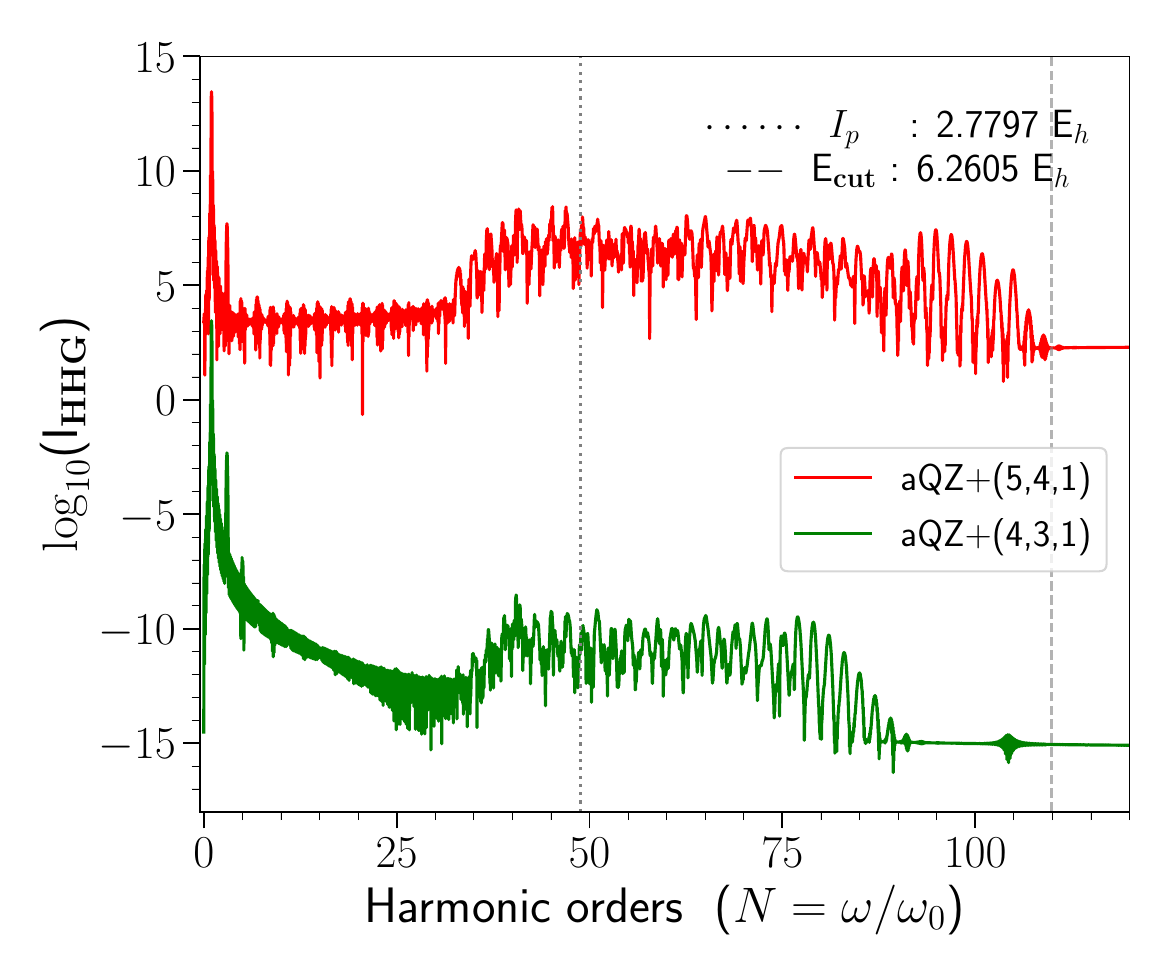}
    \includegraphics[width=.49\textwidth]{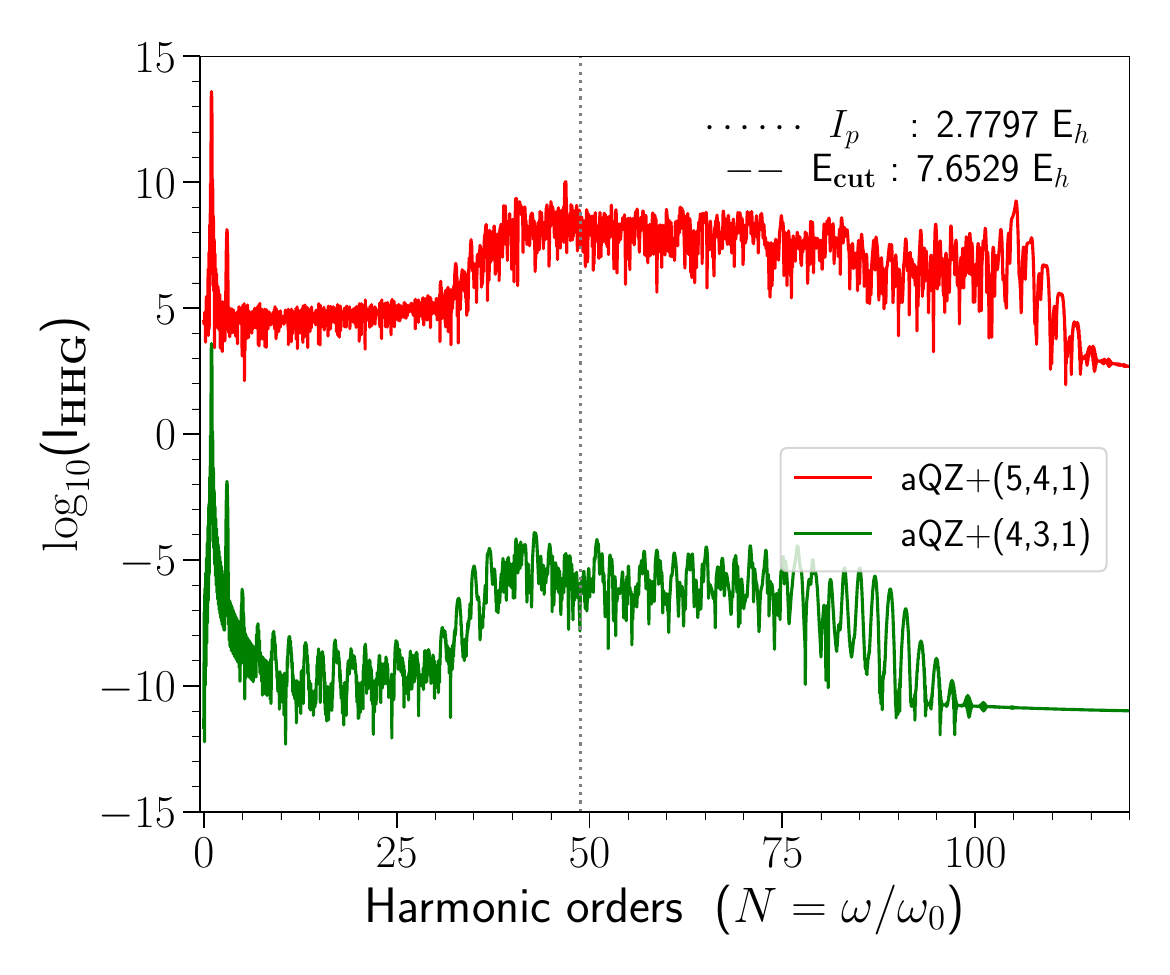}
    \includegraphics[width=.49\textwidth]{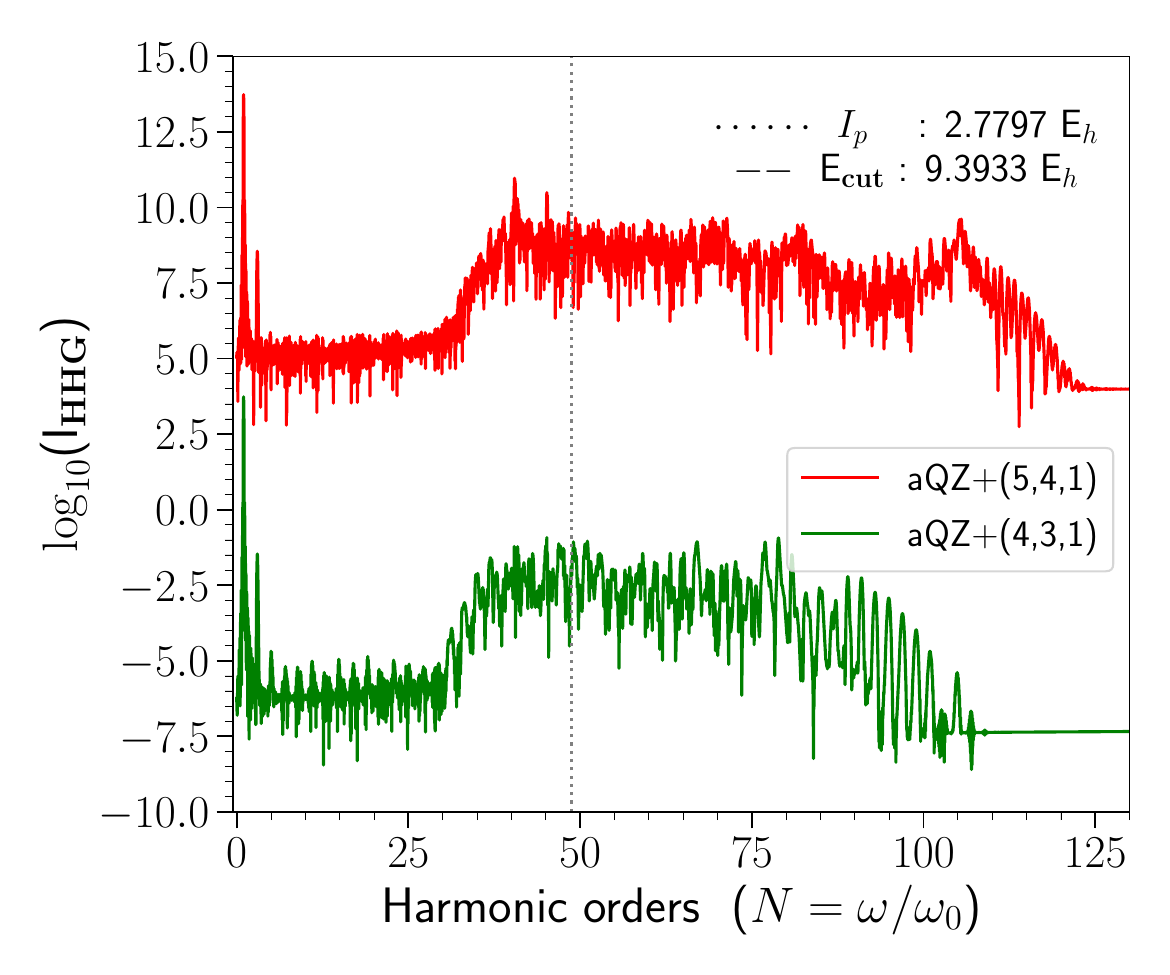}
    \caption{Comparison of HHG spectra of Lithium cation (Li$^{+}$) generated by driving laser pulses of peak intensities $\text{I}_{0} = \{5,7,9.5\}\times 10^{14}$ W/cm$^{2}$,calculated using TD-CIS with different aQZ+(N,$l$,1) basis sets. Here, spectra calculated with different basis sets are upshifted in multiples of +10 for clarity.}
    \label{fig:li_plus_nl1_hhg_comp}
\end{figure}  

\begin{figure}[h]
    \includegraphics[width=.49\textwidth]{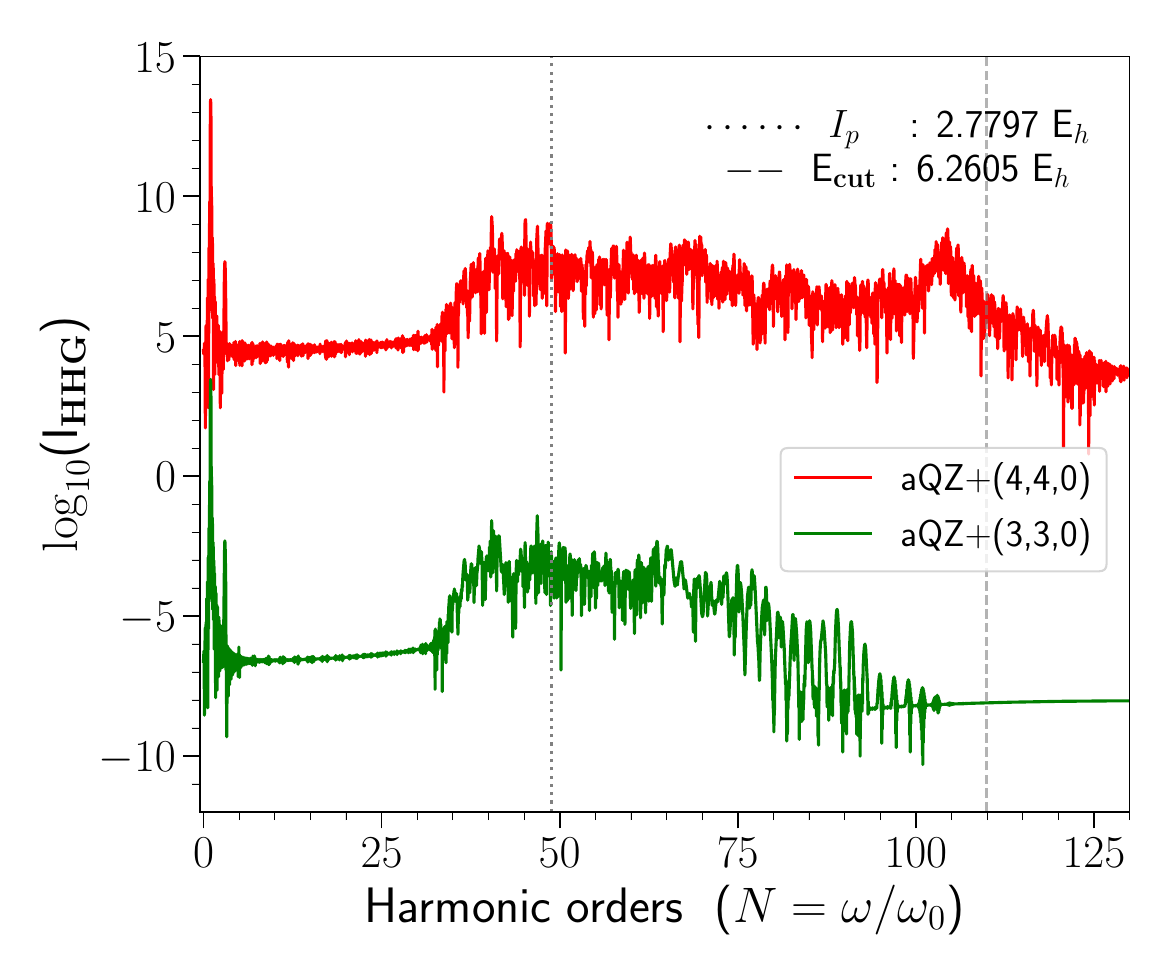}
    \includegraphics[width=.49\textwidth]{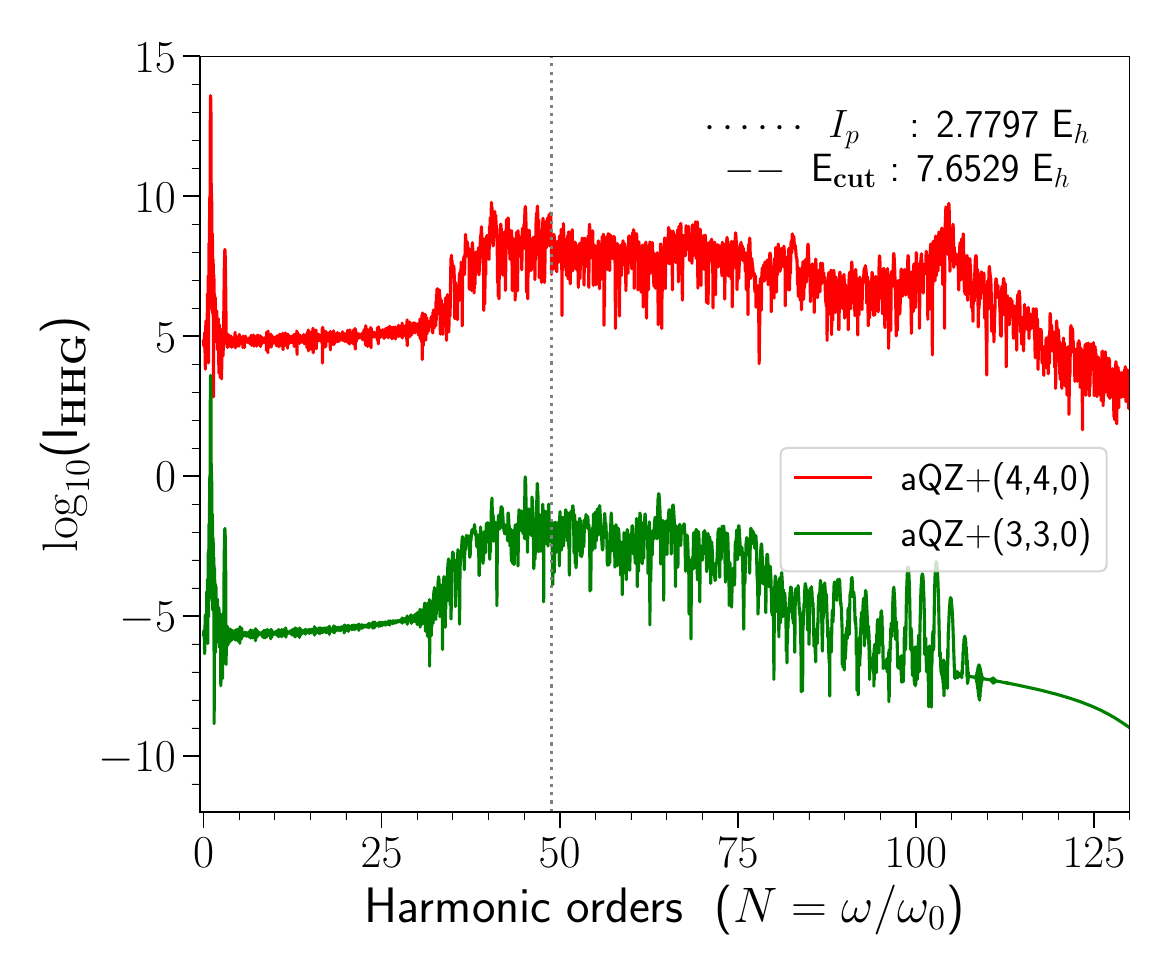}
    \includegraphics[width=.49\textwidth]{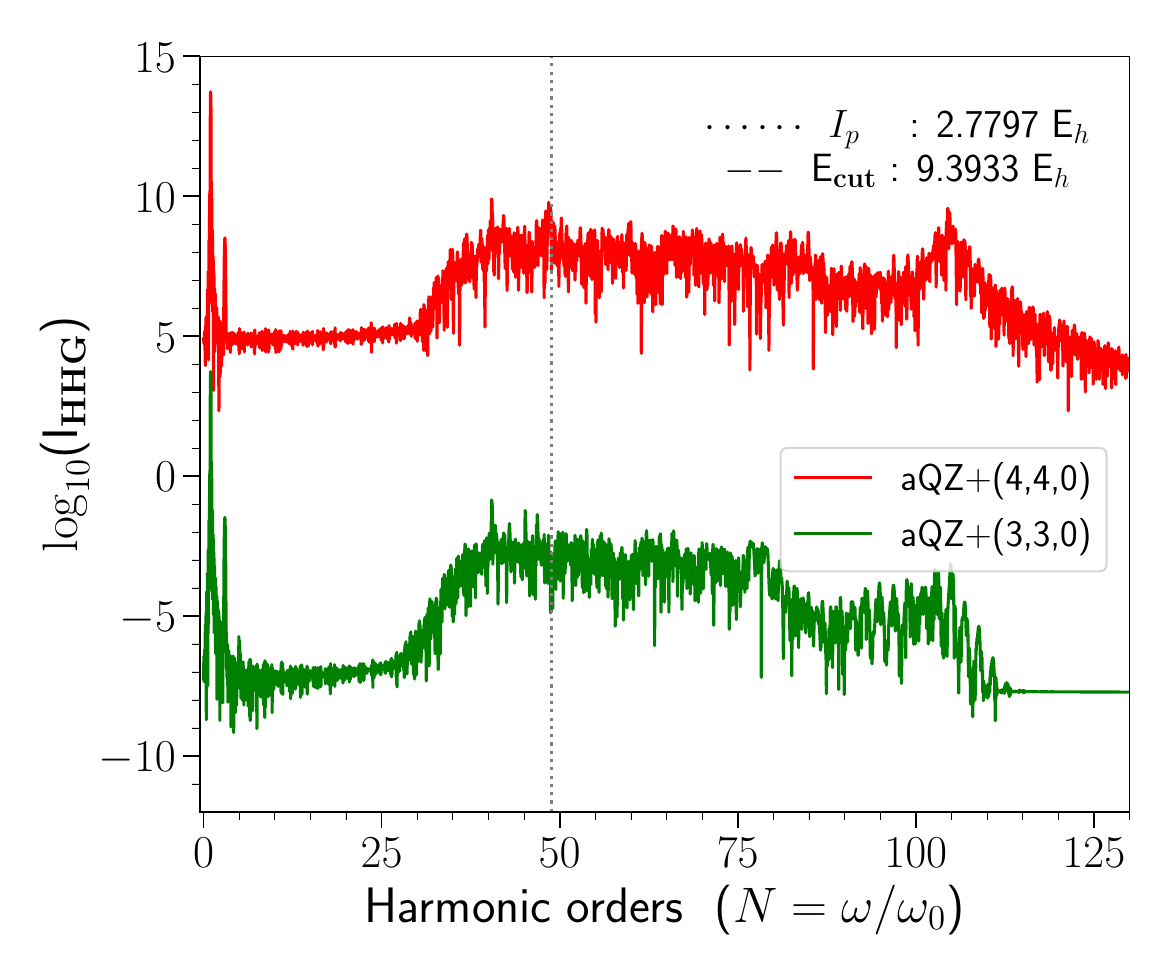}
    \caption{Comparison of HHG spectra of Lithium cation (Li$^{+}$) generated by driving laser pulses of peak intensities $\text{I}_{0} = \{5,7,9.5\}\times 10^{14}$ W/cm$^{2}$,calculated using TD-CIS with different aQZ+(N,$l$,0) basis sets. Here, spectra calculated with different basis sets are upshifted in multiples of +10 for clarity.}
    \label{fig:li_plus_nn0_hhg_comp}
\end{figure}